\documentclass[prb,twocolumn,secnumarabic,showpacs,amsmath,amssymb,floatfix,superscriptaddress,longbibliography,aps]{revtex4-1}

\usepackage{graphicx,color}
\usepackage{siunitx}
\usepackage{amsmath,amsfonts,amsthm,bm} 
\DeclareMathOperator{\D}{\,d\!}

\newcommand{\fcs}{Fe$_{1-x}$Co$_{x}$Si}
\newcommand{\cso}{Cu$_{2}$OSeO$_{3}$}
\newcommand{\czm}{Co-Zn-Mn}

\newcommand{\ozz}{$\langle100\rangle$}
\newcommand{\ooz}{$\langle110\rangle$}
\newcommand{\ooo}{$\langle111\rangle$}

\begin{document}

\title{Thermodynamic evidence of a second skyrmion lattice phase and tilted conical phase in {\cso}}

\author{M. Halder}
\affiliation{Physik Department, Technische Universit\"at M\"unchen, D-85748 Garching, Germany}

\author{A. Chacon}
\affiliation{Physik Department, Technische Universit\"at M\"unchen, D-85748 Garching, Germany}

\author{A. Bauer}
\affiliation{Physik Department, Technische Universit\"at M\"unchen, D-85748 Garching, Germany}

\author{W. Simeth}
\affiliation{Physik Department, Technische Universit\"at M\"unchen, D-85748 Garching, Germany}

\author{S. M\"uhlbauer}
\affiliation{Heinz Maier-Leibnitz Zentrum (MLZ), Technische Universit\"at M\"unchen, D-85748 Garching, Germany}

\author{H. Berger}
\affiliation{\'Ecole Polytechnique Federale de Lausanne, CH-1015 Lausanne, Switzerland}

\author{L. Heinen}
\affiliation{Institut f\"ur Theoretische Physik, Universit\"at zu K\"oln, D-50937 K\"oln, Germany}

\author{M. Garst}
\affiliation{Institut f\"ur Theoretische Physik, Technische Universit\"at Dresden, D-01062 Dresden, Germany}

\author{A. Rosch}
\affiliation{Institut f\"ur Theoretische Physik, Universit\"at zu K\"oln, D-50937 K\"oln, Germany}

\author{C. Pfleiderer}
\affiliation{Physik Department, Technische Universit\"at M\"unchen, D-85748 Garching, Germany}

\date{\today}


\begin{abstract}
Precision measurements of the magnetization and ac susceptibility of {\cso} are reported for magnetic fields along different crystallographic directions, focussing on the border between the conical and the field-polarized state for a magnetic field along the $\langle 100 \rangle$ axis, complemented by selected specific heat data. Clear signatures of the emergence of a second skyrmion phase and a tilted conical phase are observed, as recently identified by means of small-angle neutron scattering. The low-temperature skyrmion phase displays strongly hysteretic phase boundaries, but no dissipative effects. In contrast, the tilted conical phase is accompanied by strong dissipation and higher-harmonic contributions, while the transition fields are essentially nonhysteretic. The formation of the second skyrmion phase and tilted conical phase are found to be insensitive to a vanishing demagnetization factor. A quantitative estimate of the temperature dependence of the magnetocrystalline anisotropy may be consistently inferred from the magnetization and the upper critical field and agrees well with a stabilization of the low-temperature skyrmion phase and tilted conical state by conventional cubic magnetic anisotropies. 
\end{abstract}


\maketitle

\section{Introduction}

In the past decade, an abundance of magnetic materials has been discovered featuring skyrmions, i.e., topologically nontrivial spin textures.\cite{2013:Nagaosa:NN} An incomplete list of bulk systems includes cubic chiral magnets such as the B20 transition metal compounds \cite{2009:Muhlbauer:Science, 2010:Munzer:PhysRevB, 2011:Yu:NatureMater, Kanazawa:2016fd}, rhombohedral lacunar spinels \cite{2015:Kezsmarki:NM,2017:Bordacs:SR}, hexagonal M-type ferrites \cite{2011:Yu:PNAS}, perovskites \cite{2011:Ishiwata:PRB}, and tetragonal Heusler compounds.\cite{2017:Nayak:Nature} The formation of skyrmions has also been reported in epitaxially thin films of these bulk systems, as well as nanowhiskers and nanoparticles.\cite{2013:Yu:NanoLett,2017:Meynell:PRB} Further, tailored magnetic bubbles featuring skyrmion characteristics have been detected in a wide range of heterostructures \cite{2017:Fert:NatRevMat,2017:Jiang:PhysRep}, as well as in carefully selected atomically thin films.\cite{2016:Wiesendanger:NatRevMat}

While the microscopic interactions across this exceptionally wide range of materials are inherently different, all of the known systems were believed to exhibit a single temperature and field regime of the skyrmion lattice order. Thus, even though different mechanisms may allow to stabilize skyrmions, there has been no example in which two different mechanisms are sufficiently strong to cause the formation of skyrmion lattice order in the same material in disconnected temperature and field regimes. 

A first example for two such independent skyrmion phases was recently reported in {\cso}.\cite{2018:Chacon:NatPhys} Using small-angle neutron scattering, two disconnected phases could be identified driven by different mechanisms. On the one hand, {\cso} displays a well-known skyrmion phase near the onset of helimagnetic order in a small applied magnetic field.\cite{2012:Seki:Science,2012:Adams:PhysRevLett} In the following this state will be referred to as high-temperature skyrmion (HTS) phase. As a key characteristic, the temperature versus field range of the HTS phase varies only very weakly under changes of field direction. Phenomenological considerations as well as Monte-Carlo calculations establish that the HTS phase is stabilized by entropic contributions of thermal Gaussian fluctuations.\cite{2009:Muhlbauer:Science,2013:Buhrandt:PRB}

In addition, a second skyrmion phase was identified in {\cso} at the border between the conical phase and the field polarized phase in the low-temperature limit. In the following this state will be referred to as low-temperature skyrmion (LTS) phase. Experimentally, the LTS phase differs from the HTS phase in two distinct ways. First, as it stabilizes at low temperatures it does not require the effects of thermal fluctuations. Second, it exists for magnetic fields close to the crystallographic {\ozz} axis only but not for fields along {\ooz} and {\ooo}. In addition, the LTS phase is accompanied by 
another new phase in which the propagation direction of the conical state under increasing field strength increasingly tilts away from the field direction. 

The magnetic phase diagram of {\cso} including the two new phases may be fully accounted for in terms of the theoretical Ginzburg-Landau theory developed for the class of cubic chiral magnets (cf. Refs.\onlinecite{2009:Muhlbauer:Science, 2010:Munzer:PhysRevB, 2016:Bauer:Book, 2018:Chacon:NatPhys} and references therein). At the heart of this model is the notion of a hierarchical set of energy scales, comprising in decreasing strength of the ferromagnetic exchange,  Dzyaloshinsky-Moriya spin-orbit terms, dipolar interactions, and higher order crystal-field contributions \cite{Landau}. Even though {\cso} exhibits ferri- instead of ferromagnetic order with a long-wavelength modulation driven by DM interactions, this Ginzburg-Landau model appears to be in excellent agreement with experiment. Indeed, taking into account conventional cubic magnetic anisotropies of a fairly strong $\langle100\rangle$ easy magnetic axis, the LTS and tilted conical phase may be explained in excellent qualitative and semi-quantitative agreement with experiment.\cite{2018:Chacon:NatPhys}

The search for several topologically non-trivial phases in the cubic chiral magnets has a long history. For instance, data recorded in FeGe were interpreted in terms of complex mesophases. \cite{2006:Rossler:Nature,wilhelm:PRL:11,2013:Cevey:pssb} In contrast, comprehensive measurements in MnSi unambiguously established a single skyrmion phase.\cite{2012:Bauer:PRB} This motivated to revisit FeGe, where an analysis using the same set of criteria as established in MnSi provided a phase diagram identical to that observed in MnSi.\cite{2016:Bauer:Book} Further, a splitting of the skyrmion phase reported  in {\cso} under Zn doping could be attributed to chemical phase segregation.  \cite{2015:Wu:SciRep,Stefanici:arXiv}

A strong dependence on the temperature and field history of the HTS phase was found in cubic chiral magnets subject to disorder, notably {\fcs} \cite{2010:Munzer:PhysRevB,2016:Bauer:PRB} and the series of {\czm} alloys.\cite{2015:Tokunaga:NatCommun,2016:Karube:NatMater} These have been exploited to obtain detailed information on the processes of nucleation and decay of skyrmions.\cite{2013:Milde:Science,2017:Poellath:PRL} Following the discovery of two phases in {\cso}, a similar observation was reported in Co$_7$Zn$_7$Mn$_6$ with a high-temperature phase as observed in {\cso} and the B20 systems and a three-dimensionally disordered skyrmion phase at low temperatures.\cite{2018:Karube:SciAdv} However, in contrast to {\cso}, the stability of the low-temperature skyrmion phase is attributed to the interplay of DM interactions with the effects of frustration as opposed to magnetocrystalline anisotropies. 

As the observations of the LTS and tilted conical phases in {\cso} are so far based on small-angle neutron scattering (SANS), several pressing questions exist concerning the thermodynamic signatures and the sensitivity to the temperature and field history addressed in this paper. Beginning with an account of the experimental methods in section \ref{methods}, our paper proceeds in sections \ref{designations} and \ref{diagrams} with an account of the designations of the phase transitions and the magnetic phase diagrams, respectively. Typical magnetization and ac susceptibility data as recorded under zero-field cooling (ZFC) and high-field cooling (HFC) are presented in sections \ref{zfc-data} and \ref{hfc-data}, respectively. This is followed by a microscopic justification of the transition fields in section \ref{sans}, where the intensity patterns observed in neutron scattering are compared with the magnetization and susceptibility. Evidence underscoring the presence of dissipation and hysteresis is presented in section \ref{harmonics} followed by information on different sample shapes and thus demagnetizing fields in section \ref{demag}, which are compared to calculations in a Ginzburg-Landau model. Finally, a comparison of the magnetic field dependence of the specific heat and magnetization at low temperatures is presented in section \ref{specific_heat}. The discussion of the experimental data in section \ref{discussion} begins with an account of the anisotropies observed for different crystallographic orientations in section \ref{anisotropy}. The second part of the discussion presented in section \ref{energy} addresses the temperature dependence of the anisotropy energy. The field dependence of the magnetic anisotropy as reflected in the magnetization is finally discussed in section \ref{potential}. The paper closes with a brief summary in section \ref{conclusions}. 

\section{Experimental Methods}
\label{methods}

Several {\cso} single crystals were prepared from ingots grown by vapor transport. Numerous studies on samples from the same ingots have been reported in the literature.\cite{2015:Schwarze:NatMater,2016:Milde:NanoLett,2016:Zhang:NanoLett,2016:Zhang:PhysRevB,2017:Poellath:PRL,2017:Stasinopoulos:APL} This concerns in particular the small-angle neutron scattering reported in Refs.\,\cite{2012:Adams:PhysRevLett,2018:Chacon:NatPhys} as summarized below.  All single crystals investigated here displayed the same high sample quality in terms of their optical appearance, the crystallinity as inferred from the lattice mosaic spread and sharpness of diffraction spots observed in x-ray and neutron diffraction, as well as their physical properties, for instance, the value of the paramagnetic-to-helimagnetic transition temperature. In our study we focussed on five specimens, all of which were of cuboid shape. The  designations, dimensions, orientations, and demagnetizing factors of these samples are summarised in table\,\ref{table:samples}. 

\begin{table}[t!]
\vspace{-1mm}
\caption{\label{table:samples}
Designation, field direction, dimensions ($a$, $b$, and $c$), and demagnetization factor $N$ of the {\cso} single-crystal samples investigated as part of this study. The magnetic field was applied along the direction denoted $a$. The value of $N$ corresponds to this field direction. Samples are listed twice when measured for different directions of the field as stated in the table.
}
\vspace{2mm}
\begin{tabular}[t]{llllc}
\hline\hline
sample & $B \parallel $ \hspace{0mm} & $a$  $\times$  $b$ $\times$  $c$ ($\rm {mm}^3$) \hspace{1mm} & $N$ \hspace{3mm} &  data in Figs. \\ 
\hline\hline
VTG1-19  & $\langle 100 \rangle$ & $1.87 \times 1.28 \times 1.53$ & 0.28 & \ref{fig:phasediagrams} to \ref{work_temp},\ref{fig:magnetic_work} \\
VTG1-20 & $\langle 110 \rangle$ & $1.76 \times 1.27 \times 1.74$ & 0.29 & \ref{fig:phasediagrams},\ref{fig:zfc},\ref{fig:hfc},\ref{fig:harmonics},\ref{fig:cu2seo4:heat_capacity},\ref{fig:materials},\ref{work_temp},\ref{fig:magnetic_work} \\
VTG1-20 & $\langle 111 \rangle$ & $1.74 \times 1.27 \times 1.76$ & 0.30 & \ref{fig:phasediagrams},\ref{fig:zfc},\ref{fig:hfc},\ref{fig:cu2seo4:heat_capacity},\ref{fig:materials},\ref{work_temp},\ref{fig:magnetic_work} \\
\hline
VTG1-10-1 & $\langle 100 \rangle$ & $3.00 \times 0.50 \times 0.50$ & 0.07 & \ref{fig:demag} and \ref{fig:demag-N} \\
VTG1-18-2 & $\langle 100 \rangle$ & $1.85 \times 0.86 \times 0.78$ & 0.18 & \ref{fig:demag} and \ref{fig:demag-N} \\
VTG1-18-2 & $\langle 100 \rangle$ & $0.86 \times 1.85 \times 0.78$ & 0.39 & \ref{fig:demag} and \ref{fig:demag-N} \\
VTG1-18-3 & $\langle 100 \rangle$ & $0.15 \times 0.85 \times 1.88$ & 0.77 & \ref{fig:demag} and \ref{fig:demag-N} \\
\hline\hline
\end{tabular}
\end{table}

Essentially all data for the field along $\langle100\rangle$ reported in this paper were recorded in sample VTG1-19, which was closest to a cubic shape with a demagnetization factor close to $\sim\!1/3$. This way the sample shape and the associated demagnetization effects were close to the spherical sample studied in SANS \cite{2018:Chacon:NatPhys} making the differences of the corrections of demagnetizing fields tiny. For the same reasons data for different crystallographic orientations were recorded in sample VTG1-20, which was also close to cubic. All other samples served to study the effects of different demagnetization factors under a field along $\langle100\rangle$. 

The magnetization, ac susceptibility, and specific heat were measured in a Quantum Design physical properties measurement system (PPMS). The magnetization was determined with an extraction technique. All ac susceptibility measurements were performed at an excitation frequency $f = 911\,\mathrm{Hz}$ and an excitation field of $H_{ac}=0.1\,\mathrm{mT}$. It is important to note that the susceptibilities presented in our paper essentially represent the coefficients of a Fourier expansion in response to the excitation frequency. Technically speaking, these coefficients are typically referred to as \textit{harmonic} susceptibilities. In comparison, expanding the response to a magnetic field in a Taylor series, the expansion coefficients are referred to as \textit{nonlinear} susceptibilities. In principle, it is possible to convert the harmonic into the nonlinear susceptibilities as discussed in Sec.\ref{harmonics} and Ref.\,\onlinecite{mydosh:1993}. For the comparison of the magnetization and ac susceptibility with small-angle neutron scattering a data set was recorded at the beam-line SANS-1 \cite{muhlbauer:NIaMiPRSAASDaAE:16} at MLZ following the technical procedure described in Ref.\,\onlinecite{2018:Chacon:NatPhys}.

The experimental data are either shown as a function of applied field, without correction of demagnetizing effects, or as a function of internal field, where the effects of demagnetizing fields were corrected by means of an analytical expression as reported in Ref.\,\onlinecite{1998:Aharoni:JApplPhys} taking into account the cuboid sample shape. For those figures in which data are shown as a function of internal field the susceptibility calculated from the magnetization, ${\rm d}M/{\rm d}H$, as well as the real and imaginary parts of the ac susceptibility, $\chi'$ and $\chi''$, are consistently presented with respect to the internal field. That is, ${\rm d}M/{\rm d}H$ was computed with respect to the internal field, whereas measured values of $\chi'$ and $\chi''$ were converted following convention (for details see e.g. Ref.\onlinecite{2000:youssif}). 

The heat capacity was measured by means of a conventional heat pulse method using the standard PPMS set up. Data were recorded in a sequence of increasing field values, where typical heat pulses generated a temperature increase of $\sim\!1\,{\%}$. All specific heat data shown in this paper represent the average over 15 measurements recorded at the same unchanged applied field and reference temperature. 

For the scientific questions addressed in our study, the temperature and field history prove to be very important. The protocols used in our study correspond to those used in the small-angle neutron scattering study reported in Ref.\,\onlinecite{2018:Chacon:NatPhys}. However, whereas data in the SANS study were dominantly recorded as a function of temperature, essentially all of the data reported here were recorded as a function of field. 

Focusing on the formation of two new phases at low temperatures, we have confirmed that our experimental data in {\cso} may be classified in terms of data recorded under increasing and decreasing magnitude of the magnetic field denoted ``up" and ``down", respectively. In addition, we distinguish two different temperature versus field protocols, namely zero-field cooling (ZFC) and high-field cooling (HFC). For ZFC, the sample was cooled from a starting temperature well above $T_c$ in zero magnetic field with a cooling rate of $\sim\!10\,\mathrm{K\,min^{-1}}$ to the desired temperature of the field sweep. It is important to note that for the pristine helical state created this way the domain populations for different $\langle100\rangle$ directions are the same. Subsequently, data were recorded while the magnitude of the magnetic field was increased in steps of $0.5\,\mathrm{mT}$.
 
For HFC, the sample was cooled from a starting temperature well above $T_c$ in a large negative applied field of $-140\,\mathrm{mT}$, i.e., below $-H_{c2}$, with a cooling rate $\sim 10\,\mathrm{K\,min^{-1}}$ until the desired temperature of the field sweep was reached. Subsequently, data were recorded in a single field sweep in which the magnetic field was stepped from the field value at which the sample was cooled up to positive values exceeding $+H_{c2}$. Note that this field sweep comprises data under decreasing magnitude of the magnetic field (down) in the range $-H_{c2} < H < 0$ and data under increasing magnitude of the magnetic field (up) in the range $0 < H < +H_{c2}$, respectively. For fields exceeding $H_{c1}$, we find accurately the same behavior under increasing magnitude of the magnetic field no matter whether the sweep was initiated following ZFC or HFC. Discrepancies in the range $0 < H \leq +H_{c1}$ may be attributed to different domain populations in the helical state, notably for field along $\langle 100\rangle$, i.e., the easy magnetic axis, only domains along the field direction are populated after HFC. In the following, such differences below $H_{c1}$ will be pointed out where relevant.


\section{Results}
\label{results}

The presentation of the experimental results begins in section \ref{designations} with a summary of the designations and terminology used to describe the transitions of the different magnetic phases. Next, the magnetic phase diagrams are summarized in section \ref{diagrams} before typical data used to infer these phase diagrams are presented in section \ref{data}. Higher-harmonic contributions in the ac susceptibility, providing first hand information on the presence and nature of hysteresis, are reported in section \ref{harmonics}. The presentation of the experimental results continues in section \ref{demag} with data for different sample shapes and demagnetizing fields and closes in section \ref{specific_heat}, where typical specific heat data are shown.

\subsection{Designations and terminology}
\label{designations}

The temperature versus field diagram of {\cso} as determined in SANS displays five different noncollinear magnetic phases. In addition, the paramagnetic state at high temperatures and low fields and the field-polarized (ferromagnetic) state at low temperatures and high fields may be distinguished. This implies considerable complexities when inferring the magnetic phase diagram from bulk properties such as the magnetization, ac susceptibility and specific heat. In turn, it proves to be helpful to start with the designations and terminology used to address the phase boundaries before presenting the experimental data.

Among the five noncollinear magnetic phases three are well known, namely, (i) the helical order, (ii) the conical order, and (iii) the high-temperature skyrmion (HTS) phase at the border between the paramagnetic phase and the conical state. Recent SANS studies identified in addition (iv) the tilted conical (TC) state and (v) a low-temperature skyrmion (LTS) phase at the border between the conical and field-polarized state.  

Accordingly, we distinguish in the following eight transition fields as summarized in Table\,\ref{table:transition fields}. The boundaries between the helical and the conical, and the conical and field-polarized phases are denoted $H_{\rm c1}$ and $H_{\rm c2}$, respectively. The boundaries of the high-temperature skyrmion phase are denoted $H_{\rm a1}$ and $H_{\rm a2}$. Likewise, the boundaries of the low-temperature skyrmion phase and the tilted conical phase, are denoted $H_{\rm s1}$/$H_{\rm s2}$ and $H_{\rm t1}$/$H_{\rm t2}$, respectively. Values determined under increasing (up) and decreasing (down) magnetic field are labeled by the superscript ``u" and ``d", respectively. We note that for the high-temperature and low-temperature skyrmion phases as well as the tilted conical phase the subscripts 1 or 2 denote boundaries at lower and higher absolute field values, respectively. 

\begin{table}[t]
\caption{\label{table:transition fields}
Definitions of the transition fields between the different phases observed in our study. The labels ``u" and ``d" denote increasing (up) and decreasing (down) direction of the field sweep, respectively. The subscripts ``1" and ``2" denote values at lower and higher absolute field values, respectively.
}
\vspace{2mm}
\begin{tabular}[t]{ll c}
\hline\hline
fields \hspace{6mm} & sweep dir. \hspace{3mm}  & phase boundary \\
\hline\hline
\vspace{1mm}
$H_{\rm c1}^{\rm u}$, $H_{\rm c1}^{\rm d}$ & up, down & helical to conical\\
\vspace{1mm}
$H_{\rm c2}^{\rm u}$, $H_{\rm c2}^{\rm d}$ & up, down & conical to field-polarized \\
\vspace{1mm}
$H_{\rm a1}^{\rm u}$, $H_{\rm a1}^{\rm d}$ & up, down & high-temp. skyrmion, low field \\
\vspace{1mm}
$H_{\rm a2}^{\rm u}$, $H_{\rm a2}^{\rm d}$ & up, down & high-temp. skyrmion, high field \\
\hline
\vspace{1mm}
$H_{\rm s1}^{\rm u}$, $H_{\rm s1}^{\rm d}$ & up, down & low-temp. skyrmion, low field \\
\vspace{1mm}
$H_{\rm s2}^{\rm u}$, $H_{\rm s2}^{\rm d}$ & up, down & low-temp. skyrmion, high field \\
\vspace{1mm}
$H_{\rm t1}^{\rm u}$, $H_{\rm t1}^{\rm d}$ & up, down & tilted conical, low field \\
\vspace{1mm}
$H_{\rm t2}^{\rm u}$, $H_{\rm t2}^{\rm d}$ & up, down & tilted conical, high field \\
\hline\hline
\end{tabular}
\end{table}


\subsection{Magnetic phase diagrams}
\label{diagrams}

\begin{figure}[ht]
\includegraphics[width=\linewidth]{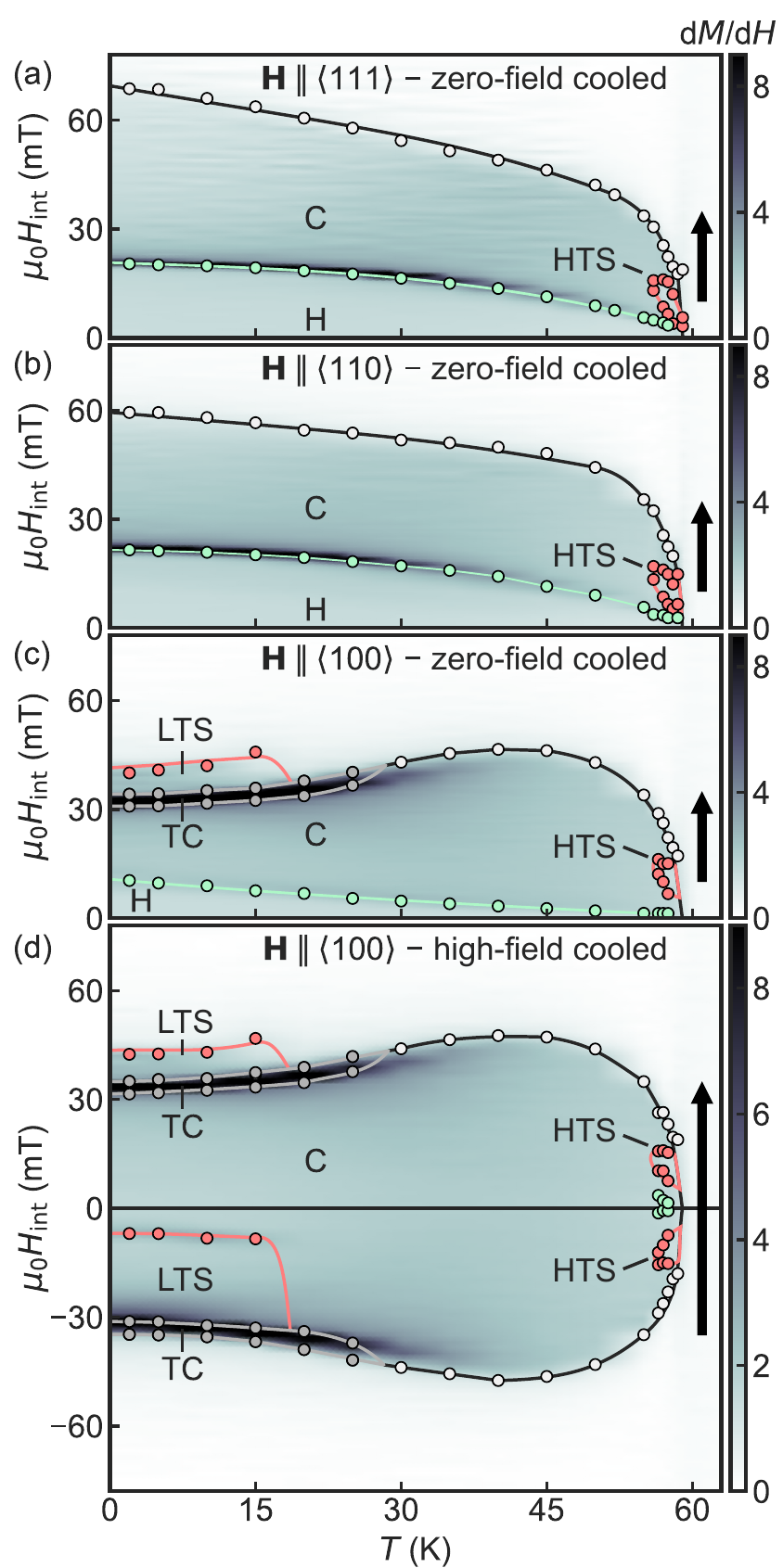}
\caption{\label{fig:phasediagrams}
Magnetic phase diagrams inferred from magnetization and susceptibility data measured in field sweeps after zero-field cooling (a)--(c) and high-field cooling (d) with fields along $\langle 111 \rangle$ (a), $\langle 110 \rangle$ (b), and $\langle 100 \rangle$ (c), (d). The following phases may be distinguished: helical (H), conical (C), field polarized (FP), tilted conical (TC), high-temperature skyrmion (HTS) and low-temperature skyrmion phase (LTS).
  }
\end{figure}

Shown in Fig.~\ref{fig:phasediagrams} are the magnetic phase diagrams for field along the major crystallographic directions, namely $\langle 111\rangle$, $\langle 110\rangle$, and $\langle 100\rangle$. All diagrams are shown as a function of internal magnetic field, following correction of demagnetizing fields. The phase boundaries, as marked here by circles, were inferred from the susceptibility data, where the definitions for the transitions are given below. The color coding shown in the background of the phase diagrams reflects the size of the susceptibility, $\mathrm{d}M/\mathrm{d}H$, calculated from the magnetization. The temperature versus field protocols used, namely zero-field cooled and high-field cooled, are stated in each panel, where the direction of the field sweeps is marked by an arrow.

For magnetic fields along $\langle 111\rangle$ and $\langle 110\rangle$, shown in Figs.~\ref{fig:phasediagrams}(a) and \ref{fig:phasediagrams}(b), the behavior reported in the literature is observed.\cite{2012:Adams:PhysRevLett} Here the phase diagrams comprise helical (H), conical (C), and HTS lattice order. With decreasing temperature both $H_{\rm c1}$ and $H_{\rm c2}$ increase monotonically. Only the ZFC phase diagrams are shown, since $H_{\rm c1}$ and $H_{\rm c2}$ exhibit very little and no hysteresis between increasing and decreasing field, respectively. It is also helpful to note that the magnetization displays a clear signature of the transition from the conical to the helical state under decreasing field (not shown). Taken together, the magnetic phase diagrams display all the characteristics that are well known from stoichiometric cubic chiral magnets, such as MnSi and FeGe.\cite{2016:Bauer:Book}

The phase diagrams for a field parallel to $\langle 100\rangle$, as shown in Figs.~\ref{fig:phasediagrams}(c) and \ref{fig:phasediagrams}(d), display the following similarities with $\langle 111\rangle$ and $\langle 110\rangle$. The HTS phase emerges in a temperature and field range, that varies somewhat with crystallographic orientation.\cite{2012:Adams:PhysRevLett} Further, the phase boundary between the conical and field-polarized state at $H_{\rm c2}$ increases with decreasing temperature down to about 40\,K. Also, under ZFC, the phase boundary between the helical and the conical state at $H_{\rm c1}$ increases, albeit the absolute value is smaller than for a field along $\langle 111\rangle$ and $\langle 110\rangle$.

In contrast to the behavior for field parallel to $\langle 111\rangle$ and $\langle 110\rangle$, $H_{c2}$ for a field parallel to $\langle 100\rangle$ reaches the highest value of $\sim\!40\,{\rm mT}$ around 40\,K followed by a gentle decrease. Along with this decrease in $H_{\rm c2}$, clear signatures in the magnetization of two new phases, identified microscopically in neutron scattering \cite{2018:Chacon:NatPhys}, may be distinguished. Namely, below $\sim\!\SI{30}{\kelvin}$ the tilted conical (TC) phase emerges. This is followed by the LTS phase, which emerges below $\sim \SI{15}{\kelvin}$. 
The absence of the tilted conical phase and LTS phase for field directions other than $\langle 100\rangle$ provides compelling evidence that magnetocrystalline anisotropies must play a decisive role in their formation. 

We note that for all temperature versus field protocols, our data are consistent with the tilted conical state appearing first, followed by the LTS phase as described in the SANS study reported in Ref.\,\onlinecite{2018:Chacon:NatPhys}. Moreover, once the tilted conical state and LTS phase have formed, the tilted conical phase disappears before the LTS phase. The precise temperature and field range are thereby subject to the temperature versus field protocol. Namely, under increasing fields [positive field values in Figs.~\ref{fig:phasediagrams}(c), \ref{fig:phasediagrams}(d)] the LTS phase extents to larger field values, whereas for decreasing fields [negative field values in Fig.~\ref{fig:phasediagrams}(d)] it persists down to lower magnetic field values. 

Another difference of field along the $\langle 100\rangle$ direction as compared with $\langle 111\rangle$ and $\langle 110\rangle$ concerns, finally, the conical to helical transition at $H_{\rm c1}$. While a clear signature is observed after ZFC under increasing field, no evidence suggesting the existence of $H_{\rm c1}$ is observed under HFC. This originates in differences of domain populations as explained above and was previously observed in other cubic chiral magnets.\cite{2016:Bauer:PRB,2017:Bauer:PRB} Namely, in this configuration, one easy axis is parallel to the magnetic field while the other easy axes are perpendicular to the field direction. Therefore, when the magnetic field is reduced from the conical to the helical state, only the domain population parallel to the field is populated leading to a single domain helical state that is indistinguishable from the conical state. 


\subsection{Experimental data}
\label{data}

The presentation of the experimental data is organized in three parts. First, typical magnetization and ac susceptibility data for different field orientations recorded under ZFC will be reported in section \ref{zfc-data}. This serves to illustrate the key features and associated definitions of the different phase transitions. It is followed by related data recorded under HFC in section \ref{hfc-data}, emphasizing the role of the temperature versus field protocol. Finally, selected susceptibility data are compared with neutron scattering data to justify the interpretation and definitions of the transition fields in section \ref{sans}.

\subsubsection{Magnetization and susceptibility under ZFC}
\label{zfc-data}

\begin{figure}[t]
\centering
\includegraphics[width=1.0\linewidth]{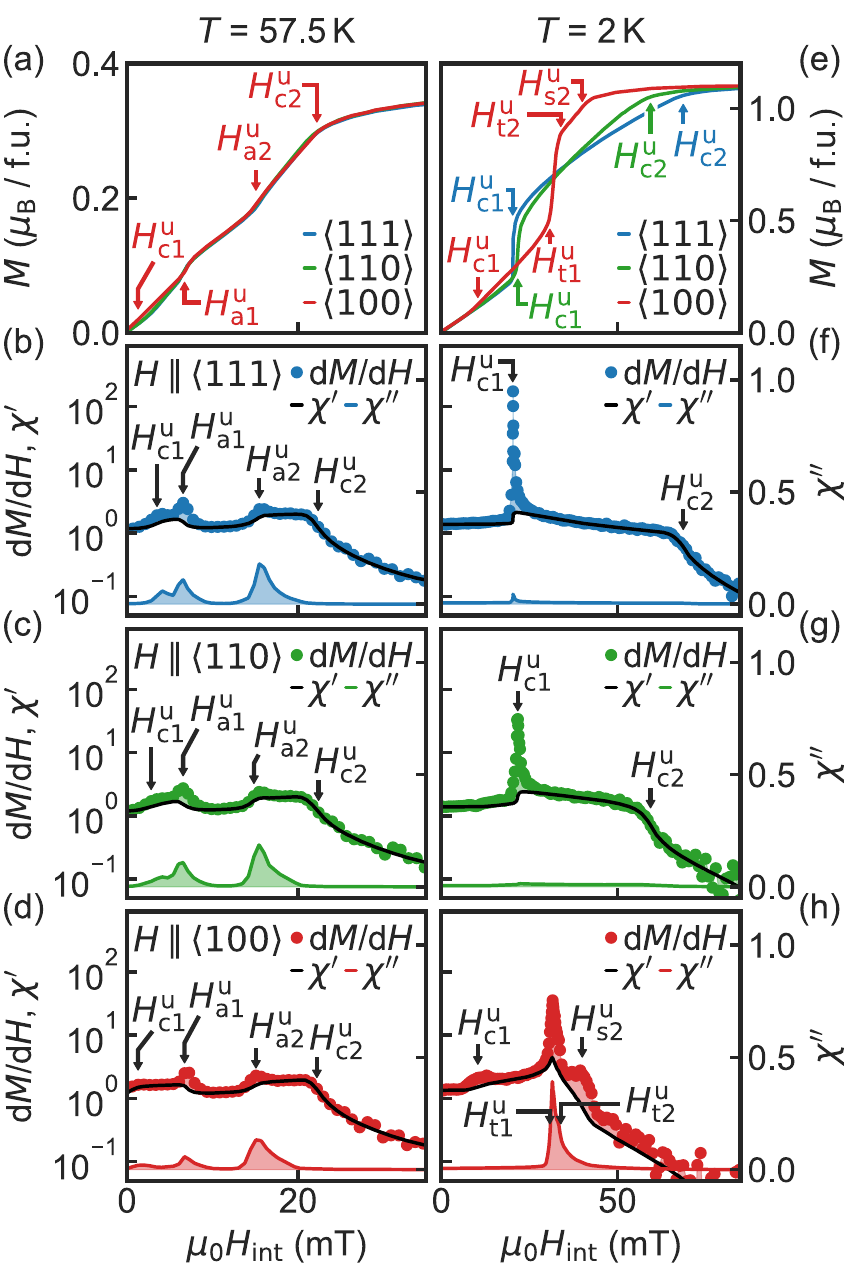}
\caption{\label{fig:zfc}
Typical magnetization and susceptibility data as a function of internal field following zero-field cooling with fields aligned along $\langle 111 \rangle$ (blue), $\langle 110 \rangle$ (green), and $\langle 100 \rangle$ (red). Data were recorded at $T=57.5\,\mathrm{K}$ (left column) and $T=2\,\mathrm{K}$ (right column). [(a) and (e)]~Magnetization data. [(b)--(d) and (f)--(h)]~Susceptibility calculated from the magnetization, $\mathrm{d}M/\mathrm{d}H$ (symbols), the real part of ac susceptibility $\chi'$ (black line) and the imaginary part of ac susceptibility $\chi''$ (colored line) for a field along $\langle 111 \rangle$ [second row: (b), (f)], field along $\langle 110 \rangle$ [third row: (c), (g)] and a field along $\langle 100 \rangle$ [fourth row: (d), (h)].
}
\end{figure}

Shown in Figs.~\ref{fig:zfc}\,(a)--\ref{fig:zfc}\,(d) are typical data of the magnetization, $M$, the susceptibility calculated from the magnetization, ${\rm d}M/{\rm d}H$, as well as the real and imaginary parts of the ac susceptibility, $\chi'$ and $\chi''$, respectively. All quantities are determined with respect to the internal field as described above. Note that the real part of the susceptibility is presented on a logarithmic scale, whereas the imaginary part is presented on a linear scale. Data are shown as a function of internal magnetic field, $H_{\rm int}$, across the high-temperature skyrmion phase for a high temperature of 57.5\,K  after ZFC, where the helimagnetic transition temperature is $T_{c}=58.5\,{\rm K}$. For fields along $\langle 111 \rangle$, $\langle 110 \rangle$, and $\langle 100 \rangle$ data are presented in blue, green, and red, respectively. In perfect agreement with the literature, the magnetization increases at low fields quasilinearly, with a distinct change of slope at $H_{\rm c2}$. Small changes of slope at low and intermediate fields are related to the helical-to-conical transition and the HTS phase, respectively.

The detailed features associated with the transitions between the different phases may be seen in $\mathrm{d}M/\mathrm{d}H$ shown in Figs.~\ref{fig:zfc}\,(b)--\ref{fig:zfc}\,(d), depicted as discrete data points. Related features are also reflected in the ac susceptibility, $\chi'$, shown as a line. With increasing field, three distinct maxima may be distinguished corresponding to the helical-to-conical transition at $H_\mathrm{c1}^\mathrm{u}$, the conical-to-skyrmion transition at $H_\mathrm{a1}^\mathrm{u}$, and the skyrmion-to-conical transition at $H_\mathrm{a2}^\mathrm{u}$. The change of slope at high fields, denoted as $H_\mathrm{c2}^\mathrm{u}$, marks the transition between the conical and the field-polarized state. 

Quantitatively, the lowest value of $H_\mathrm{c2}^\mathrm{u}$ is observed for $\mathbf{H}\!\parallel\!\langle 100 \rangle$, whereas the largest value is observed for $\mathbf{H}\!\parallel\!\langle 111 \rangle$. In comparison with the conical state, the susceptibility is reduced both in the helical and the HTS phase. Further, the real part of the ac susceptibility, $\chi'$, tracks $\mathrm{d}M/\mathrm{d}H$ except for the transition regions around $H_\mathrm{c1}^\mathrm{u}$, $H_\mathrm{a1}^\mathrm{u}$, and $H_\mathrm{a2}^\mathrm{u}$, suggesting the presence of reorientations of long-range textures and viscous processes occurring on time-scales that are slow as compared with the oscillation period of the ac fields.\cite{2012:Bauer:PRB,2016:Bauer:PRB} The underlying dissipative processes may account also for the presence of nonvanishing contributions to the imaginary part of the ac susceptibility, $\chi''$, at $H_\mathrm{a1}^\mathrm{u}$ and $H_\mathrm{a2}^\mathrm{u}$.

With decreasing temperature, the well-known signatures of the HTS phase disappear, before the characteristics shown for \SI{2}{\kelvin} in Figs.\,\ref{fig:zfc}\,(e)--\ref{fig:zfc}\,(h) emerge. For magnetic field along $\langle 111\rangle$ and $\langle 110\rangle$ two transitions may be distinguished, which may be attributed to the helical to conical transition at $H_{\rm c1}^\mathrm{u}$ and the conical to field-polarized transition at $H_{\rm c2}^\mathrm{u}$. In contrast, for $\langle 100\rangle$, the field dependence is qualitatively different and four transition fields may be discerned in the magnetization and ac susceptibility. As shown below, comparison with SANS identifies these as the helical to conical transition at $H_{\rm c1}^\mathrm{u}$, the transitions of the tilted conical state at $H_{\rm t1}^\mathrm{u}$ and $H_{\rm t2}^\mathrm{u}$, and the high field transition of the LTS phase at $H_{\rm s2}^\mathrm{u}$. As the signatures of the onset of the tilted conical state are rather pronounced, it is not possible to identify a well-defined signature of the onset of the LTS phase at $H_{\rm s1}^\mathrm{u}$.

For magnetic field between $H_{\rm c1}^\mathrm{u}$ and $H_{\rm c2}^\mathrm{u}$ distinctly anisotropic behavior may be observed between two crossing points in the field-dependence of the magnetization. Namely, at low fields the magnetization for all three directions is at first essentially identical. As a side note we remark that the signature in $M$ seen for $\langle 100\rangle$ at $H_{\rm c1}^\mathrm{u}$ is tiny. Above $H_{\rm c1}^\mathrm{u}$ the magnetization for field along $\langle 100 \rangle$ (red curve) is slightly larger. This situation changes at $H_{\rm c1}^\mathrm{u}$ for $\langle 111 \rangle$, where a pronounced jump in the magnetization for $\langle 111 \rangle$ (blue curve) defines the first crossing point. For fields between this crossing point and the onset of the tilted conical phase at $H_{\rm t1}^\mathrm{u}$, the magnetization for $\langle 111 \rangle$ (blue curve) is largest. At $H_{\rm t1}^\mathrm{u}$ the magnetization for field along $\langle 100 \rangle$ (red curve) displays a jump that defines the second crossing point. For field values above this second crossing point the magnetization for field along $\langle 100 \rangle$ (red curve) is again largest. 


Thus, between the two crossing points the easy magnetic axis appears to have changed from $\langle 100 \rangle$ to $\langle 111 \rangle$. Interestingly, closer inspection of the crossing points reveals that the magnetization is not exactly identical for all three directions. However, when calculating the magnetic work $W(M)$ considered below, a sharp crossing point is observed for all three directions. Thus, at the crossing points, the system appears to be perfectly isotropic. We return to  an account of the mechanism driving all of this behavior in Sec.\,\ref{discussion}.

Further, for a field along along $\langle 111\rangle$ and $\langle 110\rangle$ the jump in the magnetization at the helical-to-conical transition at $H_\mathrm{c1}$ results in a clear peak in $\mathrm{d}M/\mathrm{d}H$. This peak is not seen in the real part of the ac susceptibility $\chi'$, suggesting slow domain reorientations as observed in previous studies of MnSi and {\fcs} \cite{2012:Bauer:PRB,2016:Bauer:PRB,2017:Bauer:PRB} and viscous relaxation. Consistent with this evidence for dissipative processes, we find a small amount of hysteresis in $H_{\rm c1}$. In contrast, at $H_\mathrm{c2}$ we find excellent agreement between $\mathrm{d}M/\mathrm{d}H$ and the real part of the ac susceptibility $\chi'$. Also, there is essentially no hysteresis at $H_{\rm c2}$ (cf. heat capacity shown below). 

On a related note, at low temperatures, the imaginary part of the ac susceptibility, $\chi''$, displays a tiny contribution at $H_\mathrm{c1}$ for fields along $\langle 111 \rangle$ and $\langle 110 \rangle$ only. In stark contrast, we find a pronounced contribution in $\chi''$ for a field along $\langle 100 \rangle$ in the regime of the tilted conical phase. Comparison with the SANS data, presented below, reveals that $H_{\rm t1}^\mathrm{u}$ and $H_{\rm t2}^\mathrm{u}$ correspond with the points of inflection in $\chi''$. We note that  $H_{\rm t2}^\mathrm{u}$ is always smaller than $H_{\rm s2}^\mathrm{u}$. Moreover, in contrast to the high-temperature skyrmion phase, where $H_{\rm a1}$ and $H_{\rm a2}$ [cf. Figs.\,\ref{fig:zfc}\,(b)--\ref{fig:zfc}\,(d)] are accompanied by strong dissipation processes causing a large finite value of $\chi''$, no such dissipation is observed for the LTS phase in the frequency range studied. 

The evidence for dissipation may reflect the underlying microscopic processes characteristic of the different magnetic phases and/or their creation. As concerns the tilted conical phase, a change of field strength results in a change of propagation direction. In turn, it seems plausible that the strong dissipation within the tilted conical phases arises from changes of propagation direction with ac field. In contrast, the dissipation at the boundary of the HTS phase reflects processes of creation and decay of skyrmions, which originate in the creation and motion of Bloch points.\cite{2013:Milde:Science,2017:Poellath:PRL} The absence of such dissipation at the phase boundaries of the LTS may be interpreted as putative evidence for a large energy barrier that inhibits the creation and destruction of the skyrmions by means of the ac field. This might also hint at a different nucleation path of the LTS phase as compared to the HTS phase, notably by virtue of the tilted conical phase as an intermediate step.

More generally, the absence or presence of dissipation might be used to distinguish between two different scenarios for first-order transitions. The presence of dissipation indicates the coexistence of two phases and slow dynamics of phase boundaries driven by oscillating fields.  The absence of dissipation is expected both for second- and for those first-order transitions, where phase coexistence is suppressed and an oscillating field is not able to induce oscillations in the volume occupied by each phase. This may occur if there is a transition from a high-energy metastable state A to a state B with lower free energy. In this case, only transitions from A to B but not from B to A are induced. Thus, an oscillating field will not induce oscillations between the two phases. This picture explains naturally the absence of dissipative effects when leaving the low-temperature skyrmion phase, which is expected to be metastable due to its topological protection.



\subsubsection{Magnetization and susceptibility under HFC}
\label{hfc-data}

\begin{figure}[h]
\centering
\includegraphics[width=\linewidth]{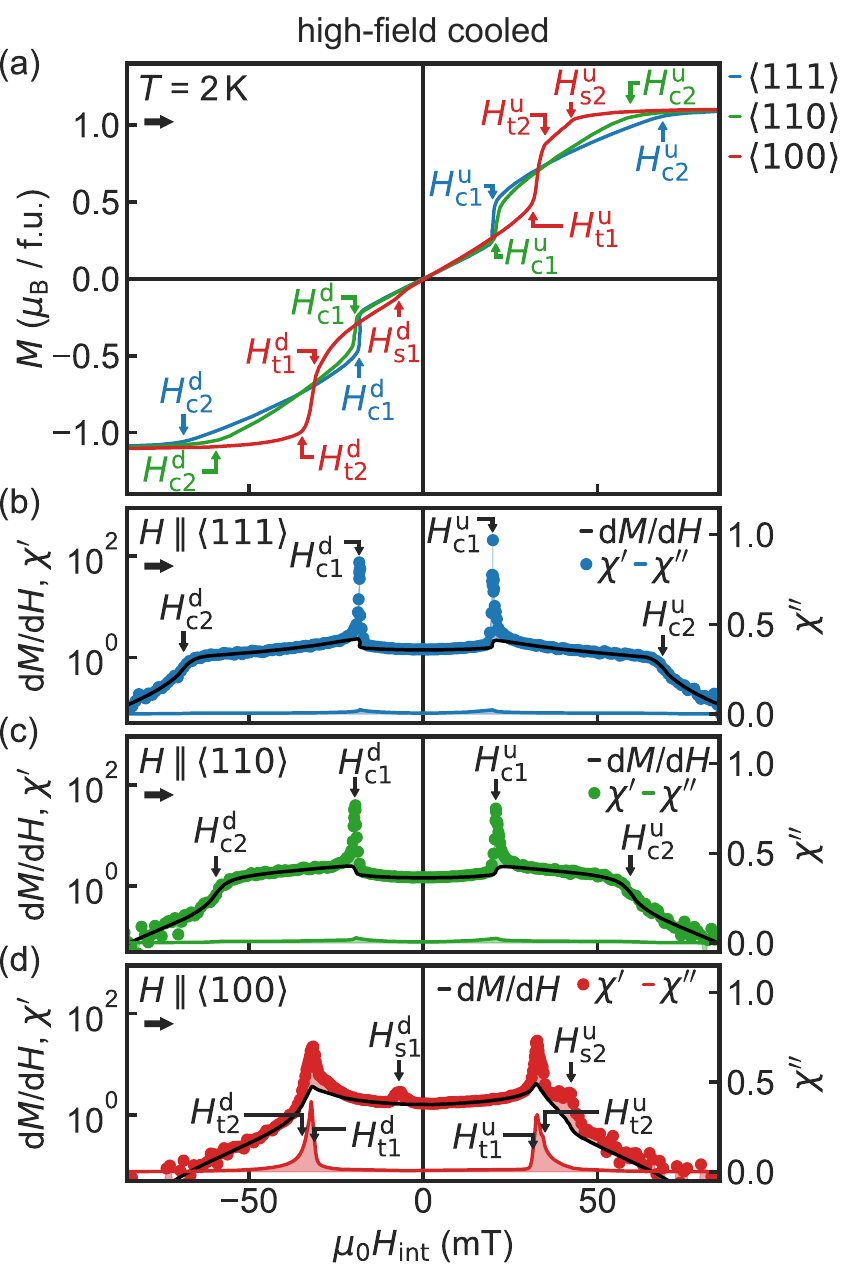}
\caption{\label{fig:hfc}
Typical magnetization and susceptibility data at $T = \SI{2}{\kelvin}$ following high-field cooling. Data were recorded in a field sweep from left to right. Data are shown as a function of internal field along $\langle 111 \rangle$ (blue), $\langle 110 \rangle$ (green), and $\langle 100 \rangle$ (red). (a) Magnetization as a function of magnetic field under HFC. (b)--(d)~Susceptibility calculated from the magnetization, $\mathrm{d}M/\mathrm{d}H$ (dots), the real part of ac susceptibility $\chi'$ (black line), and the imaginary part of ac susceptibility $\chi''$ (colored line) for fields along $\langle 111 \rangle$, $\langle 110 \rangle$, and $\langle 100 \rangle$.
}
\end{figure}

Shown in Figs.~\ref{fig:hfc}(a)--\ref{fig:hfc}(d) are the magnetization and the susceptibilities under HFC for different crystallographic directions as recorded at a temperature of \SI{2}{\kelvin}. Again the real part of the susceptibility is presented on a logarithmic scale, whereas the imaginary part is presented on a linear scale. Data are presented for field sweeps from negative to positive field values as indicated by the black arrows. Thus, data recorded at negative fields (left hand side) were determined under \textit{decreasing} absolute field strength (down) starting at $H<-H_{c2}$, whereas data observed at positive fields (right hand side) were recorded under \textit{increasing} absolute field strength (up) starting effectively at $H=0$. As emphasized above, we have confirmed in a large number of careful tests, that the data observed for positive field values (the right-hand side of Fig.~\ref{fig:hfc}), which start at $H=0$, with the exception of the signature of $H_{c1}$, agrees very well with data observed under ZFC, shown in Fig.~\ref{fig:zfc}.

It is helpful again to begin with an account of the data for fields along $\langle 111 \rangle$ and $\langle 110 \rangle$, shown in Figs.~\ref{fig:hfc}(a)--\ref{fig:hfc}(c), which correspond to the well-known properties of cubic chiral magnets reported in the literature. Essentially the same qualitative field dependence is observed under ZFC and HFC, where the magnetization as a function of field is almost point symmetric with respect to $H=0$ and $M=0$. Starting at a large negative field and reducing the field strength, a transition from the field-polarized state to the conical state at $H_{\rm c2}^{\rm d}$ is followed by the transition to the helical state at $H_{\rm c1}^{\rm d}$. Increasing the field through zero towards positive field values, the system transitions from the helical into the conical state at $H_{\rm c1}^{\rm u}$ and to the field-polarized state at $H_{\rm c2}^{\rm u}$. The susceptibilities associated with the different phases display a plateau in the conical phase, a reduced value in the helical state, and a rapid decrease when entering the field-polarized regime. Concerning the evidence of magnetic anisotropy, we note that $H_\mathrm{c1}$ is quantitatively almost identical for \ooo and \ooz, whereas $H_\mathrm{c2}$ is highest for field parallel $\langle111\rangle$ and visibly smaller for field parallel $\langle110\rangle$. 

In comparison to fields along $\langle 111 \rangle$ and  $\langle 110 \rangle$, the magnetization as a function of field for a field along $ \langle 100 \rangle$ displays a sizable asymmetry with respect to $H=0$ and $M=0$. Data recorded under increasing fields (positive field values) are equivalent to the data recorded under ZFC data with the exception of the tiny signature of $H_{\rm c1}^{\rm u}$ marked in Fig.\,\ref{fig:zfc}(h). As explained above, this reflects differences in domain population arising from the different temperature versus field histories. 

Starting at large negative fields as a function of decreasing field, the system undergoes a transition from the field-polarized to the tilted conical phase at $H^\mathrm{d}_\mathrm{t2}$ and exits the tilted conical phase at $H^\mathrm{d}_\mathrm{t1}$. The transition from the LTS phase to the conical state takes place at $H^\mathrm{d}_\mathrm{s1}$ much below $H^\mathrm{d}_\mathrm{t1}$. Continuing this field sweep through zero, the transition from the conical to the tilted conical phase is observed at $H^\mathrm{u}_\mathrm{t1}$, which vanishes at $H^\mathrm{u}_\mathrm{t2}$, followed by the transition of the LTS phase to the field-polarized state at $H^\mathrm{u}_\mathrm{s2}$. 


A key result of our study concerns the detailed behavior and field range of the tilted conical and LTS phases. Correcting the effects of demagnetizing fields, we find no hysteresis for the transition fields $H_\mathrm{t1}$ and $H_\mathrm{t2}$ of the tilted conical phase between increasing (u) and decreasing (d) field strength. In contrast, under decreasing magnetic fields (negative field values), the LTS phase appears to emerge essentially together with the tilted conical phase around $H^\mathrm{d}_\mathrm{t2}$. However, when decreasing the field further the LTS phase survives down to $H^\mathrm{d}_\mathrm{s1}$, i.e., well below the regime of the tilted conical state vanishing at $H^\mathrm{d}_\mathrm{t1}$. Strictly speaking, we infer the formation of the LTS phase from the observation of $H^\mathrm{d}_\mathrm{s1}$ and $H^\mathrm{u}_\mathrm{s2}$ in combination with what is known from the SANS study. 

Interestingly, the small maximum of $\mathrm{d}M/\mathrm{d}H$ at $H^\mathrm{d}_\mathrm{s1}$ is related to a change of the magnetization, where the magnitude of $M$ is larger in the presence of the low-temperature skyrmion phase (cf just below $H^\mathrm{d}_\mathrm{s1}$), while $M$ collapses onto the value of the magnetization observed for the other field directions just below $H^\mathrm{d}_\mathrm{s1}$. Regardless whether the data are recorded under increasing or decreasing field, the tilted conical phase always emerges first while the low temperature skyrmion phase vanishes last.

Concerning the agreement and discrepancies between $\mathrm{d}M/\mathrm{d}H$ and $\chi'$ under HFC, we find essentially the same properties observed under ZFC. Namely, analogous to the behavior at $H^\mathrm{u}_\mathrm{s2}$ described above, where $\chi'$ does not track $\mathrm{d}M/\mathrm{d}H$ and $\chi''$ remains vanishingly small, $\chi'$ does not track $\mathrm{d}M/\mathrm{d}H$ at $H^\mathrm{d}_\mathrm{s1}$ either, and there is also no evidence for dissipation in the imaginary part of the ac susceptibility. This indicates potentially important differences regarding the process of nucleation of the low-temperature skyrmion phase as compared with the high-temperature skyrmion phase that are beyond the scope of our study.



\subsubsection{Comparison with small-angle neutron scattering}
\label{sans}

\begin{figure*}
\centering
\includegraphics[width=0.8\linewidth]{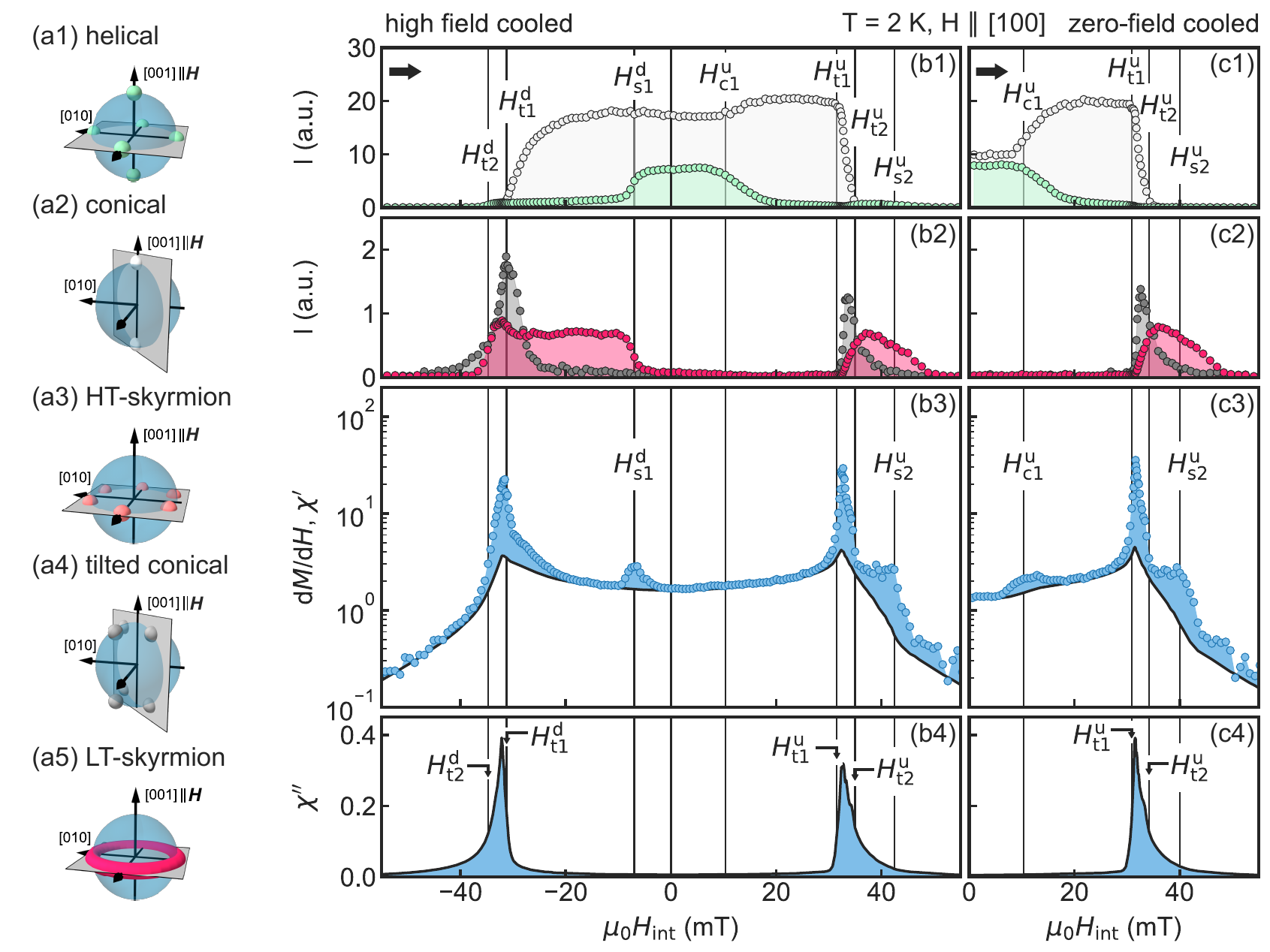}
\caption{\label{fig:SANS} Comparison of small-angle neutron scattering intensity with the susceptibility as a function of internal magnetic field. Data recorded under HFC and ZFC are shown on the left- and right-hand side, respectively. (a1)--(a5) Typical neutron scattering intensity distribution in reciprocal space for the five different modulated phases as denoted in each panel. The scattering plane is marked in gray shading. [(b1) and (c1)] Neutron scattering intensity of the helical and conical state. [(b2) and (c2)] Neutron scattering intensity of the tilted conical and LTS phase. [(b3) and (c3)] Susceptibility as calculated from the magnetization, $\mathrm{d}M/\mathrm{d}H$ (dots), and the real part of the ac susceptibility $\chi'$ (black line). [(b4), (c4)] Imaginary part of the ac susceptibility $\chi''$.}
\end{figure*}

The definitions of the transition fields introduced above may be justified by means of a direct comparison with the magnetic field dependence of the scattering intensities observed in small-angle neutron scattering shown in Fig.\,\ref{fig:SANS}. For the comparison a data set was recorded under the same conditions reported in Ref.\,\onlinecite{2018:Chacon:NatPhys}. We note that the intensities shown here represent peak intensities and not an integration over rocking scans. Therefore quantitative comparison between different phases is not possible. For further technical details, we refer to Ref.\,\onlinecite{2018:Chacon:NatPhys}. 

Great care was exercised in order to avoid systematic errors. First, the neutron scattering data were recorded for a spherical {\cso} sample prepared from the same batch of material used to prepare the samples for the magnetization and ac susceptibility measurements reported here. Second, the magnetization and susceptibility data were recorded in a sample of almost cubic shape as listed in Table\,\ref{table:samples}. This way, the demagnetization factors were similar and differences of corrections tiny. In turn, data are shown as a function of internal field, where the demagnetization correction for the SANS data was calculated from the magnetization data measured on a different sample. Third, the same ZFC and HFC protocols were followed when recording the neutron scattering data, magnetization and susceptibility to minimize systematic differences.

For what follows, it is helpful to review briefly the intensity patterns associated with the different magnetic phases shown schematically in Figs.\,\ref{fig:SANS}\,(a1)--\ref{fig:SANS}\,(a5). Taken together, all magnetic phases are characterized by scattering intensities on the surface of a small sphere in reciprocal space depicted in blue shading. The radius of this sphere reflects the characteristic length scale $|\bm{Q}|$ of the competition of ferromagnetic exchange and Dzyaloshinsky-Moriya interactions. 

The illustrations shown in Figs.\ref{fig:SANS}(a1)--\ref{fig:SANS}\,(a5) correspond to the helical phase (green), conical phase (gray), high-temperature skyrmion phase (light red), tilted conical phase (dark gray), and low-temperature skyrmion phase (dark red). In these depictions, the crystallographic $\langle100\rangle$ directions are shown as black arrows, while the detector plane is indicated by gray shading. It is important to note the orientations of the applied magnetic field with respect to the detector plane and the crystallographic crystal axes.  

In the helical state, illustrated in Fig.~\ref{fig:SANS}\,(a1),  the $\langle 100 \rangle$ easy magnetic axes result in three energetically equivalent domain populations as indicated by green-shaded dots. Under ZFC, these domains are populated equally. For sufficiently large field the conical state stabilizes. The associated scattering pattern corresponds to intensities for $\pm \mathbf{Q} \parallel \mathbf{H}$ as illustrated in Fig.~\ref{fig:SANS}\,(a2). Note that for $\mathbf{H} \parallel \langle 100 \rangle$, the conical intensity is indistinguishable from the magnetic domain of the helical state in the field direction. Further, shown in Fig.~\ref{fig:SANS}(a3) is the scattering pattern of the high-temperature skyrmion phase, which is located in a plane perpendicular to the applied magnetic field. Shown here is the sixfold scattering pattern of a single skyrmion lattice domain population, where an energetically degenerate second domain population exists in principle under a rotation of $30^{\circ}$ with respect to the field direction. 

In addition to these well-known scattering patterns, the tilted conical phase and the LTS phase are characterized by the scattering patterns illustrated in Figs.~\ref{fig:SANS}\,(a4) and \ref{fig:SANS}\,(a5). For the tilted conical phase the diffraction spots along the magnetic field split into four contributions under an angle with respect to the field direction. This tilting angle increases under increasing magnetic field. Second, in the LTS phase, scattering intensity emerges in the plane perpendicular to the applied magnetic field. While this pattern assumes the shape of a uniform ring in most experiments, characteristic of a glassy appearance of the skyrmion lattice, it was demonstrated that the ring represents an average of a randomly oriented sixfold diffraction patterns.\cite{2018:Chacon:NatPhys}

The magnetic field dependence of the different scattering intensities is shown in Figs.~\ref{fig:SANS}(b1) and (b2) for HFC, as well as Figs.~\ref{fig:SANS}\,(c1) and \ref{fig:SANS}\,(c2) for ZFC. Here, the color shading corresponds to the shading used in Figs.~\ref{fig:SANS}(a1)--\ref{fig:SANS}\,(a5). Under HFC and decreasing field, the tilted conical phase emerges at $H^\mathrm{d}_\mathrm{t2}$, defined at the point of inflection of the scattering intensity, which corresponds to the point of inflection of $\chi''$. It is not possible to identify a clear signature of the emergence of the LTS phase in the magnetization or susceptibility, which in neutron scattering clearly appears at a field value smaller than $H^\mathrm{d}_\mathrm{t2}$. The transition field at which the tilted conical phase vanishes, $H^\mathrm{d}_\mathrm{t1}$, may be defined at the point of inflection in $\chi''$ as marked. Finally, when decreasing the field further, the disappearance of the LTS phase at $H^\mathrm{d}_\mathrm{s1}$ is clearly connected with a distinct maximum in the susceptibility inferred from the magnetization, $\mathrm{d}M/\mathrm{d}H$, where neither a peak in $\chi'$ is seen nor a contribution in $\chi''$.

Increasing the magnetic field after HFC further [the right-hand side of panels (b1) -- (b4)], the conical-to-helical transition may seen at $H^\mathrm{d}_\mathrm{c1}$ (not marked for clarity), which coincides with $H^\mathrm{d}_\mathrm{s1}$, as well as $H^\mathrm{u}_\mathrm{c1}$. Yet, no signatures may be seen in the magnetization and susceptibility at $H^\mathrm{d}_\mathrm{c1}$, which why we labelled the $H^\mathrm{d}_\mathrm{s1}$ only. However, when increasing the field further, the magnetic field dependence of the scattering intensities of the tilted conical phase displays again excellent agreement with $\chi''$ at the points of inflection of both quantities, defining $H^\mathrm{u}_\mathrm{t1}$ and $H^\mathrm{u}_\mathrm{t2}$. Again, the LTS phase emerges at a field value $H^\mathrm{u}_\mathrm{s1}$, slightly larger than $H^\mathrm{u}_\mathrm{t1}$, and vanishes at a field $H^\mathrm{u}_\mathrm{s2}$ well above $H^\mathrm{u}_\mathrm{t2}$. Here $H^\mathrm{u}_\mathrm{s2}$ corresponds accurately to the location of an additional small peak in $\mathrm{d}M/\mathrm{d}H$ that is neither tracked in $\chi'$ nor visible in $\chi''$, where the latter is vanishingly small.

The characteristics observed in the neutron scattering intensity, the magnetization, and the susceptibilities observed under HFC are reproduced very well under ZFC as shown in Figs.~\ref{fig:SANS}(c1) through \ref{fig:SANS}(c4). The only difference that may be noticed here concerns the tiny signature in $\mathrm{d}M/\mathrm{d}H$ at $H^\mathrm{u}_\mathrm{c1}$, which is not present under HFC.



\subsection{Higher harmonics in the susceptibility}
\label{harmonics}

\begin{figure*}
\centering
\includegraphics[width=0.9\linewidth]{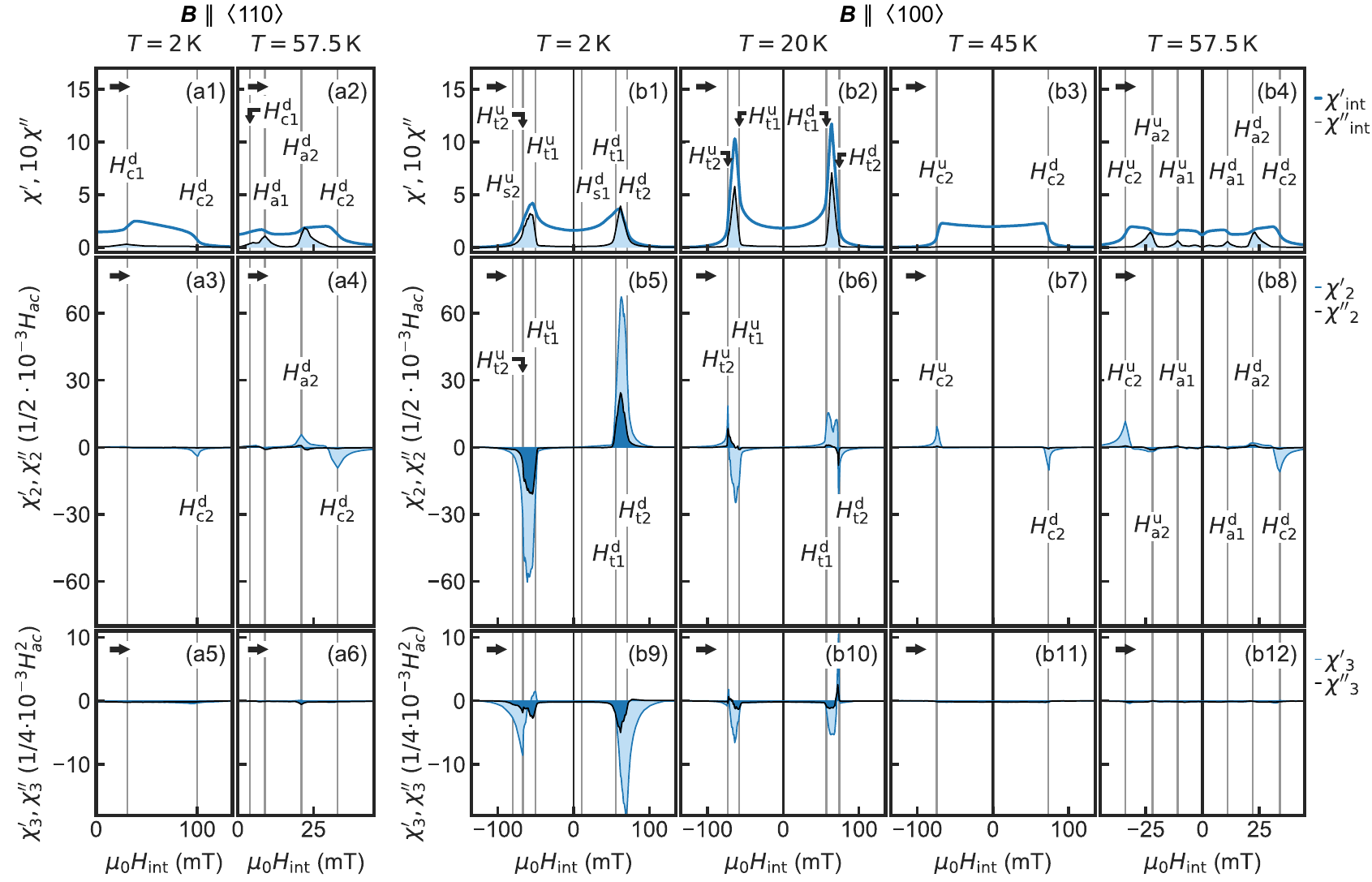}
\caption{\label{fig:harmonics}
Higher-harmonic contributions in the ac susceptibility as a function of field at selected temperatures for fields along the $\langle 110 \rangle$ and $\langle 100 \rangle$ direction. The set of panels on the left corresponds to $\mathbf{H}\parallel\langle 110 \rangle$; the set of panels on the right corresponds to $\mathbf{H}\parallel\langle 100 \rangle$. Rows display the first, second and third harmonic. Columns correspond to different temperatures as marked in the plots. Data have been recorded following high-field cooling in a field sweep from negative to positive field values. The negative branch of the \ooz data is not shown for clarity, as it provides no further information.}
\end{figure*}

Motivated by the observation of clear differences between $\mathrm{d}M/\mathrm{d}H$ and $\chi'$, as well as substantial contributions in $\chi''$, we explored the question of higher-harmonic contributions to the ac susceptibility up to third order for field along $\langle 110 \rangle$ and $\langle 100 \rangle$ at selected temperatures. Data were recorded in a single field sweep from negative to positive fields after HFC, i.e., data recorded for positive field values during these sweeps essentially corresponded to ZFC conditions. 

\begin{figure*}
\centering
\includegraphics[width=0.9\linewidth]{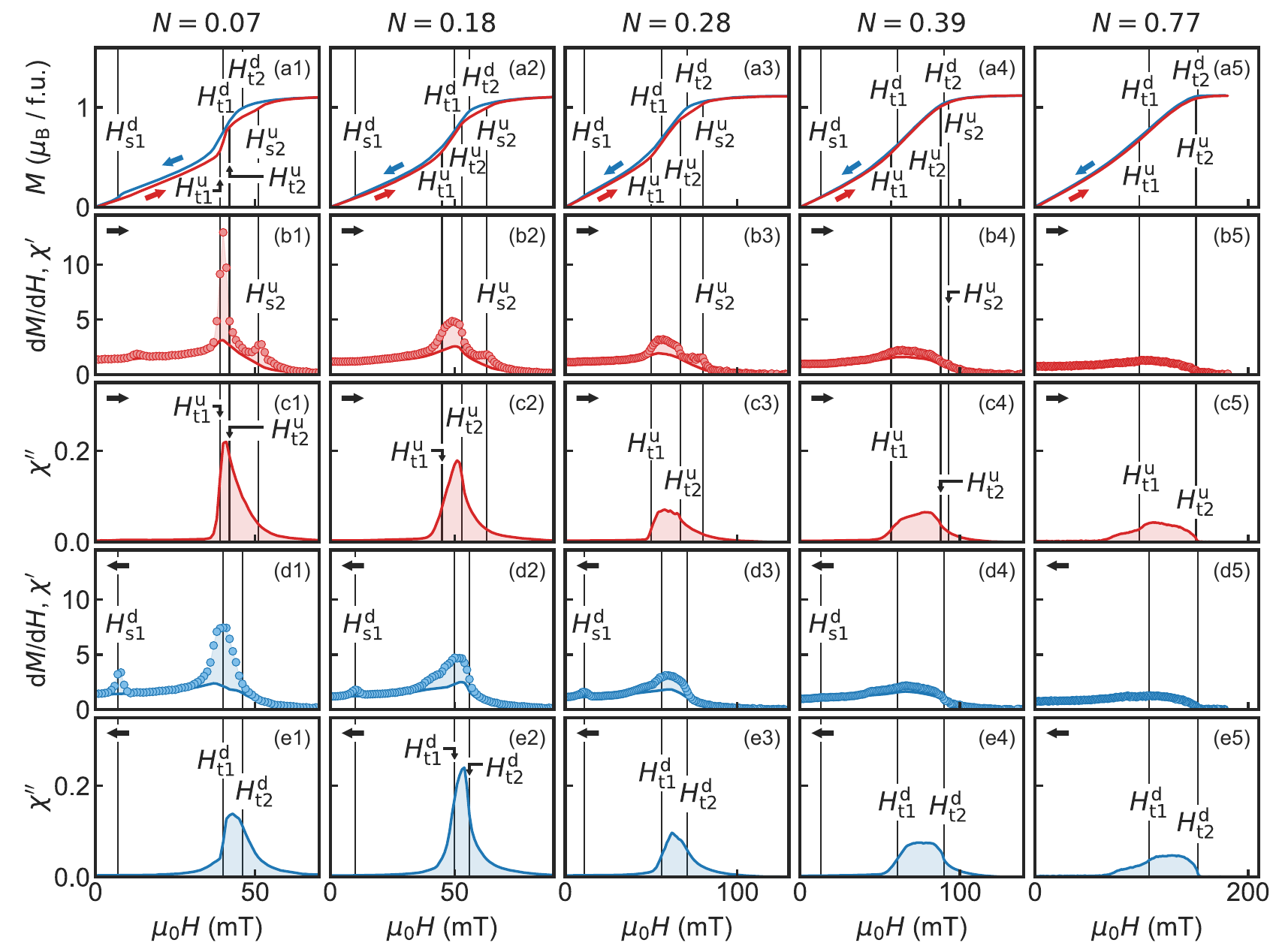}
\caption{\label{fig:demag}
Magnetization $M$, calculated susceptibility, $\mathrm{d}M/\mathrm{d}H$, as well as real and imaginary part of the ac susceptibility, $\chi'$ and $\chi''$, respectively, for different sample shapes with different demagnetization factors and field along $\langle100\rangle$. The demagnetization factors are denoted above each column (see Table\,\ref{table:samples} for further details). Data were recorded following high-field cooling in a single field sweep, where data recorded at negative fields are mirrored to positive field values. Data branches recorded under decreasing and increasing magnitude of the field are shown in blue and red shading, respectively. }
\end{figure*}

In general higher harmonics reflect the presence of strong non-linearities in $M(H)$ even for the tiny excitation amplitudes of the ac probing field (of the order 0.1\,mT). In the spirit of a brief introduction we note that the magnetic response of a sample to an external magnetic field
\begin{equation}
H(\omega, t) = H_\mathrm{dc} + H_\mathrm{ac} \cos{\omega t}
\end{equation}
may be expressed in terms of the Fourier expansion
\begin{eqnarray}
  M(\omega, t)
  &= M_0 + \sum\limits_{n=1}^{\infty} \left(M_n' \cos{n \omega t} + M_n''\sin{n \omega t}\right)\\
  &= \chi_\mathrm{dc} H_\mathrm{dc} + H_\mathrm{ac} \sum\limits_{n=1}^{\infty} \left(\theta_n' \cos{n \omega t} + \theta_n''\sin{n \omega t}\right)
\end{eqnarray}
where the Fourier coefficients represent unitless quantities given by
\begin{equation}
  \theta_n' = \frac{1}{\pi H_\mathrm{ac}} \int\limits_0^{2\pi} M(\omega, t) \cos{n \omega t} \D\omega t
\end{equation}
and
\begin{equation}
\theta_n'' = \frac{1}{\pi H_\mathrm{ac}} \int\limits_0^{2\pi} M(\omega, t) \sin{n \omega t} \D\omega t
\end{equation}
As mentioned in Sec.\,\ref{methods}, in the literature, the coefficients $\theta_n'$ and $\theta_n''$ are also known as \textit{harmonic} susceptibilities. In comparison, when developing the response to an excitation in a Taylor series the expansion coefficients, denoted $ \chi_n'$ and $ \chi_n''$, are referred to as \textit{nonlinear} susceptibilities. For mathematically well behaved properties it is possible to infer the real and imaginary parts of the nonlinear susceptibilities, $\chi'_n$ and $\chi''_n$ from these Fourier coefficients $ \theta_n'$ and $ \theta_n''$  (see, e.g., Ref.\,\onlinecite{mydosh:1993} for details). However, since higher order contributions rapidly decrease in strength, we approximate the nonlinear susceptibility by the leading order contributions of the harmonic susceptibilities, namely, $\chi=\chi_1\approx\theta_1$, $\chi_2\approx2\,H_\mathrm{ac}^{-1}\theta_2$ and $\chi_3\approx4\,H_\mathrm{ac}^{-2}\theta_3$.

In view that the magnetization curve for field along {\ozz} is not point symmetric with respect to $M=0$ and $H=0$, the second harmonic may not be point symmetric either.  Further, for the sharp changes of slope of the magnetization curve the second harmonic may be expected to be finite subject to the precise excitation amplitude. 

Shown in Fig.\,\ref{fig:harmonics} are typical higher order contributions of the harmonic susceptibility, notably $\chi$, $\theta_2$, and $\theta_3$. In our measurements, we recorded the properties up to the third harmonic at selected temperatures following HFC. Data are shown as a function of internal field for field along $\langle 110 \rangle$ (left panels) and field along $\langle 100 \rangle$ (right panels), where rows correspond to first, second and third harmonic susceptibility, while columns correspond to different temperatures. Data for negative fields for {\ooz} are not shown for clarity and because they are symmetrical to the data at positive field values, thus not adding any additional information. 

For a field along $\langle 110 \rangle$ and $T=\SI{2}{\kelvin}$, the second harmonic is essentially zero except for a negative peak of the real part, $\chi'_2$, at $H_\mathrm{c2}$. At $T=\SI{57.5}{\kelvin}$, tiny additional peaks are present at $H_\mathrm{a2}$ and two very weak peaks at $H_\mathrm{a1}$ and $H_\mathrm{c1}$ (hardly visible on the scale chosen here). Moreover, the third harmonic, $\chi_3$, is essentially zero at both $T = \SI{2}{\kelvin}$ and $\SI{57.5}{\kelvin}$.

In comparison, for a field along $\langle 100 \rangle$, the behavior at high temperatures is similar to that for field along $\langle 110\rangle$. Namely, at $T=\SI{57.5}{\kelvin}$, the second harmonic, $\chi'_2$, displays a negative peak associated with $H_\mathrm{c2}$ as well as additional tiny peaks at $H_\mathrm{a2}$,  $H_\mathrm{a1}$, and  $H_\mathrm{c1}$ (barely resolved on the scale chosen here). Moreover, the third harmonic, $\chi_3$, is essentially zero. Similarly, at $T=\SI{45}{\kelvin}$ only the peak associated with $H_\mathrm{c2}$ is observed in $\chi_2$, where the data are point symmetric with respect to $H=0$ and $\chi_2=0$. Again the third harmonic, $\chi_3$, is essentially zero.

In contrast, at $T=\SI{2}{\kelvin}$, a huge second harmonic is observed in the field range of the tilted conical phase for $\mathrm{H} \parallel \langle 100 \rangle$. Interestingly, in comparison with the second harmonic signal at high temperatures associated with $H_\mathrm{c2}$ the new contribution has the opposite sign. Also, a huge third harmonic is observed in the field region of the tilted conical phase. In fact, key features of the second and third harmonic are already present at $T=\SI{20}{\kelvin}$, where they are weaker with additional fine structure (double peak). 

The highly nonlinear response we observe in {\cso} represents a key characteristic of the tilted conical state rather than the LTS phase. We speculate that the strong nonlinear and dissipative response arises from domain walls and other inhomogeneous textures generated when the tilted phase is nucleated. In the literature, similar properties are typically reported for superconductors, and molecular and low-dimensional magnets\cite{2013:balanda}, where specific models are required for the interpretation. Unfortunately, it is at present not clear how the non-linearities are generated. In the future, a Cole-Cole plot of the interdependence of the real and imaginary parts of $\chi$ for different frequencies may, for instance, provide information on the characteristic activation energies-\cite{2013:balanda} However, this level of experimental exploration and analysis is beyond the scope of the study presented here. 

\subsection{Demagnetizing Effects}
\label{demag}

In view of the importance of the temperature versus field protocol and the rather subtle features in the magnetization and susceptibilities associated with the various phase transitions we have recorded the magnetization and ac susceptibility for a series of samples with different demagnetization factors. For samples with large demagnetization factors, we find the reconstruction of the intrinsic behavior to be prohibitively difficult as small effects are irreversibly smeared out. Also, for an irregular sample shape, the nucleation processes may actually be altered. Moreover, theoretical modeling reported in Ref.\, \onlinecite{2018:Chacon:NatPhys} revealed that dipolar interactions play an important role for the details of the magnetic phase diagram as discussed below. 

Shown in Fig.\,\ref{fig:demag} are typical magnetization and susceptibility data as a function of applied field for various samples with different demagnetizing factors (see also Table\,\ref{table:samples}). Each column corresponds to a different demagnetization factor as stated at the top of the figure. Data were measured in a single field sweep corresponding to HFC, i.e., analogous to the data shown in Fig.\,\ref{fig:hfc}. For lack of space, data at negative fields are shown as a function of positive field values. Therefore data shown in rows (b) and (c) were recorded under increasing field, corresponding to the right-hand side of Fig.\,\ref{fig:hfc}, i.e., positive field values. They are essentially equivalent to the behavior observed under ZFC. In comparison, data shown in rows (d) and (e) were recorded under decreasing field, corresponding to the left-hand side of Fig.\,\ref{fig:hfc}, i.e., negative field values.

\begin{figure*}
\centering
\includegraphics[width=1.0\linewidth]{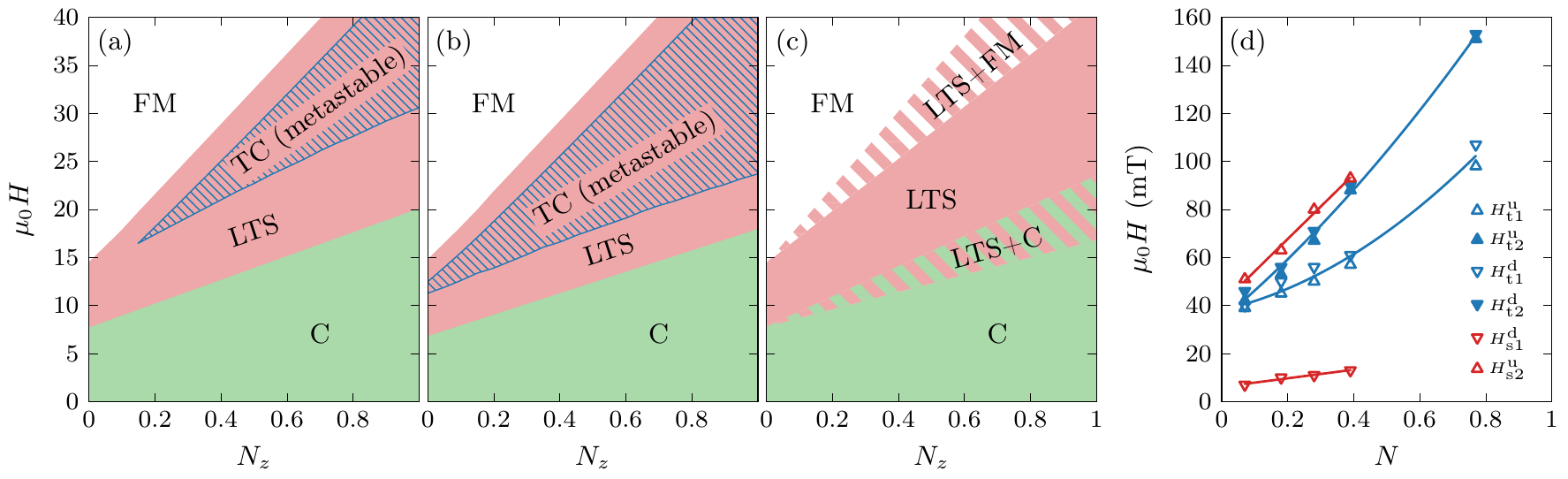}
\caption{\label{fig:demag-N}
Magnetic phase diagrams as a function of demagnetization factor $N$ and applied magnetic field $H$ applied along $\langle 100\rangle$. Note that different units are used for $H$ in the theoretical calculations as compared with experiment. (a) Theoretical phase diagram obtained from a Ginzburg-Landau model in the presence of the cubic anisotropy of Eq.~\eqref{CubicAniso}, as described in Ref.~\onlinecite{2018:Chacon:NatPhys}. Here we assumed that only single-domain states exist. We also display the metastable tilted conical state in the regime where its energy is higher than the LTS phase but lower than both the conical and ferromagnetic phase. It vanishes for small $N_z$. Parameters and units correspond to those used in Ref.~\onlinecite{2018:Chacon:NatPhys}. (b) Phase diagram of (a) when including the exchange anisotropy $c\sum_i(\nabla_iM_i)^2$ with $c=-0.06$, as described in the main text. The metastable tilted conical state exists even for $N_z \to 0$. (c) Thermodynamic phase diagram [parameters as in (a)] including regions where demagnetization effects stabilize phase coexistence, see main text. Note that the experimental regions of phase coexistence can be much larger due to hysteresis effects in the region where phases are metastable. (d) Summary of the experimental critical fields for samples with different demagnetization factor $N$, as detailed in Fig.\,\ref{fig:demag}, with data for negative fields mirrored to positive fields.
}
\end{figure*}

Assuming uniform demagnetization fields across the sample volume a shearing of the data towards larger field values with increasing demagnetization factor is expected. Moreover, the bar-shaped sample geometry results in a distribution of internal field directions that generates a smearing out of all features. As shown in the first column of Fig.\,\ref{fig:demag} for a tiny demagnetizing factor, $N=0.07$, all features observed under ZFC and HFC described in detail above may be readily identified. The two most important characteristics concern the presence of hysteresis in the magnetization due to the LTS phase, and the observation of dissipation in the imaginary part of the susceptibility $\chi''$, associated with the tilted conical phase. 

For increasing demagnetization factor the steep rise (jump) in the magnetization associated with the onset of the tilted conical phase turns into a shallow rise. Nonetheless, a non-vanishing contribution in $\chi''$ is observed in a well-defined field range. This suggests that the tilted conical phase continues to form in a finite field range. In contrast, the signatures of the LTS phase are smeared out and vanish. Without microscopic information it is unfortunately not possible to infer any further information on the LTS phase. 

In order to illustrate the effects of demagnetizing fields on a theoretical level, we consider the Ginzburg-Landau model described in Ref.~\onlinecite{2018:Chacon:NatPhys} taking into account the cubic magnetocrystalline anisotropy, cf. Eq.~\eqref{CubicAniso} shown below. For a given applied field $\mathbf{H}$, the magnetic moments are subject to an internal field $\mathbf{H}_{\rm int} = \mathbf{H} - {\bf N} \mathbf{M}$ that depends on the magnetization $\mathbf{M}$ as well as the demagnetization tensor ${\bf N}$. This results in three main physical effects: (i) a shift of phase boundaries due to the change of the internal fields, (ii) a possible tilt of the internal field in the tilted conical phase, and (iii) the stabilization of coexistence regimes.

Shown in Fig.\,\ref{fig:demag-N}\,(a) and (b) are phase diagrams under an applied field $\mathbf{H}$ along $\langle 100\rangle$ as a function of demagnetization factor $N_z$. The two phase diagrams only consider uniform phases, ignoring the possibility of phase coexistence discussed below. The thermodynamically stable phases are the conical, the LTS and the ferromagnetic state. We also show the metastable tilted conical phase in the regime where it has lower energy than both the ferromagnetic and conical phase. For the chosen parameters, its energy is, however, always higher than the energy of the LTS state. Note that we do not show a possible skyrmion square lattice phase. In some part of the phase diagram (see Ref.~\onlinecite{2018:Chacon:NatPhys}) its energy is very close (identical within numerical errors and model uncertainties) to the triangular skyrmion lattice.

\begin{figure*}
\centering
\includegraphics{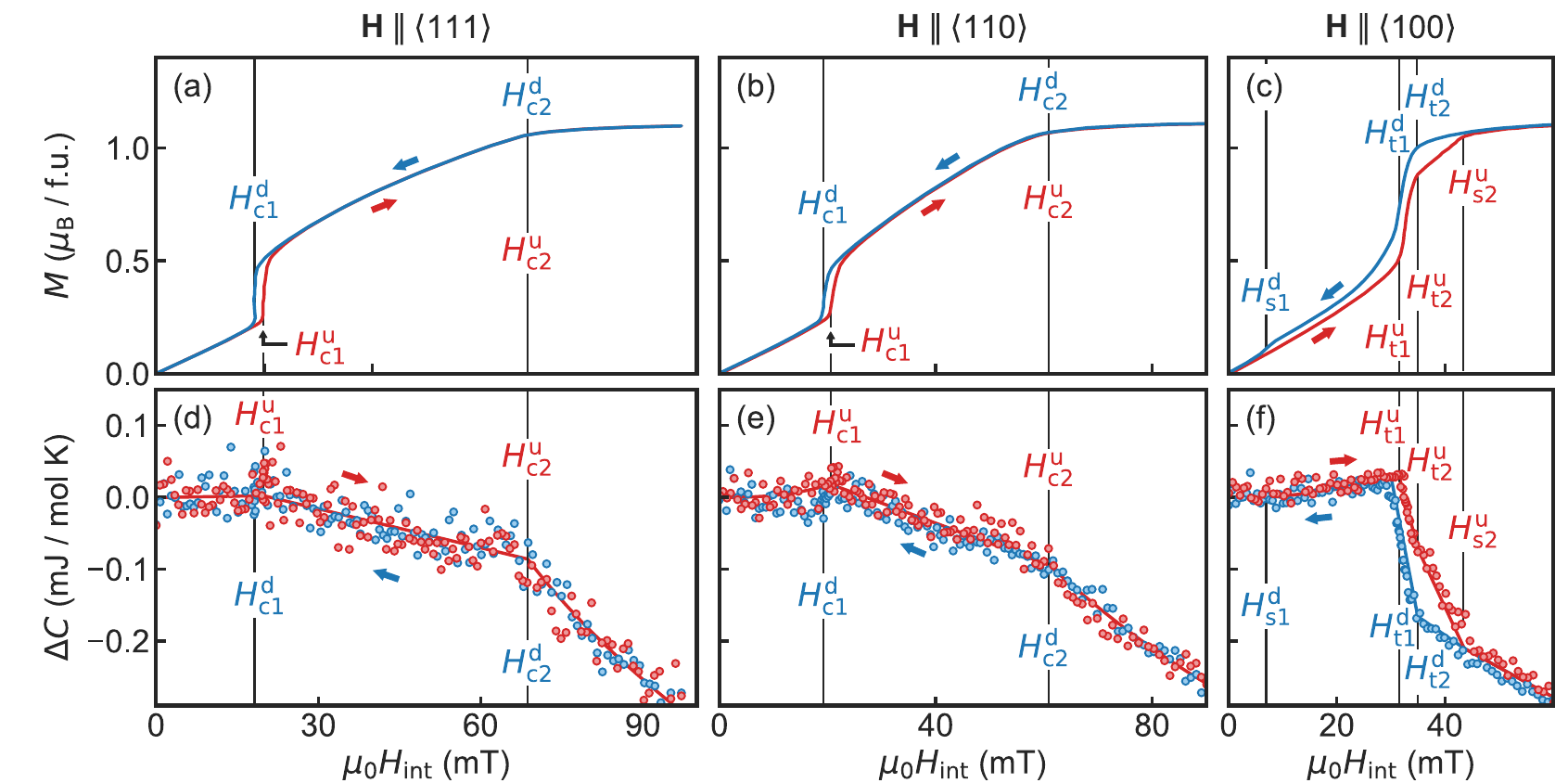}
\caption{\label{fig:cu2seo4:heat_capacity} Magnetization (first row) and specific heat (second row) as a function of internal field at \SI{2}{\kelvin} for all major directions.  Data were recorded following high-field cooling in a single field sweep from negative to positive field values, where data recorded at negative fields are mirrored to positive field values. Data branches recorded under decreasing and increasing are indicated in blue and red shading, respectively.
}
\end{figure*}

For increasing $N_z$ the stability range of all phase boundaries shifts to larger magnetic fields as the demagnetization factor reduces internal magnetic fields. The energetics of the metastable tilted conical phase is affected in more subtle ways as the tilting can induce internal field components perpendicular to the external magnetic field (the calculation assumes a single domain). Similar to the pitch of the modulation, also the uniform magnetization is tilted away from the applied magnetic field, notably $\mathbf{H} \parallel \langle 100\rangle$. Thus, for the single domain state assumed here, the internal field, comprising the applied field, the uniform magnetization of the tilted conical state and the demagnetization field, increasingly deviates from $\langle 100\rangle$. For $N_z\to 0$ this effect is enhanced such that the tilted conical phase eventually becomes energetically unfavorable compared to the field-polarized state, as the general condition $N_x+N_y+N_z=1$ implies that the demagnetizing field of the transverse field components gets boosted.

For the parameters used in Ref.~\onlinecite{2018:Chacon:NatPhys} the tilted conical phase vanishes for $N_z \to 0$, see Fig.\,\ref{fig:demag-N}\,(a). Our experimental data display, in contrast, the signatures of the tilted conical phase even for tiny values of $N$, as summarized in Fig.\,\ref{fig:demag-N}\,(d). However, there are different possibilities (or combinations thereof) to account for this seeming discrepancy. First, additional anisotropies, such as exchange contributions $c[(\nabla_x M_x)^2+(\nabla_y M_y)^2+(\nabla_z M_z)^2]$, could stabilize the tilted conical phase further such that it remains metastable even for $N_z \to 0$. This is illustrated in Fig.\,\ref{fig:demag-N}\,(b), where the additional exchange anisotropy term given here has been taken into account as compared with the model evaluated in Fig.\,\ref{fig:demag-N}\,(a). Furthermore, the suppression of the tilted conical phase by internal field components perpendicular to the external field can be reduced when several domains with different tilting directions exist [cf.\,Fig.\,\ref{fig:SANS}\,(a4)]. Thus the total transverse magnetization might average out. We have checked that this effect indeed enlarges the metastable tilted phase but not up to $N_z=0$.

Both phase diagrams, Figs.\,\ref{fig:demag-N}\,(a) and (b), were generated in finding a uniform phase with minimal energy for a given external field $\mathbf{H}$. At the first order phase transitions, however, the two states in question have different magnetization. In this case demagnetization effects stabilize regions of phase coexistence. Boundaries of regions of phase-coexistence (ignoring hysteresis effects and domain wall formation assuming a perfectly ``soft" magnet and smooth transitions into the coexistence regimes) are determined from the condition that the internal magnetic field has to match the critical field at $N_z=0$. The resulting phase diagram (without the metastable titled phase) is shown in  Fig.\,\ref{fig:demag-N}\,(c).

The experimentally observed regions of phase coexistence are much larger, see Fig.\,\ref{fig:demag-N}\,(d). This can be attributed to strong hysteresis effects: after their formation, topologically protected skyrmions remain locally stable over a large field range.\cite{2017:Wild:SciAdv}


\subsection{Specific heat}
\label{specific_heat}

For temperatures exceeding $\sim10\,{\rm K}$, the specific heat $C$ is dominated by contribution of phonons. In turn, a detailed investigation of the temperature dependence of $C$ at low temperatures was beyond the scope of the work presented. In order to gain some insights into possible signatures of the two new magnetic phases, we focused instead on the magnetic field dependence of $C$ at a single low temperature. 

Shown in Fig.\,\ref{fig:cu2seo4:heat_capacity} is a comparison of the magnetization, $M$, and the specific heat $C$ as a function of magnetic field along {\ooo}, {\ooz} and {\ozz} at a constant temperature of $\SI{2}{\kelvin}$. For clarity only the change of the specific heat, $\Delta C = C(B)-C(H=0)$ is shown. Data were recorded following high-field cooling in a single field sweep from negative to positive field values, where data recorded at negative fields are mirrored to positive field values. Data branches recorded under decreasing and increasing field are indicated in blue and red shading, respectively. Data recorded under increasing and decreasing field as marked by the arrows, corresponding essentially to ZFC and HFC conditions, respectively. The magnetization, shown in Figs.\,\ref{fig:cu2seo4:heat_capacity}(a), \ref{fig:cu2seo4:heat_capacity}(b) and \ref{fig:cu2seo4:heat_capacity}(c) for field along {\ooo}, {\ooz} and {\ozz}, respectively, displays the key characteristics associated with the helical phase, the conical phase, the tilted conical phase, the low-temperature skyrmion phase and the field polarized phase. 

The specific heat for field along {\ooo} and {\ooz} as a function of magnetic field, shown in Figs.\,\ref{fig:cu2seo4:heat_capacity}(d) and \ref{fig:cu2seo4:heat_capacity}(e) exhibits three regimes of different slope, $\mathrm{d}C/\mathrm{d}H$. In the helical state, the specific heat is essentially constant, followed by a linear decrease in the conical state, which becomes steeper in the field-polarized state. Apart from a small amount of hysteresis in $M$ at $H_\mathrm{c1}$, there is no evidence for further hysteresis.

In contrast, there are clear differences between increasing and decreasing field for field along {\ozz} shown in Fig.\,\ref{fig:cu2seo4:heat_capacity}\,(f). Namely, under increasing magnetic field (red), three distinct kinks corresponding to the transitions at $H_\mathrm{t1}^\mathrm{u}$, $H_\mathrm{t2}^\mathrm{u}$, and $H_\mathrm{s2}^\mathrm{u}$ are visible. At low magnetic fields up to $H_\mathrm{t1}^\mathrm{u}$, the specific heat is essentially unchanged, $\Delta C(B)\approx0$, where we note that this includes the helical and the conical phase. Between $H_\mathrm{t1}^\mathrm{u}$ and $H_\mathrm{t2}^\mathrm{u}$, $\Delta C(B)$ drops sharply, coinciding with the steep increase in $M$. Between $H_\mathrm{t2}^\mathrm{u}$ and $H_\mathrm{s2}^\mathrm{u}$, the rate of change, $\mathrm{d}C/\mathrm{d}H$, decreases, followed by another decrease of $\mathrm{d}C/\mathrm{d}H$ above $H_\mathrm{s2}^\mathrm{u}$. In comparison, under decreasing field (blue), only two changes of slope corresponding to $H_\mathrm{t1}^\mathrm{d}$ and $H_\mathrm{t2}^\mathrm{d}$ may be observed. Thus, there is clearly hysteresis between $H_\mathrm{t1}^\mathrm{d}$ and $H_\mathrm{s2}^\mathrm{u}$. When further decreasing the field, the specific heat coincides with the data recorded under increasing field. In particular, there is very faint evidence of $H_\mathrm{s1}^\mathrm{d}$ in the specific heat, which is clearly present in the magnetization.

As the tilted conical phase and the low-temperature skyrmion phase emerge as a function of magnetic field, changes of the entropy to be expected between the different phases are rather tiny. Nonetheless, we observe a suppression of the specific heat with increasing magnetic field for all three field directions. The observation of clear signatures in the specific heat at the transition fields of the tilted conical phase and low-temperature skyrmion phase, as well as the observation of hysteresis provide clear signatures of thermodynamically distinct phases. However, some caution is necessary as the strongly hysteretic behavior observed in neutron scattering, the magnetization and the specific heat are characteristic of strong first order transitions. Both the size of the effects as well as the experimental method we use here, notably small heat pulse relaxation calorimetry, make the detection of a latent heat for a weakly field dependent phase transition line essentially impossible.


\section{Discussion}
\label{discussion}
The changes of the magnetic phase diagram of {\cso} as a function of crystallographic direction inferred from our magnetization, susceptibility and specific heat data reflect the presence of distinct magnetic anisotropies. The discussion of these anisotropies is organized in three parts. In section \ref{anisotropy} a comparison of the anisotropies of the magnetization of {\cso} is presented with those observed in the related compounds {\fcs} ($x=0.2$) and MnSi. This is followed by a quantitative estimate of the strength of the cubic anisotropies, presented in section \ref{energy}. The discussion closes in section \ref{potential} with qualitative considerations of the effective anisotropy potential as a function of applied magnetic field, and how these connect with the anisotropy of the magnetization and the different magnetic phases.  

\subsection{Anisotropy of the magnetization}
\label{anisotropy}

\begin{figure}
\centering
\includegraphics[width=0.9\linewidth]{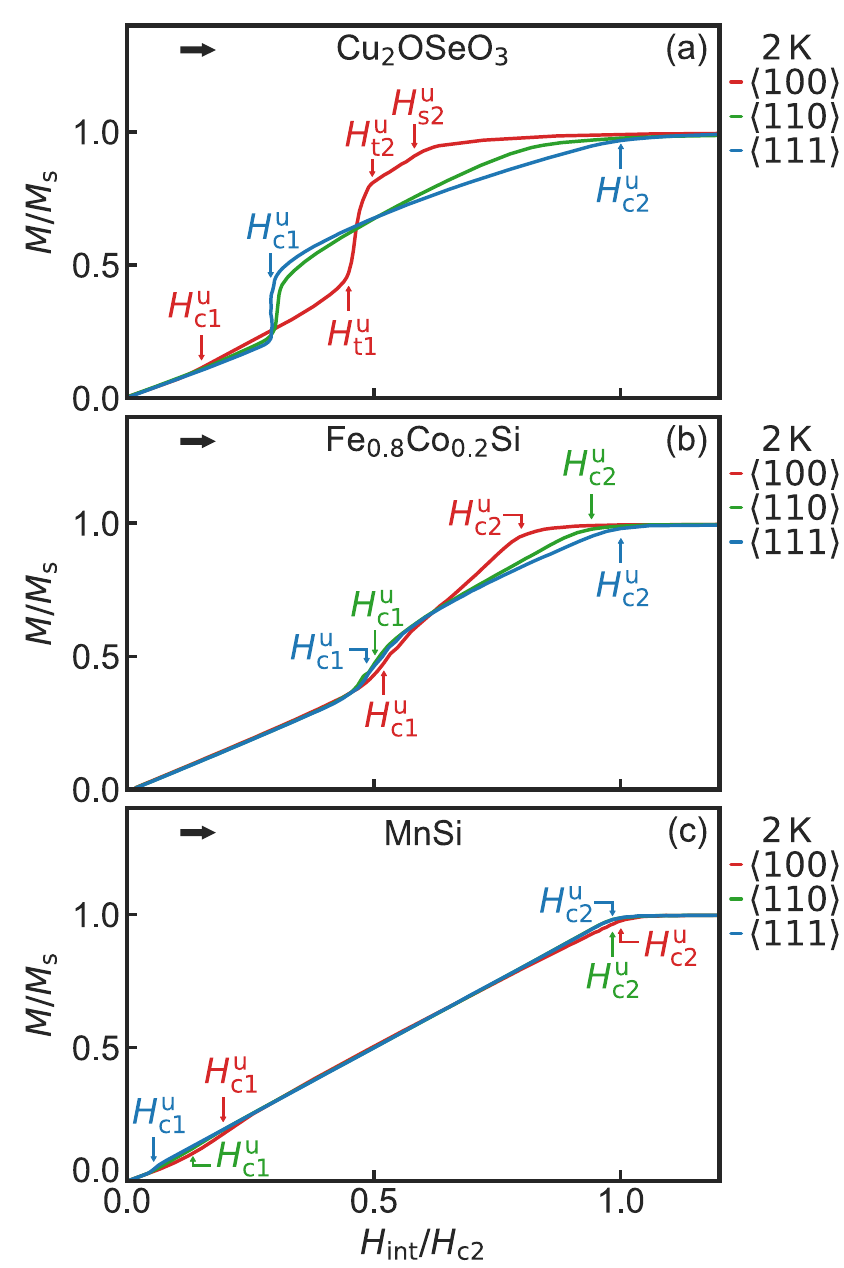}
\caption{\label{fig:materials}
Magnetization of {\cso}, {\fcs} ($x=0.2$), and MnSi, illustrating the different strength of the cubic magnetic anisotropies. Data are shown as a function of internal magnetic field in units of the upper critical field, $H_{\rm c2}$, determined in the direction exhibiting the largest value of $H_{\rm c2}$.  The magnetization is presented in units of the magnetization at high fields. (a) Magnetization of {\cso}, exhibiting strong $\langle 100\rangle$ easy-axes anisotropy. (b) Magnetization of {\fcs} ($x=0.2$), exhibiting moderate $\langle 100\rangle$ easy-axes anisotropy. (c) Magnetization of MnSi, exhibiting very weak $\langle 111\rangle$ easy-axes anisotropy.
}
\end{figure}

Shown in Fig.\,\ref{fig:materials} is a comparison of the magnetization of {\cso}, {\fcs} ($x=0.2$), and MnSi for all major crystallographic directions.  Data are shown as a function of internal magnetic field in units of the upper critical field $H_{\rm c2}$, as determined in the direction exhibiting the largest value of $H_{\rm c2}$.  The magnetization is presented in units of the magnetization at high fields.   

As emphasized in section \ref{data}, data for {\cso}, shown in Fig.\,\ref{fig:materials}\,(a), exhibit pronounced magnetic anisotropy.  While the magnetization appears to be essentially isotropic up to $H_{\rm c1}$, we find that the magnetization for {\ooo} is larger than for any other direction between $H_{\rm c1}$ for {\ooo} and $H_{\rm t1}$ for {\ozz}. Above $H_{\rm t1}$, the magnetization for {\ozz} is largest. The behavior at intermediate fields may be interpreted in terms of a field induced inversion of the magnetic anisotropy, where the {\ooo} axes appear to be the easy axes for a small field range. However, as shown in further detail below, the cubic anisotropy (sign of $K$) does not change as a function field and the {\ozz} axes remain the easy axes, consistent with $H_{\rm c2}$ being smallest and largest for {\ozz} and {\ooo}, respectively. Instead, the behavior observed here reflects the property of a modulated spin structure that remains essentially rigid in a cubic anisotropy potential.

The magnetization of {\fcs} ($x=0.2$) shown in Fig.\,\ref{fig:materials}\,(b) is reminiscent of {\cso} and characteristic of a weak $\langle 100\rangle$ easy-axes anisotropy, even though the variations in {\fcs} are not as pronounced. In comparison to {\cso}, the lower critical field $H_\mathrm{c1}$ seems essentially isotropic. It is important to emphasize that in particular the helical-to-conical transition in {\fcs} may be dominated by disorder and defect-related pinning, accounting for the large value of $H_{\rm c1}$ as compared to $H_{\rm c2}$.\cite{2016:Bauer:PRB} The fairly small changes of $H_{\rm c2}$ as a function of orientation reflect a weak magnetic anisotropy. Also, the magnetization for fields above $H_{\rm c1}$ is qualitatively different for different field directions, where the magnetization for {\ozz} is steeper than for the other directions up to $H_{\rm c2}$. For fields above $H_{\rm c1}$, the slope is initially larger than just below $H_{\rm c2}$. This corresponds to the jump in $M$ at $H_{\rm c1}$ in {\cso} characteristic of a broadened first order transition. Thus, in comparison to {\cso}, the magnetic anisotropy as analyzed in terms of the magnetic work presented below is by far not as pronounced. An exciting unresolved question for {\fcs} concerns, whether a tilted conical phase and LTS phase exist at least for some $x$.

The field dependence of the magnetization of MnSi shown in Fig.\,\ref{fig:materials}\,(c) exhibits, finally, the signatures of a very weak $\langle 111\rangle$ easy-axes anisotropy. Here differences between the different crystallographic directions are difficult to discern. In comparison to {\cso} and {\fcs} ($x=0.2$), the ratio of $H_{\rm c2}$ to $H_{\rm c1}$ is largest, whereas the variation of $H_{\rm c2}$ with orientation is smallest (see Ref.\,\onlinecite{2017:Bauer:PRB} for further details). Moreover, the changes of slope are tiny between the different phases for all crystallographic directions. Thus, taken together, the comparison of the magnetization shown here underscores that the anisotropy is clearly strongest in {\cso} with {\ozz} easy axes, followed by {\fcs}, which displays also {\ozz} easy axes, whereas MnSi exhibits the well-known, very weak {\ooo} easy-axes anisotropy.


\subsection{Magnetocrystalline Anisotropy}
\label{energy}

The LTS and tilted conical phases arise in the presence of cubic magnetocrystalline anisotropies provided that they are sufficiently strong.\cite{2018:Chacon:NatPhys,2018:Qian:arxiv} The energy density associated with the leading anisotropy may be expressed in terms of the unit vector $\hat M$ describing the orientation of the magnetization
\begin{equation} \label{CubicAniso}
\mathcal{F}_a = K (\hat M_x^4 + \hat M_y^4 + \hat M_z^4)
\end{equation}
where $K$ is the anisotropy constant. The definition of $\mathcal{F}_a$ used here is identical to a recent ferromagnetic resonance study \cite{2017:Stasinopoulos:APL} and differs from the SANS study \cite{2018:Chacon:NatPhys} in terms of the sign of $K$ and the units of the magnetization. As shown in Ref.~\onlinecite{2018:Chacon:NatPhys}, the cubic anisotropy is sufficient to give rise to a stable LTS and a metastable tilted conical phase for a magnetic field along $\langle 100\rangle$, provided that $K$ is negative and exceeds a critical value $K_{c}$. An estimate of the critical ansisotropy value is given by $K_{c} \approx - 0.07 \mu_0 H^{\rm int}_{c2} M_s$, where $H^{\rm int}_{c2}$ is the internal critical field of the conical to field-polarized transition (evaluated for $K = 0$).  (Note that we use a different notation as compared to Ref.~\onlinecite{2018:Chacon:NatPhys}.) However, the stability of the LTS and tilted conical phases may be reduced or enhanced by additional contributions to the free energy arising, for example, from other anisotropy terms.

\begin{figure}
\centering
\includegraphics[width=0.9\linewidth]{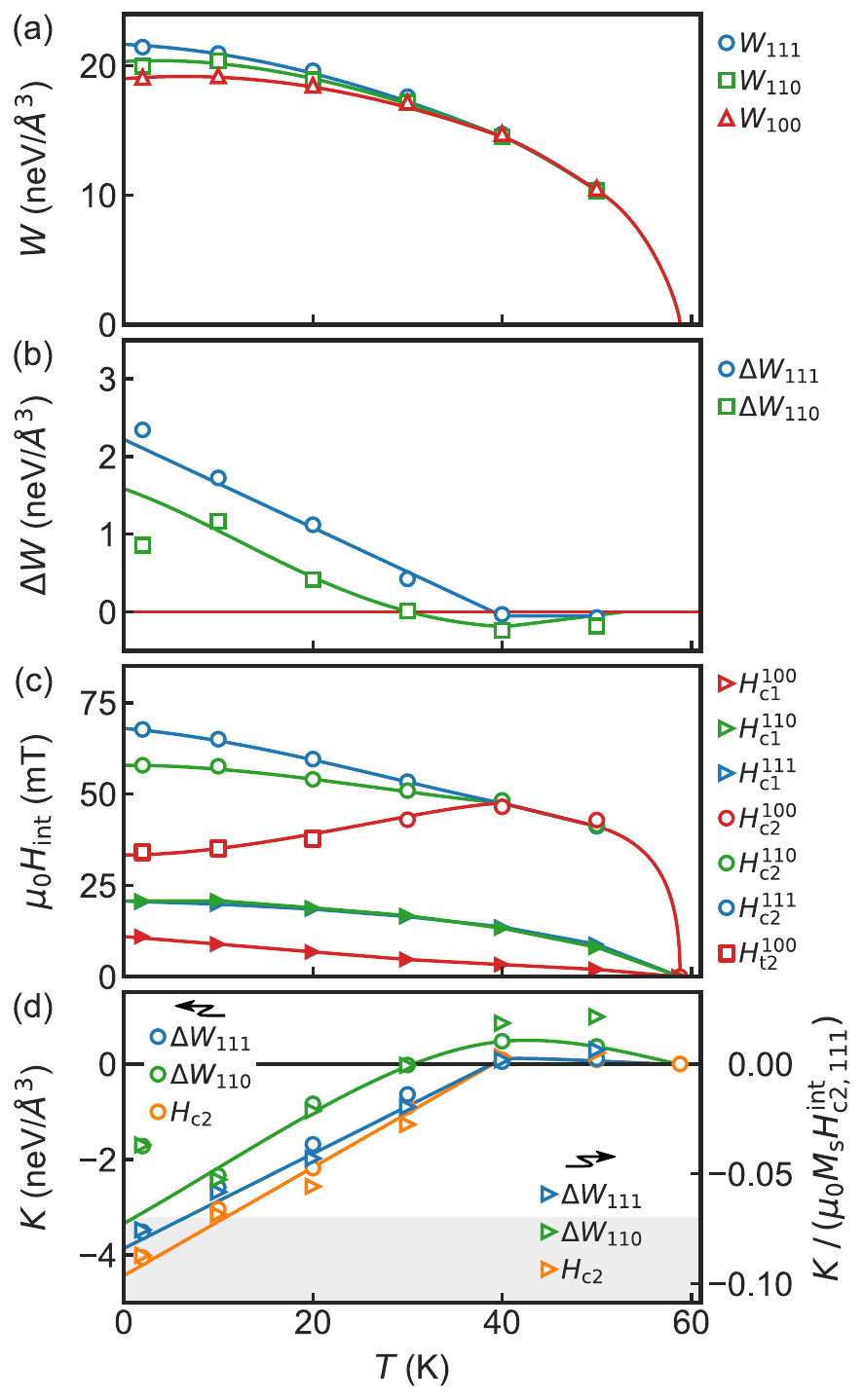}
\caption{\label{work_temp} 
Magnetic work $W$, transition fields $H_{\rm c1}$ and $H_{\rm c2}$, and anisotropy constant $K$ as inferred from the magnetization and the transition fields as a function of temperature. (a) Temperature dependence of the magnetic work $W$ for all major crystallographic orientations under magnetic field saturating the magnetization. (b) Difference of the magnetic work $\Delta W$ with respect to field along {\ozz}. (c) Transition fields $H_{\rm c1}$ and $H_{\rm c2}$ for field along {\ooo} (blue), {\ooz} (green), and {\ozz} (red) as a function temperature. (d) Anisotropy constant $K$ as a function of temperature, as extracted from the magnetic work (green, blue), and the upper critical field $H_\mathrm{c2}$ (orange). Shown in gray shading is the regime in which an LTS phase becomes favorable ($K\leq-0.07$).
}
\end{figure}

Following Ref.~\onlinecite{1937:Akulov} the magnetocrystalline anisotropy may be estimated by considering the magnetic work 
\begin{equation}
W_{\hat H} = \mu_0 \int_{0}^{M_s}  dM H
\end{equation}
to magnetize the system to the saturated state $M_s$ with a field pointing along the direction determined by the unit vector $\hat H$. The subscript $\hat H$ serves to indicate the dependence on the specific crystallographic direction. From the experimental magnetization data, we may determine $W_{\hat H}$ for fields along high symmetry directions, i.e., $\langle 100 \rangle$, $\langle 110 \rangle$, and $ \langle 111 \rangle$. Values as a function of temperature determined this way are shown in Fig.\ref{work_temp}\,(a). With decreasing temperature, $W$ increases monotonically below $T_{\rm c}$. At high temperatures, $W$ does not change as a function of field direction. However, towards lower temperatures a small but distinct anisotropy emerges below $\sim\SI{30}{\kelvin}$. 
In the zero temperature limit, we find $W\approx \SI{20}{\nano\electronvolt\angstrom^{-3}}$, which is in quantitative agreement with density functional calculations predicting the value $\SI{23.8}{\nano\electronvolt\angstrom^{-3}}$.\cite{2012:Yang:PRL} 
 
In the field-polarized state the magnetocrystalline anisotropy simplifies for magnetic fields along  high symmetry directions
\begin{equation}
\mathcal{F}_a =  K  \cdot
\begin{cases}
1   &   \mathbf{M} \parallel \langle 100 \rangle\\
1/2 &   \mathbf{M} \parallel \langle 110 \rangle\\
1/3 &   \mathbf{M} \parallel \langle 111 \rangle.
\end{cases}
\end{equation}
One may, for example, estimate the anisotropy constant by considering the differences of the magnetic work between different orientations: \cite{1937:Akulov}
\begin{eqnarray}
\label{diff-1}
W_{\langle 110 \rangle} - W_{\langle 100 \rangle} & = -\frac{1}{2} K\\
\label{diff-2}
W_{\langle 111 \rangle} - W_{\langle 100 \rangle} & = -\frac{2}{3} K.
\end{eqnarray}
These differences are shown in Fig.\,\ref{work_temp}(b). While the curves are guides to the eye, it is interesting to note the quantitative consistency of the different directions with Eqs.\,\ref{diff-1} and \ref{diff-2}.  The resulting anisotropy constant $K$ as a function of temperature derived from these differences is  shown by the blue and green curves in Fig.\ref{work_temp}\,(d). 

In fact, the critical field $H^{\rm int}_{c2}$ also displays a directional dependence in the presence of the anisotropy $K$ as reported in Ref.\,\onlinecite{2015:Grigoriev:PRB}. The difference of critical fields along $\langle 110 \rangle$ and $\langle 111 \rangle$ is proportional to $K$ and may be expressed as
\begin{equation}
\begin{split}
H^{\rm int}_{c2, {\langle 110\rangle}} &- H^{\rm int}_{c2, {\langle 111\rangle}} = \\
& -\frac{5}{3} \frac{K}{\mu_0 M_s} \left[1+
 \mathcal{O}\left(\frac{ K_{\sigma}}{\mu_0 H_{c2} M_s} \right) \right]
\end{split},
\end{equation}
where the correction, $ \mathcal{O}(...)$, may be neglected for small $K$. This allows to determine the value of $K$ with the help of the upper critical fields $H_{c2}$ for the three crystallographic directions {\ozz}, {\ooz}, and {\ooo} as shown in Figs.\,\ref{work_temp}(c) and \ref{work_temp}(d). At high temperatures, just below $T_{\rm c}$ essentially no anisotropy may be observed. However, with decreasing temperature, the values of $H_{\rm c2}$ for different directions separate below $\sim\!\SI{40}{\kelvin}$, becoming distinctly different. 

Values of the anisotropy constant $K$ inferred from $H_{c2}$ are shown as orange curve in Fig.\,\ref{work_temp}(d). With decreasing temperature $K$ deviates from zero and decreases below $\sim\!\SI{35}{\kelvin}$. Within the accuracy of the estimates carried out here it is not possible to rule out a change of sign and small positive value of $K$ between $T_c$ and $\sim\SI{35}{\kelvin}$. This might hint at the existence of further, weaker anisotropy terms. The putative existence of such terms may be clarified with the help of the anisotropy of $H_{c1}$. However, it is important to emphasize that despite a possible change of sign of $K$ as a function of temperature the easy magnetic axes remain the {\ozz} axes throughout. In the limit of low temperatures, the values of $K$ as calculated from magnetization and the upper critical fields are, finally, quantitatively consistent with the anisotropy constant determined at $T = 5$ K with the help of ferromagnetic resonance measurements, $K_{\rm FMR} = - 0.6 \times 10^3$  J/m$^3 \approx - 3.7$ neV/\AA$^3$ (see supplement of Ref.~\onlinecite{2017:Stasinopoulos:APL} for details). 

Also shown in Fig.\,\ref{work_temp}(d) are the anisotropy constants in dimensionless units, $K/(\mu_0 M_s H^{\rm int}_{c2,\langle111 \rangle}) \approx K/(\mu_0 M_s H^{\rm int}_{c2}|_{K=0})$ (ordinate on the right-hand side), where $M_s$ and $H^{\rm int}_{c2}$ are temperature dependent. As reported in Ref.~\onlinecite{2018:Chacon:NatPhys}  the cubic anisotropy may be expected to stabilize the LTS as a metatstable state without need for further exchange anisotropies when the ratio $K/(\mu_0 M_s H^{\rm int}_{c2,\langle111 \rangle})$ falls below a threshold of $-0.07$.  This condition is indicated by gray shading in Fig.\,\ref{fig:magnetic_work}(d). Indeed, below $\sim\!\SI{15}{\kelvin}$, where the LTS phase is observed for $\mathbf{H} \parallel \langle 100\rangle$, the ratio $K/(\mu_0 M_s H^{\rm int}_{c2,\langle111 \rangle})$ reaches this strength suggesting that the cubic anisotropy $K$ represents the main stabilization mechanism for the additional phases. 


\subsection{Effective energy landscape}
\label{potential}

For weak magnetic anisotropies it is very well established experimentally and theoretically, that the direction of the conical helix aligns with the magnetic field. While this alignment is on the expense of magnetic anisotropy energy, it allows to minimize Zeeman energy as the magnetic moments cant most effectively towards the field direction. In turn, this raises the question why a sufficiently strong anisotropy stabilizes  a skyrmion phase and a tilted conical phase in which the moments twist around propagation directions that are not aligned with the magnetic field. In the following, we present a qualitative discussion in order to provide an intuitive picture of the competition between Zeeman energy and magnetocrystalline anisotropy for modulated structures. A quantitative numerical analysis has been reported in Ref.~\onlinecite{2018:Chacon:NatPhys}.

For what follows, it is helpful to consider the magnetization of a right-handed conical modulation,
\begin{equation} \label{ConHelix}
\begin{split}
\mathbf{M}(\mathbf{r})/M_s  = & \cos \alpha\, \hat e_3 \\\nonumber
&+ \sin \alpha \,\, (\hat e_1 \cos(k \hat e_3 \mathbf{r}) 
+ \hat e_2 \sin(k \hat e_3 \mathbf{r}) )
\end{split},
\end{equation}
where $M_s$ is the saturated magnetization, $\alpha$ is the cone angle with $\cos \alpha \approx H/H_{\rm c2}$ and $k$ is the magnitude of the pitch vector that is oriented along the unit vector $\hat k = \hat e_3$ with the orthonormal right-handed basis $\hat e_1 \times \hat e_2 = \hat e_3$. The uniform magnetization component of the conical helix points into the same direction than the pitch vector, namely $\hat e_3$. Using the pristine conical helix as the starting point of a variational calculation in the magnetocrystalline potential of Eq.~\eqref{CubicAniso}, an effective anisotropy potential for the orientation of the pitch vector $\hat k$ may be obtained given by
\begin{equation}
\mathcal{V}_{\rm a}(\hat k) = K_{\rm eff}(\alpha) \left(\hat k_x^4 + \hat k_y^4 + \hat k_z^4 \right).
\end{equation} 
We note that this approximation neglects distortions of the conical helix, which are of the order of a few \% as observed in the SANS study (see supplement of Ref.~\onlinecite{2018:Chacon:NatPhys}; see also discussion in Ref.~\onlinecite{2017:Bauer:PRB}). The effective anisotropy, $K_{\rm eff}$, depends on the cone angle $\alpha$ where
\begin{equation} \label{EffAniso}
K_{\rm eff}(\alpha) = \frac{K}{64} \left(9 + 20 \cos(2\alpha) + 35 \cos(4\alpha) \right).
\end{equation}
As a key observation, $K_{\rm eff}$ changes sign as a function of $\alpha$. Whereas the sign of $K$ and $K_{\rm eff}$ are the same for small $\alpha$ and cone angles close to $\pi/2$, they possess opposite sign at intermediate cone angles, $30^\circ \lesssim \alpha \lesssim 70^\circ$.  

\begin{figure}[t]
\centering
\includegraphics[width=1.0\linewidth]{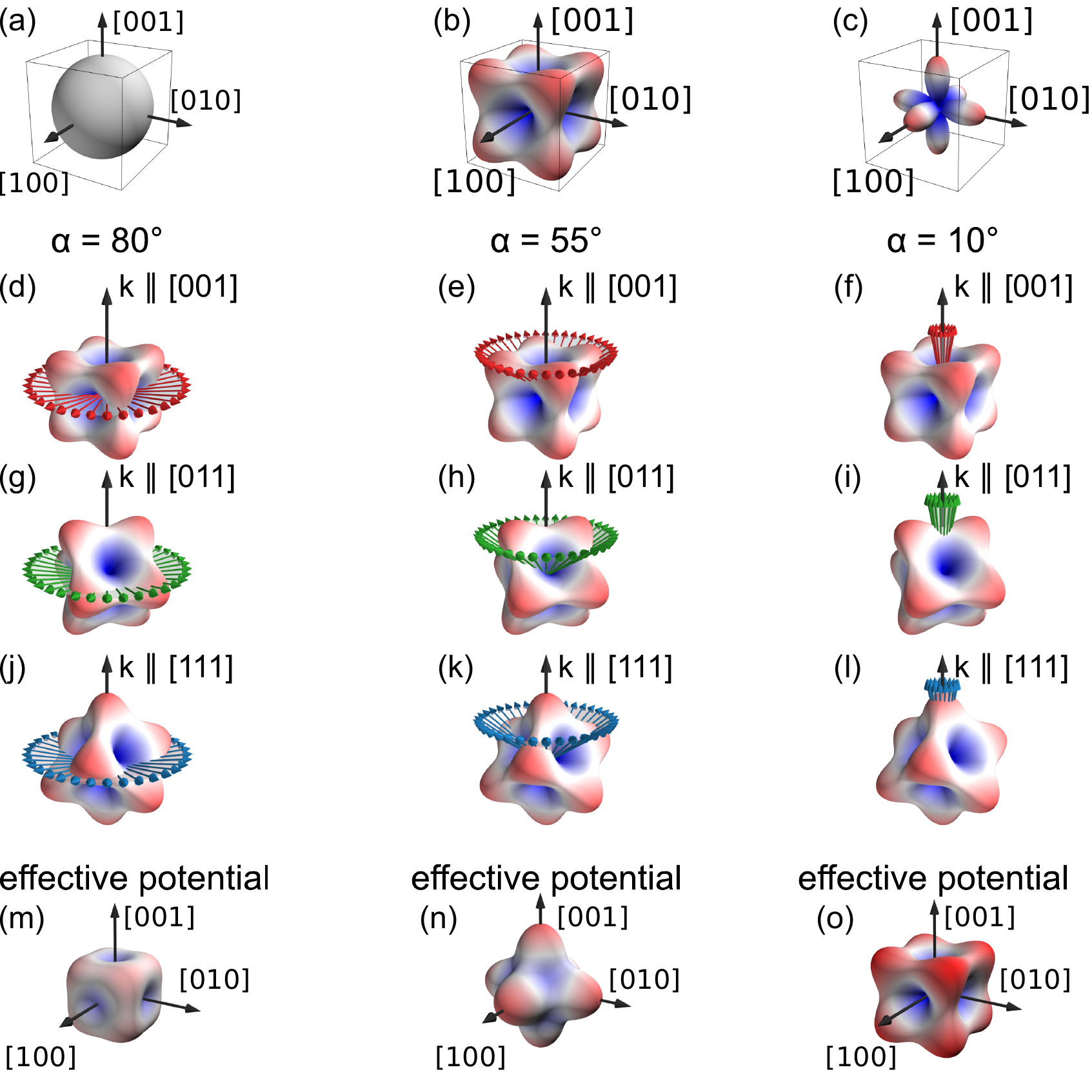}
\caption{\label{fig:energy_surface}
Illustration of the effective anisotropy that arises when the conical helix, Eq\,\eqref{ConHelix}, is embedded in the magnetocrystalline energy landscape of Eq.~\eqref{CubicAniso}. Large energies are shown in red shading, whereas low energies are shown in blue shading. (a)--(c) explain the representation of the orientational dependence of the energetic cost with $K=0$, $K<0$ and $K>0$, respectively. For $K<0$ the magnetic easy axes are $\langle 100\rangle$, whereas they are the hard axes for $K>0$. (d)--(l) illustrate the orientations that are covered by the magnetic moments of the conical helix with a certain cone angle $\alpha$ for different field orientations of the pitch vector $\mathbf{k}$. (m)--(o) display the effective anisotropy potential of Eq.~\eqref{EffAniso} that changes sign as function of cone angle $\alpha$.
}
\end{figure}

The effective anisotropy potential is illustrated in Fig.\,\ref{fig:energy_surface}. Large energies are shown in red shading, whereas low energies are shown in blue shading. The first three panels represent a cubic potential (a) with $K = 0$, (b) with $K<0$ favoring the cubic $\langle 100\rangle$ axes, and (c) with $K>0$ favouring the cubic $\langle 111\rangle$ axes. The panels in the next three rows, (d)--(l), indicate the energy landscape of the magnetocrystalline potential, Eq.~\eqref{CubicAniso} with $K<0$, that is scanned by the conical helix state of Eq.~\eqref{ConHelix} for different cone angles $\alpha$ and different orientations of the pitch vector $\hat k$. Finally, the last three panels represent the effective anisotropy potential for the pitch orientation $\hat k$, i.e, the function $\mathcal{V}_{\rm a}(\hat k)$ for three different cone angles. Whereas $K_{\rm eff}(\alpha) < 0$ for $\alpha = 10^\circ$ in panel (m) and for $\alpha = 80^\circ$ in panel (o), it is positive $K_{\rm eff}(\alpha) > 0$ for $\alpha = 55^\circ$ in panel (n).

When the magnetic field is applied along a $\langle 100 \rangle$ direction, both, the Zeeman energy as well as the effective anisotropy energy are minimized for $K_{\rm eff}(\alpha) < 0$. However, they compete for intermediate cone angles where $K_{\rm eff}(\alpha) > 0$. Whereas the Zeeman energy favors to align the helix with the field, the effective anisotropy supports a pitch vector that is oriented away from the field, since, otherwise, the magnetic moments of the helical state would dominantly point into hard \ooo\ axes resulting in a large energy penalty, see Fig.~\ref{fig:energy_surface}(e). If the cubic anisotropy exceeds a threshold value $K < K_c$ with $K_c \approx - 0.07 \mu_0 M_s H^{\rm int}_{c2}$, the anisotropy dominates such that even a LTS phase becomes favorable. A numerical assessment \cite{2018:Chacon:NatPhys} reveals, moreover, that the LTS phase comprising modulations perpendicular to the field, even forms the ground state at intermediate magnetic fields. 

\begin{figure}
  \centering
  \includegraphics[width=1.0\linewidth]{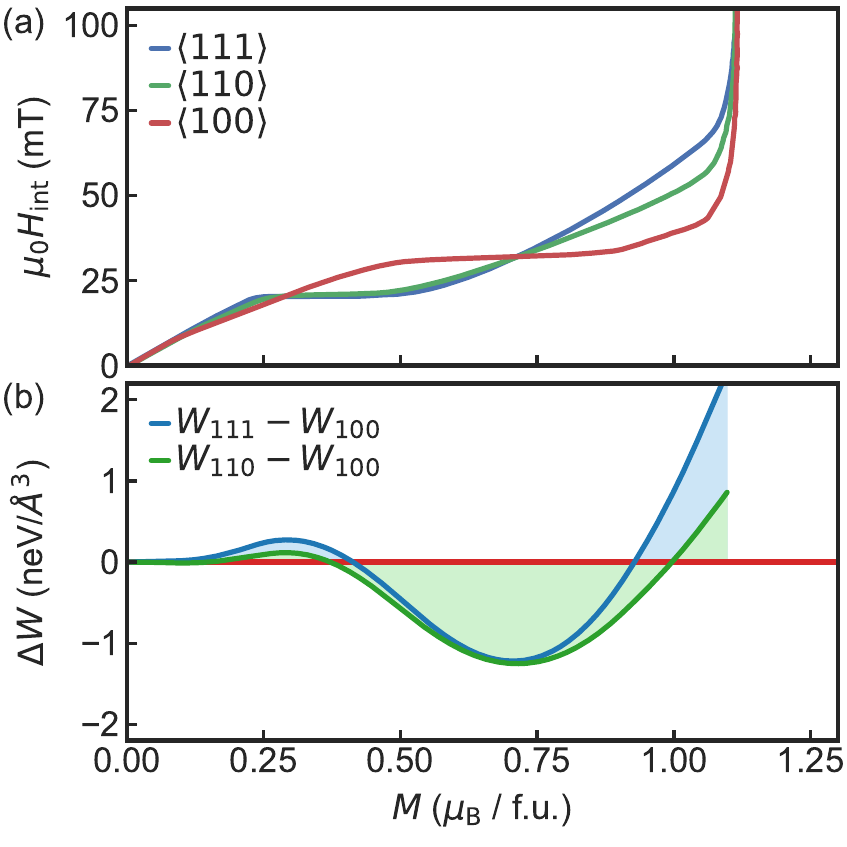}
  \caption{\label{fig:magnetic_work}
  Magnetization and magnetic work illustrating the field dependence of the effective anisotropy.
(a)~Internal magnetic field as a function of magnetization for different directions.
(b)~Magnetic work, inferred from magnetization data relative to the \ozz\ orientation as a function of the magnetization.
  }
\end{figure}

The notion of an effective magnetocrystalline anisotropy, Eq.~\eqref{EffAniso}, may also be confirmed by considering the magnetization process. Fig.~\ref{fig:magnetic_work}(a) displays typical magnetization data in terms of the internal field $H_{\rm int}$ versus $M$ for three crystallographic directions. The data shown here were measured at a temperature of \SI{2}{\kelvin} following ZFC with field along \ooo\ (blue), \ooz\ (green), and \ozz\ (red). As pointed out first in Sec.\,\ref{zfc-data} the magnetization displays two intersections of the \ooo\ and \ooz\ directions. 
The magnetic work $W_{\hat H}(M) = \mu_0 \int_{0}^{M}  dM' H(M')$ for the $\langle 111\rangle$ and $\langle 110 \rangle$ direction as inferred from the magnetization data is shown in Fig.~\ref{fig:magnetic_work}(b) relative to the magnetic work for the \ozz\ orientation. It reflects the behavior of the effective anisotropy shown in Figs.\,\ref{fig:energy_surface}(m) through \ref{fig:energy_surface}(o). 

Close to saturation as well as for small values of $M$ both  \ooo\ and \ooz\ are energetically larger than \ozz\ identifying the latter as the easy axis. For intermediate magnetization, however, the energy of the \ooo\ and \ooz\ orientations drops below the \ozz\ orientation, equivalent to a change of the effective anisotropy, i.e., a sign change of $K_\mathrm{eff}$. The magnetization interval where the inversion occurs indeed corresponds to the magnetization range in which the tilted conical phase is observed.

\section{Conclusions}
\label{conclusions}

In conclusion, we reported a comprehensive study of the magnetization and ac susceptibility of the magnetic phase diagram of {\cso}. For magnetic field parallel to the $\langle 100\rangle$ axis in the cubic crystal structure and low temperatures we found clear evidence of the formation of two new phases, a tilted conical state and a LTS state identified recently in a SANS study. The magnetization and susceptibility are thereby in remarkable agreement with the SANS data, providing clear thermodynamic signatures of these two new phases. Complementary selected specific heat data support these results. A detailed analysis of the strength of the magnetic anisotropy establishes that the conventional quartic contribution to the free energy by itself is sufficient to stabilize the LTS phase. Detailed measurements exploring the role of different sample shapes shed new insights on the role of demagnetizing fields in the stabilization of the tilted conical state. Taken together, we find that the LTS phase in {\cso} represents a thermodynamically stable ground state driven by cubic magnetocrystalline aisotropies, whereas the tilted conical state exists as a metastable phase even for tiny demagnetization factors. 

It is finally instructive to speculate on the more general importance of our observations in {\cso}. As discussed in our paper the magnetocrystalline anisotropies in the B20 compounds MnSi and {\fcs} do not appear to be sufficient to drive the formation of a LTS phase. In contrast, recent reports in Co$_7$Zn$_7$Mn$_6$ identified a second skyrmion lattice phase at low temperatures interpreted as a three-dimensional order arising from the interplay of DM interactions with the effects of frustration.\cite{2018:Karube:SciAdv} Judging from the literature this material exhibits also a pronounced magnetocrystalline anisotropy for the $\langle 100 \rangle$ axes similar to {\cso}. It is therefore tempting to speculate, whether the low temperature skyrmion phase in Co$_7$Zn$_7$Mn$_6$ originates, in fact, in the same mechanism we identify in {\cso}, however in the presence large amounts of disorder. This speculation finds further support by the observation that the effects of frustration, which must be present in {\cso} for the sake of the same analogy, do not seem to appear essential for stabilizing the LTS in {\cso}.
\\\\
\begin{acknowledgments}
We wish to thank S. Mayr for support. 
AC, MH, and WS acknowledge financial support through the TUM Graduate School. 
This project has received funding from the European Research Council (ERC) under the European Union's Horizon 2020 research and innovation programme (grant agreement No 788031).
LH and AR acknowledge financial support through DFG CRC1238 (project C02). 
MG acknowledges financial support through DFG CRC 1143 and DFG Grant GR1072/5. 
MH, AC, AB, and CP acknowledge support through DFG TRR80 (projects E1, F2 and F7) and
ERC-AdG (291079 TOPFIT). 
MG, AR and CP acknowledge support through DFG SPP2137 (Skyrmionics).
\end{acknowledgments}


\begin{thebibliography}{51}%
\makeatletter
\providecommand \@ifxundefined [1]{%
 \@ifx{#1\undefined}
}%
\providecommand \@ifnum [1]{%
 \ifnum #1\expandafter \@firstoftwo
 \else \expandafter \@secondoftwo
 \fi
}%
\providecommand \@ifx [1]{%
 \ifx #1\expandafter \@firstoftwo
 \else \expandafter \@secondoftwo
 \fi
}%
\providecommand \natexlab [1]{#1}%
\providecommand \enquote  [1]{``#1''}%
\providecommand \bibnamefont  [1]{#1}%
\providecommand \bibfnamefont [1]{#1}%
\providecommand \citenamefont [1]{#1}%
\providecommand \href@noop [0]{\@secondoftwo}%
\providecommand \href [0]{\begingroup \@sanitize@url \@href}%
\providecommand \@href[1]{\@@startlink{#1}\@@href}%
\providecommand \@@href[1]{\endgroup#1\@@endlink}%
\providecommand \@sanitize@url [0]{\catcode `\\12\catcode `\$12\catcode
  `\&12\catcode `\#12\catcode `\^12\catcode `\_12\catcode `\%12\relax}%
\providecommand \@@startlink[1]{}%
\providecommand \@@endlink[0]{}%
\providecommand \url  [0]{\begingroup\@sanitize@url \@url }%
\providecommand \@url [1]{\endgroup\@href {#1}{\urlprefix }}%
\providecommand \urlprefix  [0]{URL }%
\providecommand \Eprint [0]{\href }%
\providecommand \doibase [0]{http://dx.doi.org/}%
\providecommand \selectlanguage [0]{\@gobble}%
\providecommand \bibinfo  [0]{\@secondoftwo}%
\providecommand \bibfield  [0]{\@secondoftwo}%
\providecommand \translation [1]{[#1]}%
\providecommand \BibitemOpen [0]{}%
\providecommand \bibitemStop [0]{}%
\providecommand \bibitemNoStop [0]{.\EOS\space}%
\providecommand \EOS [0]{\spacefactor3000\relax}%
\providecommand \BibitemShut  [1]{\csname bibitem#1\endcsname}%
\let\auto@bib@innerbib\@empty
\bibitem [{\citenamefont {Nagaosa}\ and\ \citenamefont
  {Tokura}({2013})}]{2013:Nagaosa:NN}%
  \BibitemOpen
  \bibfield  {author} {\bibinfo {author} {\bibfnamefont {Naoto}\ \bibnamefont
  {Nagaosa}}\ and\ \bibinfo {author} {\bibfnamefont {Yoshinori}\ \bibnamefont
  {Tokura}},\ }\bibfield  {title} {\enquote {\bibinfo {title} {{Topological
  properties and dynamics of magnetic skyrmions}},}\ }\href {\doibase
  {10.1038/NNANO.2013.243}} {\bibfield  {journal} {\bibinfo  {journal} {{Nature
  Nano.}}\ }\textbf {\bibinfo {volume} {{8}}},\ \bibinfo {pages} {{899--911}}
  (\bibinfo {year} {{2013}})}\BibitemShut {NoStop}%
\bibitem [{\citenamefont {M{\"u}hlbauer}\ \emph {et~al.}(2009)\citenamefont
  {M{\"u}hlbauer}, \citenamefont {Binz}, \citenamefont {Jonietz}, \citenamefont
  {Pfleiderer}, \citenamefont {Rosch}, \citenamefont {Neubauer}, \citenamefont
  {Georgii},\ and\ \citenamefont {B{\"o}ni}}]{2009:Muhlbauer:Science}%
  \BibitemOpen
  \bibfield  {author} {\bibinfo {author} {\bibfnamefont {S.}~\bibnamefont
  {M{\"u}hlbauer}}, \bibinfo {author} {\bibfnamefont {B.}~\bibnamefont {Binz}},
  \bibinfo {author} {\bibfnamefont {F.}~\bibnamefont {Jonietz}}, \bibinfo
  {author} {\bibfnamefont {C.}~\bibnamefont {Pfleiderer}}, \bibinfo {author}
  {\bibfnamefont {A.}~\bibnamefont {Rosch}}, \bibinfo {author} {\bibfnamefont
  {A.}~\bibnamefont {Neubauer}}, \bibinfo {author} {\bibfnamefont
  {R.}~\bibnamefont {Georgii}}, \ and\ \bibinfo {author} {\bibfnamefont
  {P.}~\bibnamefont {B{\"o}ni}},\ }\bibfield  {title} {\enquote {\bibinfo
  {title} {{Skyrmion Lattice in a Chiral Magnet}},}\ }\href {\doibase
  10.1126/science.1166767} {\bibfield  {journal} {\bibinfo  {journal}
  {Science}\ }\textbf {\bibinfo {volume} {323}},\ \bibinfo {pages} {915--919}
  (\bibinfo {year} {2009})}\BibitemShut {NoStop}%
\bibitem [{\citenamefont {M\"{u}nzer}\ \emph {et~al.}(2010)\citenamefont
  {M\"{u}nzer}, \citenamefont {Neubauer}, \citenamefont {Adams}, \citenamefont
  {M\"{u}hlbauer}, \citenamefont {Franz}, \citenamefont {Jonietz},
  \citenamefont {Georgii}, \citenamefont {B\"{o}ni}, \citenamefont {Pedersen},
  \citenamefont {Schmidt}, \citenamefont {Rosch},\ and\ \citenamefont
  {Pfleiderer}}]{2010:Munzer:PhysRevB}%
  \BibitemOpen
  \bibfield  {author} {\bibinfo {author} {\bibfnamefont {W.}~\bibnamefont
  {M\"{u}nzer}}, \bibinfo {author} {\bibfnamefont {A.}~\bibnamefont
  {Neubauer}}, \bibinfo {author} {\bibfnamefont {T.}~\bibnamefont {Adams}},
  \bibinfo {author} {\bibfnamefont {S.}~\bibnamefont {M\"{u}hlbauer}}, \bibinfo
  {author} {\bibfnamefont {C.}~\bibnamefont {Franz}}, \bibinfo {author}
  {\bibfnamefont {F.}~\bibnamefont {Jonietz}}, \bibinfo {author} {\bibfnamefont
  {R.}~\bibnamefont {Georgii}}, \bibinfo {author} {\bibfnamefont
  {P.}~\bibnamefont {B\"{o}ni}}, \bibinfo {author} {\bibfnamefont
  {B.}~\bibnamefont {Pedersen}}, \bibinfo {author} {\bibfnamefont
  {M.}~\bibnamefont {Schmidt}}, \bibinfo {author} {\bibfnamefont
  {A.}~\bibnamefont {Rosch}}, \ and\ \bibinfo {author} {\bibfnamefont
  {C.}~\bibnamefont {Pfleiderer}},\ }\bibfield  {title} {\enquote {\bibinfo
  {title} {{Skyrmion lattice in the doped semiconductor
  Fe$_{1-x}$Co$_{x}$Si}},}\ }\href {\doibase 10.1103/PhysRevB.81.041203}
  {\bibfield  {journal} {\bibinfo  {journal} {Phys. Rev. B}\ }\textbf {\bibinfo
  {volume} {81}},\ \bibinfo {pages} {041203(R)} (\bibinfo {year}
  {2010})}\BibitemShut {NoStop}%
\bibitem [{\citenamefont {Yu}\ \emph {et~al.}(2011)\citenamefont {Yu},
  \citenamefont {Kanazawa}, \citenamefont {Onose}, \citenamefont {Kimoto},
  \citenamefont {Zhang}, \citenamefont {Ishiwata}, \citenamefont {Matsui},\
  and\ \citenamefont {Tokura}}]{2011:Yu:NatureMater}%
  \BibitemOpen
  \bibfield  {author} {\bibinfo {author} {\bibfnamefont {X.~Z.}\ \bibnamefont
  {Yu}}, \bibinfo {author} {\bibfnamefont {N.}~\bibnamefont {Kanazawa}},
  \bibinfo {author} {\bibfnamefont {Y.}~\bibnamefont {Onose}}, \bibinfo
  {author} {\bibfnamefont {K.}~\bibnamefont {Kimoto}}, \bibinfo {author}
  {\bibfnamefont {W.~Z.}\ \bibnamefont {Zhang}}, \bibinfo {author}
  {\bibfnamefont {S.}~\bibnamefont {Ishiwata}}, \bibinfo {author}
  {\bibfnamefont {Y.}~\bibnamefont {Matsui}}, \ and\ \bibinfo {author}
  {\bibfnamefont {Y.}~\bibnamefont {Tokura}},\ }\bibfield  {title} {\enquote
  {\bibinfo {title} {{Near room-temperature formation of a skyrmion crystal in
  thin-films of the helimagnet FeGe}},}\ }\href {\doibase 10.1038/nmat2916}
  {\bibfield  {journal} {\bibinfo  {journal} {Nature Materials}\ }\textbf
  {\bibinfo {volume} {10}},\ \bibinfo {pages} {106--109} (\bibinfo {year}
  {2011})}\BibitemShut {NoStop}%
\bibitem [{\citenamefont {Kanazawa}\ \emph {et~al.}(2016)\citenamefont
  {Kanazawa}, \citenamefont {Nii}, \citenamefont {Zhang}, \citenamefont
  {Mishchenko}, \citenamefont {De~Filippis}, \citenamefont {Kagawa},
  \citenamefont {Iwasa}, \citenamefont {Nagaosa},\ and\ \citenamefont
  {Tokura}}]{Kanazawa:2016fd}%
  \BibitemOpen
  \bibfield  {author} {\bibinfo {author} {\bibfnamefont {N}~\bibnamefont
  {Kanazawa}}, \bibinfo {author} {\bibfnamefont {Y}~\bibnamefont {Nii}},
  \bibinfo {author} {\bibfnamefont {X~X}\ \bibnamefont {Zhang}}, \bibinfo
  {author} {\bibfnamefont {A~S}\ \bibnamefont {Mishchenko}}, \bibinfo {author}
  {\bibfnamefont {G}~\bibnamefont {De~Filippis}}, \bibinfo {author}
  {\bibfnamefont {F}~\bibnamefont {Kagawa}}, \bibinfo {author} {\bibfnamefont
  {Y}~\bibnamefont {Iwasa}}, \bibinfo {author} {\bibfnamefont {Naoto}\
  \bibnamefont {Nagaosa}}, \ and\ \bibinfo {author} {\bibfnamefont
  {Y}~\bibnamefont {Tokura}},\ }\bibfield  {title} {\enquote {\bibinfo {title}
  {{Critical phenomena of emergent magnetic monopoles in a chiral magnet}},}\
  }\href@noop {} {\bibfield  {journal} {\bibinfo  {journal} {Nature Commun.}\
  }\textbf {\bibinfo {volume} {7}},\ \bibinfo {pages} {11622} (\bibinfo {year}
  {2016})}\BibitemShut {NoStop}%
\bibitem [{\citenamefont {Kezsmarki}\ \emph {et~al.}({2015})\citenamefont
  {Kezsmarki}, \citenamefont {Bordacs}, \citenamefont {Milde}, \citenamefont
  {Neuber}, \citenamefont {Eng}, \citenamefont {White}, \citenamefont {Ronnow},
  \citenamefont {Dewhurst}, \citenamefont {Mochizuki}, \citenamefont {Yanai},
  \citenamefont {Nakamura}, \citenamefont {Ehlers}, \citenamefont {Tsurkan},\
  and\ \citenamefont {Loidl}}]{2015:Kezsmarki:NM}%
  \BibitemOpen
  \bibfield  {author} {\bibinfo {author} {\bibfnamefont {I.}~\bibnamefont
  {Kezsmarki}}, \bibinfo {author} {\bibfnamefont {S.}~\bibnamefont {Bordacs}},
  \bibinfo {author} {\bibfnamefont {P.}~\bibnamefont {Milde}}, \bibinfo
  {author} {\bibfnamefont {E.}~\bibnamefont {Neuber}}, \bibinfo {author}
  {\bibfnamefont {L.~M.}\ \bibnamefont {Eng}}, \bibinfo {author} {\bibfnamefont
  {J.~S.}\ \bibnamefont {White}}, \bibinfo {author} {\bibfnamefont {H.~M.}\
  \bibnamefont {Ronnow}}, \bibinfo {author} {\bibfnamefont {C.~D.}\
  \bibnamefont {Dewhurst}}, \bibinfo {author} {\bibfnamefont {M.}~\bibnamefont
  {Mochizuki}}, \bibinfo {author} {\bibfnamefont {K.}~\bibnamefont {Yanai}},
  \bibinfo {author} {\bibfnamefont {H.}~\bibnamefont {Nakamura}}, \bibinfo
  {author} {\bibfnamefont {D.}~\bibnamefont {Ehlers}}, \bibinfo {author}
  {\bibfnamefont {V.}~\bibnamefont {Tsurkan}}, \ and\ \bibinfo {author}
  {\bibfnamefont {A.}~\bibnamefont {Loidl}},\ }\bibfield  {title} {\enquote
  {\bibinfo {title} {{Neel-type skyrmion lattice with confined orientation in
  the polar magnetic semiconductor GaV$_4$S$_8$}},}\ }\href {\doibase
  {10.1038/NMAT4402}} {\bibfield  {journal} {\bibinfo  {journal} {{Nature
  Materials}}\ }\textbf {\bibinfo {volume} {{14}}},\ \bibinfo {pages}
  {{1116--1122}} (\bibinfo {year} {{2015}})}\BibitemShut {NoStop}%
\bibitem [{\citenamefont {Bordacs}\ \emph {et~al.}({2017})\citenamefont
  {Bordacs}, \citenamefont {Butykai}, \citenamefont {Szigeti}, \citenamefont
  {White}, \citenamefont {Cubitt}, \citenamefont {Leonov}, \citenamefont
  {Widmann}, \citenamefont {Ehlers}, \citenamefont {von Nidda}, \citenamefont
  {Tsurkan}, \citenamefont {Loidl},\ and\ \citenamefont
  {Kezsmarki}}]{2017:Bordacs:SR}%
  \BibitemOpen
  \bibfield  {author} {\bibinfo {author} {\bibfnamefont {S.}~\bibnamefont
  {Bordacs}}, \bibinfo {author} {\bibfnamefont {A.}~\bibnamefont {Butykai}},
  \bibinfo {author} {\bibfnamefont {B.~G.}\ \bibnamefont {Szigeti}}, \bibinfo
  {author} {\bibfnamefont {J.~S.}\ \bibnamefont {White}}, \bibinfo {author}
  {\bibfnamefont {R.}~\bibnamefont {Cubitt}}, \bibinfo {author} {\bibfnamefont
  {A.~O.}\ \bibnamefont {Leonov}}, \bibinfo {author} {\bibfnamefont
  {S.}~\bibnamefont {Widmann}}, \bibinfo {author} {\bibfnamefont
  {D.}~\bibnamefont {Ehlers}}, \bibinfo {author} {\bibfnamefont {H.~A.~Krug}\
  \bibnamefont {von Nidda}}, \bibinfo {author} {\bibfnamefont {V.}~\bibnamefont
  {Tsurkan}}, \bibinfo {author} {\bibfnamefont {A.}~\bibnamefont {Loidl}}, \
  and\ \bibinfo {author} {\bibfnamefont {I.}~\bibnamefont {Kezsmarki}},\
  }\bibfield  {title} {\enquote {\bibinfo {title} {{Equilibrium Skyrmion
  Lattice Ground State in a Polar Easy-plane Magnet}},}\ }\href {\doibase
  {10.1038/s41598-017-07996-x}} {\bibfield  {journal} {\bibinfo  {journal}
  {{Sci. Rep.}}\ }\textbf {\bibinfo {volume} {{7}}},\ \bibinfo {pages} {{7584}}
  (\bibinfo {year} {{2017}})}\BibitemShut {NoStop}%
\bibitem [{\citenamefont {Yu}\ \emph {et~al.}({2012})\citenamefont {Yu},
  \citenamefont {Mostovoy}, \citenamefont {Tokunaga}, \citenamefont {Zhang},
  \citenamefont {Kimoto}, \citenamefont {Matsui}, \citenamefont {Kaneko},
  \citenamefont {Nagaosa},\ and\ \citenamefont {Tokura}}]{2011:Yu:PNAS}%
  \BibitemOpen
  \bibfield  {author} {\bibinfo {author} {\bibfnamefont {Xiuzhen}\ \bibnamefont
  {Yu}}, \bibinfo {author} {\bibfnamefont {Maxim}\ \bibnamefont {Mostovoy}},
  \bibinfo {author} {\bibfnamefont {Yusuke}\ \bibnamefont {Tokunaga}}, \bibinfo
  {author} {\bibfnamefont {Weizhu}\ \bibnamefont {Zhang}}, \bibinfo {author}
  {\bibfnamefont {Koji}\ \bibnamefont {Kimoto}}, \bibinfo {author}
  {\bibfnamefont {Yoshio}\ \bibnamefont {Matsui}}, \bibinfo {author}
  {\bibfnamefont {Yoshio}\ \bibnamefont {Kaneko}}, \bibinfo {author}
  {\bibfnamefont {Naoto}\ \bibnamefont {Nagaosa}}, \ and\ \bibinfo {author}
  {\bibfnamefont {Yoshinori}\ \bibnamefont {Tokura}},\ }\bibfield  {title}
  {\enquote {\bibinfo {title} {{Magnetic stripes and skyrmions with helicity
  reversals}},}\ }\href {\doibase {10.1073/pnas.1118496109}} {\bibfield
  {journal} {\bibinfo  {journal} {{Proc. Nat. Acad. Sci.}}\ }\textbf {\bibinfo
  {volume} {{109}}},\ \bibinfo {pages} {{8856--8860}} (\bibinfo {year}
  {{2012}})}\BibitemShut {NoStop}%
\bibitem [{\citenamefont {Ishiwata}\ \emph {et~al.}(2011)\citenamefont
  {Ishiwata}, \citenamefont {Tokunaga}, \citenamefont {Kaneko}, \citenamefont
  {Okuyama}, \citenamefont {Tokunaga}, \citenamefont {Wakimoto}, \citenamefont
  {Kakurai}, \citenamefont {Arima}, \citenamefont {Taguchi},\ and\
  \citenamefont {Tokura}}]{2011:Ishiwata:PRB}%
  \BibitemOpen
  \bibfield  {author} {\bibinfo {author} {\bibfnamefont {S.}~\bibnamefont
  {Ishiwata}}, \bibinfo {author} {\bibfnamefont {M.}~\bibnamefont {Tokunaga}},
  \bibinfo {author} {\bibfnamefont {Y.}~\bibnamefont {Kaneko}}, \bibinfo
  {author} {\bibfnamefont {D.}~\bibnamefont {Okuyama}}, \bibinfo {author}
  {\bibfnamefont {Y.}~\bibnamefont {Tokunaga}}, \bibinfo {author}
  {\bibfnamefont {S.}~\bibnamefont {Wakimoto}}, \bibinfo {author}
  {\bibfnamefont {K.}~\bibnamefont {Kakurai}}, \bibinfo {author} {\bibfnamefont
  {T.}~\bibnamefont {Arima}}, \bibinfo {author} {\bibfnamefont
  {Y.}~\bibnamefont {Taguchi}}, \ and\ \bibinfo {author} {\bibfnamefont
  {Y.}~\bibnamefont {Tokura}},\ }\bibfield  {title} {\enquote {\bibinfo {title}
  {Versatile helimagnetic phases under magnetic fields in cubic perovskite
  {SrFeO$_3$}},}\ }\href {\doibase 10.1103/PhysRevB.84.054427} {\bibfield
  {journal} {\bibinfo  {journal} {Phys. Rev. B}\ }\textbf {\bibinfo {volume}
  {84}},\ \bibinfo {pages} {054427} (\bibinfo {year} {2011})}\BibitemShut
  {NoStop}%
\bibitem [{\citenamefont {Nayak}\ \emph {et~al.}({2017})\citenamefont {Nayak},
  \citenamefont {Kumar}, \citenamefont {Ma}, \citenamefont {Werner},
  \citenamefont {Pippel}, \citenamefont {Sahoo}, \citenamefont {Damay},
  \citenamefont {Roessler}, \citenamefont {Felser},\ and\ \citenamefont
  {Parkin}}]{2017:Nayak:Nature}%
  \BibitemOpen
  \bibfield  {author} {\bibinfo {author} {\bibfnamefont {Ajaya~K.}\
  \bibnamefont {Nayak}}, \bibinfo {author} {\bibfnamefont {Vivek}\ \bibnamefont
  {Kumar}}, \bibinfo {author} {\bibfnamefont {Tianping}\ \bibnamefont {Ma}},
  \bibinfo {author} {\bibfnamefont {Peter}\ \bibnamefont {Werner}}, \bibinfo
  {author} {\bibfnamefont {Eckhard}\ \bibnamefont {Pippel}}, \bibinfo {author}
  {\bibfnamefont {Roshnee}\ \bibnamefont {Sahoo}}, \bibinfo {author}
  {\bibfnamefont {Francoise}\ \bibnamefont {Damay}}, \bibinfo {author}
  {\bibfnamefont {Ulrich~K.}\ \bibnamefont {Roessler}}, \bibinfo {author}
  {\bibfnamefont {Claudia}\ \bibnamefont {Felser}}, \ and\ \bibinfo {author}
  {\bibfnamefont {Stuart S.~P.}\ \bibnamefont {Parkin}},\ }\bibfield  {title}
  {\enquote {\bibinfo {title} {{Magnetic antiskyrmions above room temperature
  in tetragonal Heusler materials}},}\ }\href {\doibase {10.1038/nature23466}}
  {\bibfield  {journal} {\bibinfo  {journal} {{Nature}}\ }\textbf {\bibinfo
  {volume} {{548}}},\ \bibinfo {pages} {{561--566}} (\bibinfo {year}
  {{2017}})}\BibitemShut {NoStop}%
\bibitem [{\citenamefont {Yu}\ \emph {et~al.}(2013)\citenamefont {Yu},
  \citenamefont {DeGrave}, \citenamefont {Hara}, \citenamefont {Hara},
  \citenamefont {Jin},\ and\ \citenamefont {Tokura}}]{2013:Yu:NanoLett}%
  \BibitemOpen
  \bibfield  {author} {\bibinfo {author} {\bibfnamefont {X.}~\bibnamefont
  {Yu}}, \bibinfo {author} {\bibfnamefont {J.P.}\ \bibnamefont {DeGrave}},
  \bibinfo {author} {\bibfnamefont {Y.}~\bibnamefont {Hara}}, \bibinfo {author}
  {\bibfnamefont {T.}~\bibnamefont {Hara}}, \bibinfo {author} {\bibfnamefont
  {S.}~\bibnamefont {Jin}}, \ and\ \bibinfo {author} {\bibfnamefont
  {Y.}~\bibnamefont {Tokura}},\ }\bibfield  {title} {\enquote {\bibinfo {title}
  {{Observation of the Magnetic Skyrmion Lattice in a MnSi Nanowire by Lorentz
  TEM}},}\ }\href {\doibase 10.1021/nl401687d} {\bibfield  {journal} {\bibinfo
  {journal} {Nano Letters}\ }\textbf {\bibinfo {volume} {13}},\ \bibinfo
  {pages} {3755} (\bibinfo {year} {2013})}\BibitemShut {NoStop}%
\bibitem [{\citenamefont {Meynell}\ \emph {et~al.}(2017)\citenamefont
  {Meynell}, \citenamefont {Wilson}, \citenamefont {Krycka}, \citenamefont
  {Kirby}, \citenamefont {Fritzsche},\ and\ \citenamefont
  {Monchesky}}]{2017:Meynell:PRB}%
  \BibitemOpen
  \bibfield  {author} {\bibinfo {author} {\bibfnamefont {S.~A.}\ \bibnamefont
  {Meynell}}, \bibinfo {author} {\bibfnamefont {M.~N.}\ \bibnamefont {Wilson}},
  \bibinfo {author} {\bibfnamefont {K.~L.}\ \bibnamefont {Krycka}}, \bibinfo
  {author} {\bibfnamefont {B.~J.}\ \bibnamefont {Kirby}}, \bibinfo {author}
  {\bibfnamefont {H.}~\bibnamefont {Fritzsche}}, \ and\ \bibinfo {author}
  {\bibfnamefont {T.~L.}\ \bibnamefont {Monchesky}},\ }\bibfield  {title}
  {\enquote {\bibinfo {title} {Neutron study of in-plane skyrmions in mnsi thin
  films},}\ }\href {\doibase 10.1103/PhysRevB.96.054402} {\bibfield  {journal}
  {\bibinfo  {journal} {Phys. Rev. B}\ }\textbf {\bibinfo {volume} {96}},\
  \bibinfo {pages} {054402} (\bibinfo {year} {2017})}\BibitemShut {NoStop}%
\bibitem [{\citenamefont {Fert}\ \emph {et~al.}(2017)\citenamefont {Fert},
  \citenamefont {Reyren},\ and\ \citenamefont {Cros}}]{2017:Fert:NatRevMat}%
  \BibitemOpen
  \bibfield  {author} {\bibinfo {author} {\bibfnamefont {A.}~\bibnamefont
  {Fert}}, \bibinfo {author} {\bibfnamefont {N.}~\bibnamefont {Reyren}}, \ and\
  \bibinfo {author} {\bibfnamefont {V.}~\bibnamefont {Cros}},\ }\bibfield
  {title} {\enquote {\bibinfo {title} {{Magnetic skyrmions: advances in physics
  and potential applications}},}\ }\href {\doibase 10.1038/natrevmats.2017.31}
  {\bibfield  {journal} {\bibinfo  {journal} {Nature Reviews Materials}\
  }\textbf {\bibinfo {volume} {2}},\ \bibinfo {pages} {17031} (\bibinfo {year}
  {2017})}\BibitemShut {NoStop}%
\bibitem [{\citenamefont {Jiang}\ \emph {et~al.}(2017)\citenamefont {Jiang},
  \citenamefont {Chen}, \citenamefont {Liu}, \citenamefont {Zang},
  \citenamefont {te~Velthuis},\ and\ \citenamefont
  {Hoffmann}}]{2017:Jiang:PhysRep}%
  \BibitemOpen
  \bibfield  {author} {\bibinfo {author} {\bibfnamefont {Wanjun}\ \bibnamefont
  {Jiang}}, \bibinfo {author} {\bibfnamefont {Gong}\ \bibnamefont {Chen}},
  \bibinfo {author} {\bibfnamefont {Kai}\ \bibnamefont {Liu}}, \bibinfo
  {author} {\bibfnamefont {Jiadong}\ \bibnamefont {Zang}}, \bibinfo {author}
  {\bibfnamefont {S.G.E.}\ \bibnamefont {te~Velthuis}}, \ and\ \bibinfo
  {author} {\bibfnamefont {A.}~\bibnamefont {Hoffmann}},\ }\bibfield  {title}
  {\enquote {\bibinfo {title} {{Skyrmions in magnetic multilayers}},}\ }\href
  {\doibase 10.1016/j.physrep.2017.08.001} {\bibfield  {journal} {\bibinfo
  {journal} {Physics Reports}\ }\textbf {\bibinfo {volume} {704}},\ \bibinfo
  {pages} {1 -- 49} (\bibinfo {year} {2017})}\BibitemShut {NoStop}%
\bibitem [{\citenamefont {Wiesendanger}(2016)}]{2016:Wiesendanger:NatRevMat}%
  \BibitemOpen
  \bibfield  {author} {\bibinfo {author} {\bibfnamefont {R.}~\bibnamefont
  {Wiesendanger}},\ }\bibfield  {title} {\enquote {\bibinfo {title} {{Nanoscale
  magnetic skyrmions in metallic films and multilayers: a new twist for
  spintronics}},}\ }\href {\doibase 10.1038/natrevmats.2016.44} {\bibfield
  {journal} {\bibinfo  {journal} {Nature Reviews Materials}\ }\textbf {\bibinfo
  {volume} {1}},\ \bibinfo {pages} {16044} (\bibinfo {year}
  {2016})}\BibitemShut {NoStop}%
\bibitem [{\citenamefont {Chacon}\ \emph {et~al.}(2018)\citenamefont {Chacon},
  \citenamefont {Heinen}, \citenamefont {Halder}, \citenamefont {Bauer},
  \citenamefont {Simeth}, \citenamefont {M{\"u}hlbauer}, \citenamefont
  {Berger}, \citenamefont {Garst}, \citenamefont {Rosch},\ and\ \citenamefont
  {Pfleiderer}}]{2018:Chacon:NatPhys}%
  \BibitemOpen
  \bibfield  {author} {\bibinfo {author} {\bibfnamefont {A.}~\bibnamefont
  {Chacon}}, \bibinfo {author} {\bibfnamefont {L.}~\bibnamefont {Heinen}},
  \bibinfo {author} {\bibfnamefont {M.}~\bibnamefont {Halder}}, \bibinfo
  {author} {\bibfnamefont {A.}~\bibnamefont {Bauer}}, \bibinfo {author}
  {\bibfnamefont {W.}~\bibnamefont {Simeth}}, \bibinfo {author} {\bibfnamefont
  {S.}~\bibnamefont {M{\"u}hlbauer}}, \bibinfo {author} {\bibfnamefont
  {H.}~\bibnamefont {Berger}}, \bibinfo {author} {\bibfnamefont
  {M.}~\bibnamefont {Garst}}, \bibinfo {author} {\bibfnamefont
  {A.}~\bibnamefont {Rosch}}, \ and\ \bibinfo {author} {\bibfnamefont
  {C.}~\bibnamefont {Pfleiderer}},\ }\bibfield  {title} {\enquote {\bibinfo
  {title} {{Observation of two independent skyrmion phases in a chiral magnetic
  material}},}\ }\href {\doibase DOI: 10.1038/s41567-018-0184-y} {\bibfield
  {journal} {\bibinfo  {journal} {Nature Physics}\ }\textbf {\bibinfo {volume}
  {14}},\ \bibinfo {pages} {936} (\bibinfo {year} {2018})},\ \bibinfo {note}
  {advance online publication on 22 June 2018}\BibitemShut {NoStop}%
\bibitem [{\citenamefont {Seki}\ \emph {et~al.}(2012)\citenamefont {Seki},
  \citenamefont {Yu}, \citenamefont {Ishiwata},\ and\ \citenamefont
  {Tokura}}]{2012:Seki:Science}%
  \BibitemOpen
  \bibfield  {author} {\bibinfo {author} {\bibfnamefont {S.}~\bibnamefont
  {Seki}}, \bibinfo {author} {\bibfnamefont {X.~Z.}\ \bibnamefont {Yu}},
  \bibinfo {author} {\bibfnamefont {S.}~\bibnamefont {Ishiwata}}, \ and\
  \bibinfo {author} {\bibfnamefont {Y.}~\bibnamefont {Tokura}},\ }\bibfield
  {title} {\enquote {\bibinfo {title} {{Observation of Skyrmions in a
  Multiferroic Material}},}\ }\href {\doibase 10.1126/science.1214143}
  {\bibfield  {journal} {\bibinfo  {journal} {Science}\ }\textbf {\bibinfo
  {volume} {336}},\ \bibinfo {pages} {198--201} (\bibinfo {year}
  {2012})}\BibitemShut {NoStop}%
\bibitem [{\citenamefont {Adams}\ \emph {et~al.}(2012)\citenamefont {Adams},
  \citenamefont {Chacon}, \citenamefont {Wagner}, \citenamefont {Bauer},
  \citenamefont {Brandl}, \citenamefont {Pedersen}, \citenamefont {Berger},
  \citenamefont {Lemmens},\ and\ \citenamefont
  {Pfleiderer}}]{2012:Adams:PhysRevLett}%
  \BibitemOpen
  \bibfield  {author} {\bibinfo {author} {\bibfnamefont {T.}~\bibnamefont
  {Adams}}, \bibinfo {author} {\bibfnamefont {A.}~\bibnamefont {Chacon}},
  \bibinfo {author} {\bibfnamefont {M.}~\bibnamefont {Wagner}}, \bibinfo
  {author} {\bibfnamefont {A.}~\bibnamefont {Bauer}}, \bibinfo {author}
  {\bibfnamefont {G.}~\bibnamefont {Brandl}}, \bibinfo {author} {\bibfnamefont
  {B.}~\bibnamefont {Pedersen}}, \bibinfo {author} {\bibfnamefont
  {H.}~\bibnamefont {Berger}}, \bibinfo {author} {\bibfnamefont
  {P.}~\bibnamefont {Lemmens}}, \ and\ \bibinfo {author} {\bibfnamefont
  {C.}~\bibnamefont {Pfleiderer}},\ }\bibfield  {title} {\enquote {\bibinfo
  {title} {{Long-Wavelength Helimagnetic Order and Skyrmion Lattice Phase in
  Cu$_{2}$OSeO$_{3}$}},}\ }\href {\doibase 10.1103/PhysRevLett.108.237204}
  {\bibfield  {journal} {\bibinfo  {journal} {Phys. Rev. Lett.}\ }\textbf
  {\bibinfo {volume} {108}},\ \bibinfo {pages} {237204} (\bibinfo {year}
  {2012})}\BibitemShut {NoStop}%
\bibitem [{\citenamefont {Buhrandt}\ and\ \citenamefont
  {Fritz}(2013)}]{2013:Buhrandt:PRB}%
  \BibitemOpen
  \bibfield  {author} {\bibinfo {author} {\bibfnamefont {Stefan}\ \bibnamefont
  {Buhrandt}}\ and\ \bibinfo {author} {\bibfnamefont {Lars}\ \bibnamefont
  {Fritz}},\ }\bibfield  {title} {\enquote {\bibinfo {title} {Skyrmion lattice
  phase in three-dimensional chiral magnets from monte carlo simulations},}\
  }\href {\doibase 10.1103/PhysRevB.88.195137} {\bibfield  {journal} {\bibinfo
  {journal} {Phys. Rev. B}\ }\textbf {\bibinfo {volume} {88}},\ \bibinfo
  {pages} {195137} (\bibinfo {year} {2013})}\BibitemShut {NoStop}%
\bibitem [{\citenamefont {Bauer}\ and\ \citenamefont
  {Pfleiderer}(2016{\natexlab{a}})}]{2016:Bauer:Book}%
  \BibitemOpen
  \bibfield  {author} {\bibinfo {author} {\bibfnamefont {A.}~\bibnamefont
  {Bauer}}\ and\ \bibinfo {author} {\bibfnamefont {C.}~\bibnamefont
  {Pfleiderer}},\ }\enquote {\bibinfo {title} {{Topological Structures in
  Ferroic Materials: Domain Walls, Vortices and Skyrmions}},}\ \ (\bibinfo
  {publisher} {Springer International Publishing},\ \bibinfo {year} {2016})\
  Chap.\ \bibinfo {chapter} {{Generic Aspects of Skyrmion Lattices in Chiral
  Magnets}}, p.~\bibinfo {pages} {1}\BibitemShut {NoStop}%
\bibitem [{\citenamefont {Landau}\ and\ \citenamefont
  {Lifshitz}(1980)}]{Landau}%
  \BibitemOpen
  \bibfield  {author} {\bibinfo {author} {\bibfnamefont {L.~D.}\ \bibnamefont
  {Landau}}\ and\ \bibinfo {author} {\bibfnamefont {E.~M.}\ \bibnamefont
  {Lifshitz}},\ }\href@noop {} {\emph {\bibinfo {title} {Course of theoretical
  physics, vol. 8}}}\ (\bibinfo  {publisher} {Pergamon Press},\ \bibinfo {year}
  {1980})\BibitemShut {NoStop}%
\bibitem [{\citenamefont {R{\"o}{\ss}ler}\ \emph {et~al.}(2006)\citenamefont
  {R{\"o}{\ss}ler}, \citenamefont {Bogdanov},\ and\ \citenamefont
  {Pfleiderer}}]{2006:Rossler:Nature}%
  \BibitemOpen
  \bibfield  {author} {\bibinfo {author} {\bibfnamefont {U.~K.}\ \bibnamefont
  {R{\"o}{\ss}ler}}, \bibinfo {author} {\bibfnamefont {A.~N.}\ \bibnamefont
  {Bogdanov}}, \ and\ \bibinfo {author} {\bibfnamefont {C.}~\bibnamefont
  {Pfleiderer}},\ }\bibfield  {title} {\enquote {\bibinfo {title} {Spontaneous
  skyrmion ground states in magnetic metals},}\ }\href@noop {} {\bibfield
  {journal} {\bibinfo  {journal} {Nature}\ }\textbf {\bibinfo {volume} {442}},\
  \bibinfo {pages} {797} (\bibinfo {year} {2006})}\BibitemShut {NoStop}%
\bibitem [{\citenamefont {Wilhelm}\ \emph {et~al.}(2011)\citenamefont
  {Wilhelm}, \citenamefont {Baenitz}, \citenamefont {Schmidt}, \citenamefont
  {R{\"o}{\ss}ler}, \citenamefont {Leonov},\ and\ \citenamefont
  {Bogdanov}}]{wilhelm:PRL:11}%
  \BibitemOpen
  \bibfield  {author} {\bibinfo {author} {\bibfnamefont {H.}~\bibnamefont
  {Wilhelm}}, \bibinfo {author} {\bibfnamefont {M.}~\bibnamefont {Baenitz}},
  \bibinfo {author} {\bibfnamefont {M.}~\bibnamefont {Schmidt}}, \bibinfo
  {author} {\bibfnamefont {U.~K.}\ \bibnamefont {R{\"o}{\ss}ler}}, \bibinfo
  {author} {\bibfnamefont {A.~A.}\ \bibnamefont {Leonov}}, \ and\ \bibinfo
  {author} {\bibfnamefont {A.~N.}\ \bibnamefont {Bogdanov}},\ }\bibfield
  {title} {\enquote {\bibinfo {title} {Precursor {{Phenomena}} at the
  {{Magnetic Ordering}} of the {{Cubic Helimagnet FeGe}}},}\ }\href {\doibase
  10.1103/PhysRevLett.107.127203} {\bibfield  {journal} {\bibinfo  {journal}
  {Phys. Rev. Lett.}\ }\textbf {\bibinfo {volume} {107}},\ \bibinfo {pages}
  {127203} (\bibinfo {year} {2011})}\BibitemShut {NoStop}%
\bibitem [{\citenamefont {Cevey}\ \emph {et~al.}(2013)\citenamefont {Cevey},
  \citenamefont {Wilhelm}, \citenamefont {Schmidt},\ and\ \citenamefont
  {Lortz}}]{2013:Cevey:pssb}%
  \BibitemOpen
  \bibfield  {author} {\bibinfo {author} {\bibfnamefont {L.}~\bibnamefont
  {Cevey}}, \bibinfo {author} {\bibfnamefont {H.}~\bibnamefont {Wilhelm}},
  \bibinfo {author} {\bibfnamefont {M.}~\bibnamefont {Schmidt}}, \ and\
  \bibinfo {author} {\bibfnamefont {R.}~\bibnamefont {Lortz}},\ }\bibfield
  {title} {\enquote {\bibinfo {title} {Thermodynamic investigations in the
  precursor region of {FeGe}},}\ }\href {\doibase 10.1002/pssb.201200632}
  {\bibfield  {journal} {\bibinfo  {journal} {Phys. Status Solidi B}\ }\textbf
  {\bibinfo {volume} {250}},\ \bibinfo {pages} {650} (\bibinfo {year}
  {2013})}\BibitemShut {NoStop}%
\bibitem [{\citenamefont {Bauer}\ and\ \citenamefont
  {Pfleiderer}(2012)}]{2012:Bauer:PRB}%
  \BibitemOpen
  \bibfield  {author} {\bibinfo {author} {\bibfnamefont {A.}~\bibnamefont
  {Bauer}}\ and\ \bibinfo {author} {\bibfnamefont {C.}~\bibnamefont
  {Pfleiderer}},\ }\bibfield  {title} {\enquote {\bibinfo {title} {{Magnetic
  phase diagram of MnSi inferred from magnetization and ac susceptibility}},}\
  }\href {\doibase 10.1103/PhysRevB.85.214418} {\bibfield  {journal} {\bibinfo
  {journal} {Phys. Rev. B}\ }\textbf {\bibinfo {volume} {85}},\ \bibinfo
  {pages} {214418} (\bibinfo {year} {2012})}\BibitemShut {NoStop}%
\bibitem [{\citenamefont {Wu}\ \emph {et~al.}({2015})\citenamefont {Wu},
  \citenamefont {Wei}, \citenamefont {Chandrasekhar}, \citenamefont {Chen},
  \citenamefont {Berger},\ and\ \citenamefont {Yang}}]{2015:Wu:SciRep}%
  \BibitemOpen
  \bibfield  {author} {\bibinfo {author} {\bibfnamefont {H.~C.}\ \bibnamefont
  {Wu}}, \bibinfo {author} {\bibfnamefont {T.~Y.}\ \bibnamefont {Wei}},
  \bibinfo {author} {\bibfnamefont {K.~D.}\ \bibnamefont {Chandrasekhar}},
  \bibinfo {author} {\bibfnamefont {T.~Y.}\ \bibnamefont {Chen}}, \bibinfo
  {author} {\bibfnamefont {H.}~\bibnamefont {Berger}}, \ and\ \bibinfo {author}
  {\bibfnamefont {H.~D.}\ \bibnamefont {Yang}},\ }\bibfield  {title} {\enquote
  {\bibinfo {title} {{Unexpected observation of splitting of skyrmion phase in
  Zn doped {Cu$_2$OSeO$_3$}}},}\ }\href {\doibase {10.1038/srep13579}}
  {\bibfield  {journal} {\bibinfo  {journal} {{Scientific Reports}}\ }\textbf
  {\bibinfo {volume} {{5}}} (\bibinfo {year} {{2015}}),\
  {10.1038/srep13579}}\BibitemShut {NoStop}%
\bibitem [{Ste(2018)}]{Stefanici:arXiv}%
  \BibitemOpen
  \href@noop {} {} (\bibinfo {year} {2018}),\ \bibinfo {note} {a. Stefancic, S.
  Moody, T.J. Hicken, T.M. Birch, G. Balakrishnan, S.A. Barnett, M. Crisanti,
  J.S.O. Evans, S.J.R. Holt, K.J.A. Franke, P.D. Hatton, B.M. Huddart, M.R.
  Lees, F.L. Pratt, C.C. Tang, M.N. Wilson, F. Xiao, T. Lancaster, Origin of
  skyrmion lattice phase splitting in Zn-substituted Cu$_2$OSeO$_3$,
  unpublished, arXiv/1807.04641}\BibitemShut {NoStop}%
\bibitem [{\citenamefont {Bauer}\ \emph {et~al.}(2016)\citenamefont {Bauer},
  \citenamefont {Garst},\ and\ \citenamefont {Pfleiderer}}]{2016:Bauer:PRB}%
  \BibitemOpen
  \bibfield  {author} {\bibinfo {author} {\bibfnamefont {A.}~\bibnamefont
  {Bauer}}, \bibinfo {author} {\bibfnamefont {M.}~\bibnamefont {Garst}}, \ and\
  \bibinfo {author} {\bibfnamefont {C.}~\bibnamefont {Pfleiderer}},\ }\bibfield
   {title} {\enquote {\bibinfo {title} {{History dependence of the magnetic
  properties of single-crystal Fe$_{1-x}$Co$_{x}$Si}},}\ }\href {\doibase
  10.1103/PhysRevB.93.235144} {\bibfield  {journal} {\bibinfo  {journal} {Phys.
  Rev. B}\ }\textbf {\bibinfo {volume} {93}},\ \bibinfo {pages} {235144}
  (\bibinfo {year} {2016})}\BibitemShut {NoStop}%
\bibitem [{\citenamefont {Tokunaga}\ \emph {et~al.}(2015)\citenamefont
  {Tokunaga}, \citenamefont {Yu}, \citenamefont {White}, \citenamefont
  {R\o{}nnow}, \citenamefont {Morikawa}, \citenamefont {Taguchi},\ and\
  \citenamefont {Tokura}}]{2015:Tokunaga:NatCommun}%
  \BibitemOpen
  \bibfield  {author} {\bibinfo {author} {\bibfnamefont {Y.}~\bibnamefont
  {Tokunaga}}, \bibinfo {author} {\bibfnamefont {X.~Z.}\ \bibnamefont {Yu}},
  \bibinfo {author} {\bibfnamefont {J.~S.}\ \bibnamefont {White}}, \bibinfo
  {author} {\bibfnamefont {H.~M.}\ \bibnamefont {R\o{}nnow}}, \bibinfo {author}
  {\bibfnamefont {D.}~\bibnamefont {Morikawa}}, \bibinfo {author}
  {\bibfnamefont {Y.}~\bibnamefont {Taguchi}}, \ and\ \bibinfo {author}
  {\bibfnamefont {Y.}~\bibnamefont {Tokura}},\ }\bibfield  {title} {\enquote
  {\bibinfo {title} {{A new class of chiral materials hosting magnetic
  skyrmions beyond room temperature}},}\ }\href {\doibase 10.1038/ncomms8638}
  {\bibfield  {journal} {\bibinfo  {journal} {Nature Commun.}\ }\textbf
  {\bibinfo {volume} {6}},\ \bibinfo {pages} {7638} (\bibinfo {year}
  {2015})}\BibitemShut {NoStop}%
\bibitem [{\citenamefont {Karube}\ \emph {et~al.}(2016)\citenamefont {Karube},
  \citenamefont {White}, \citenamefont {Reynolds}, \citenamefont {Gavilano},
  \citenamefont {Oike}, \citenamefont {Kikkawa}, \citenamefont {Kagawa},
  \citenamefont {Tokunaga}, \citenamefont {Ronnow}, \citenamefont {Tokura},\
  and\ \citenamefont {Taguchi}}]{2016:Karube:NatMater}%
  \BibitemOpen
  \bibfield  {author} {\bibinfo {author} {\bibfnamefont {K.}~\bibnamefont
  {Karube}}, \bibinfo {author} {\bibfnamefont {J.~S.}\ \bibnamefont {White}},
  \bibinfo {author} {\bibfnamefont {N.}~\bibnamefont {Reynolds}}, \bibinfo
  {author} {\bibfnamefont {J.~L.}\ \bibnamefont {Gavilano}}, \bibinfo {author}
  {\bibfnamefont {H.}~\bibnamefont {Oike}}, \bibinfo {author} {\bibfnamefont
  {A.}~\bibnamefont {Kikkawa}}, \bibinfo {author} {\bibfnamefont
  {F.}~\bibnamefont {Kagawa}}, \bibinfo {author} {\bibfnamefont
  {Y.}~\bibnamefont {Tokunaga}}, \bibinfo {author} {\bibfnamefont {H.~M.}\
  \bibnamefont {Ronnow}}, \bibinfo {author} {\bibfnamefont {Y.}~\bibnamefont
  {Tokura}}, \ and\ \bibinfo {author} {\bibfnamefont {Y.}~\bibnamefont
  {Taguchi}},\ }\bibfield  {title} {\enquote {\bibinfo {title} {{Robust
  metastable skyrmions and their triangular--square lattice structural
  transition in a high-temperature chiral magnet}},}\ }\href {\doibase
  10.1038/NMAT4752} {\bibfield  {journal} {\bibinfo  {journal} {Nature
  Materials}\ }\textbf {\bibinfo {volume} {15}},\ \bibinfo {pages} {1237--1242}
  (\bibinfo {year} {2016})}\BibitemShut {NoStop}%
\bibitem [{\citenamefont {Milde}\ \emph {et~al.}(2013)\citenamefont {Milde},
  \citenamefont {K{\"o}hler}, \citenamefont {Seidel}, \citenamefont {Eng},
  \citenamefont {Bauer}, \citenamefont {Chacon}, \citenamefont {Kindervater},
  \citenamefont {M{\"u}hlbauer}, \citenamefont {Pfleiderer}, \citenamefont
  {Buhrandt}, \citenamefont {Sch{\"u}tte},\ and\ \citenamefont
  {Rosch}}]{2013:Milde:Science}%
  \BibitemOpen
  \bibfield  {author} {\bibinfo {author} {\bibfnamefont {P.}~\bibnamefont
  {Milde}}, \bibinfo {author} {\bibfnamefont {D.}~\bibnamefont {K{\"o}hler}},
  \bibinfo {author} {\bibfnamefont {J.}~\bibnamefont {Seidel}}, \bibinfo
  {author} {\bibfnamefont {L.~M.}\ \bibnamefont {Eng}}, \bibinfo {author}
  {\bibfnamefont {A.}~\bibnamefont {Bauer}}, \bibinfo {author} {\bibfnamefont
  {A.}~\bibnamefont {Chacon}}, \bibinfo {author} {\bibfnamefont
  {J.}~\bibnamefont {Kindervater}}, \bibinfo {author} {\bibfnamefont
  {S.}~\bibnamefont {M{\"u}hlbauer}}, \bibinfo {author} {\bibfnamefont
  {C.}~\bibnamefont {Pfleiderer}}, \bibinfo {author} {\bibfnamefont
  {S.}~\bibnamefont {Buhrandt}}, \bibinfo {author} {\bibfnamefont
  {C.}~\bibnamefont {Sch{\"u}tte}}, \ and\ \bibinfo {author} {\bibfnamefont
  {A.}~\bibnamefont {Rosch}},\ }\bibfield  {title} {\enquote {\bibinfo {title}
  {{Unwinding of a Skyrmion Lattice by Magnetic Monopoles}},}\ }\href {\doibase
  10.1126/science.1234657} {\bibfield  {journal} {\bibinfo  {journal}
  {Science}\ }\textbf {\bibinfo {volume} {340}},\ \bibinfo {pages} {1076--1080}
  (\bibinfo {year} {2013})}\BibitemShut {NoStop}%
\bibitem [{\citenamefont {P\"ollath}\ \emph {et~al.}(2017)\citenamefont
  {P\"ollath}, \citenamefont {Wild}, \citenamefont {Heinen}, \citenamefont
  {Meier}, \citenamefont {Kronseder}, \citenamefont {Tutsch}, \citenamefont
  {Bauer}, \citenamefont {Berger}, \citenamefont {Pfleiderer}, \citenamefont
  {Zweck}, \citenamefont {Rosch},\ and\ \citenamefont
  {Back}}]{2017:Poellath:PRL}%
  \BibitemOpen
  \bibfield  {author} {\bibinfo {author} {\bibfnamefont {S.}~\bibnamefont
  {P\"ollath}}, \bibinfo {author} {\bibfnamefont {J.}~\bibnamefont {Wild}},
  \bibinfo {author} {\bibfnamefont {L.}~\bibnamefont {Heinen}}, \bibinfo
  {author} {\bibfnamefont {T.~N.~G.}\ \bibnamefont {Meier}}, \bibinfo {author}
  {\bibfnamefont {M.}~\bibnamefont {Kronseder}}, \bibinfo {author}
  {\bibfnamefont {L.}~\bibnamefont {Tutsch}}, \bibinfo {author} {\bibfnamefont
  {A.}~\bibnamefont {Bauer}}, \bibinfo {author} {\bibfnamefont
  {H.}~\bibnamefont {Berger}}, \bibinfo {author} {\bibfnamefont
  {C.}~\bibnamefont {Pfleiderer}}, \bibinfo {author} {\bibfnamefont
  {J.}~\bibnamefont {Zweck}}, \bibinfo {author} {\bibfnamefont
  {A.}~\bibnamefont {Rosch}}, \ and\ \bibinfo {author} {\bibfnamefont {C.~H.}\
  \bibnamefont {Back}},\ }\bibfield  {title} {\enquote {\bibinfo {title}
  {Dynamical defects in rotating magnetic skyrmion lattices},}\ }\href
  {\doibase 10.1103/PhysRevLett.118.207205} {\bibfield  {journal} {\bibinfo
  {journal} {Phys. Rev. Lett.}\ }\textbf {\bibinfo {volume} {118}},\ \bibinfo
  {pages} {207205} (\bibinfo {year} {2017})}\BibitemShut {NoStop}%
\bibitem [{\citenamefont {Karube}\ \emph {et~al.}(2018)\citenamefont {Karube},
  \citenamefont {White}, \citenamefont {Morikawa}, \citenamefont {Dewhurst},
  \citenamefont {Cubitt}, \citenamefont {Kikkawa}, \citenamefont {Yu},
  \citenamefont {Tokunaga}, \citenamefont {Arima}, \citenamefont {Ronnow},
  \citenamefont {Tokura},\ and\ \citenamefont {Taguchi}}]{2018:Karube:SciAdv}%
  \BibitemOpen
  \bibfield  {author} {\bibinfo {author} {\bibfnamefont {K.}~\bibnamefont
  {Karube}}, \bibinfo {author} {\bibfnamefont {J.~S.}\ \bibnamefont {White}},
  \bibinfo {author} {\bibfnamefont {D.}~\bibnamefont {Morikawa}}, \bibinfo
  {author} {\bibfnamefont {C.~D.}\ \bibnamefont {Dewhurst}}, \bibinfo {author}
  {\bibfnamefont {R.}~\bibnamefont {Cubitt}}, \bibinfo {author} {\bibfnamefont
  {A.}~\bibnamefont {Kikkawa}}, \bibinfo {author} {\bibfnamefont
  {X.}~\bibnamefont {Yu}}, \bibinfo {author} {\bibfnamefont {Y.}~\bibnamefont
  {Tokunaga}}, \bibinfo {author} {\bibfnamefont {T.}~\bibnamefont {Arima}},
  \bibinfo {author} {\bibfnamefont {H.~M.}\ \bibnamefont {Ronnow}}, \bibinfo
  {author} {\bibfnamefont {Y.}~\bibnamefont {Tokura}}, \ and\ \bibinfo {author}
  {\bibfnamefont {Y.}~\bibnamefont {Taguchi}},\ }\bibfield  {title} {\enquote
  {\bibinfo {title} {{Disordered skymrion lattice phase stabilized by magnetic
  frustration in a chiral magnet}},}\ }\href {\doibase
  {10.1126/sciadv.aar7043}} {\bibfield  {journal} {\bibinfo  {journal} {Science
  Advances}\ }\textbf {\bibinfo {volume} {4}},\ \bibinfo {pages} {eaar7043}
  (\bibinfo {year} {2018})}\BibitemShut {NoStop}%
\bibitem [{\citenamefont {Schwarze}\ \emph {et~al.}(2015)\citenamefont
  {Schwarze}, \citenamefont {Waizner}, \citenamefont {Garst}, \citenamefont
  {Bauer}, \citenamefont {Stasinopoulos}, \citenamefont {Berger}, \citenamefont
  {Pfleiderer},\ and\ \citenamefont {Grundler}}]{2015:Schwarze:NatMater}%
  \BibitemOpen
  \bibfield  {author} {\bibinfo {author} {\bibfnamefont {T.}~\bibnamefont
  {Schwarze}}, \bibinfo {author} {\bibfnamefont {J.}~\bibnamefont {Waizner}},
  \bibinfo {author} {\bibfnamefont {M.}~\bibnamefont {Garst}}, \bibinfo
  {author} {\bibfnamefont {A.}~\bibnamefont {Bauer}}, \bibinfo {author}
  {\bibfnamefont {I.}~\bibnamefont {Stasinopoulos}}, \bibinfo {author}
  {\bibfnamefont {H.}~\bibnamefont {Berger}}, \bibinfo {author} {\bibfnamefont
  {C.}~\bibnamefont {Pfleiderer}}, \ and\ \bibinfo {author} {\bibfnamefont
  {D.}~\bibnamefont {Grundler}},\ }\bibfield  {title} {\enquote {\bibinfo
  {title} {{Universal helimagnon and skyrmion excitations in metallic,
  semiconducting and insulating chiral magnets}},}\ }\href {\doibase
  10.1038/nmat4223} {\bibfield  {journal} {\bibinfo  {journal} {Nature
  Materials}\ }\textbf {\bibinfo {volume} {14}},\ \bibinfo {pages} {478--483}
  (\bibinfo {year} {2015})}\BibitemShut {NoStop}%
\bibitem [{\citenamefont {Milde}\ \emph {et~al.}(2016)\citenamefont {Milde},
  \citenamefont {Neuber}, \citenamefont {Bauer}, \citenamefont {Pfleiderer},
  \citenamefont {Berger},\ and\ \citenamefont {Eng}}]{2016:Milde:NanoLett}%
  \BibitemOpen
  \bibfield  {author} {\bibinfo {author} {\bibfnamefont {P.}~\bibnamefont
  {Milde}}, \bibinfo {author} {\bibfnamefont {E.}~\bibnamefont {Neuber}},
  \bibinfo {author} {\bibfnamefont {A.}~\bibnamefont {Bauer}}, \bibinfo
  {author} {\bibfnamefont {C.}~\bibnamefont {Pfleiderer}}, \bibinfo {author}
  {\bibfnamefont {H.}~\bibnamefont {Berger}}, \ and\ \bibinfo {author}
  {\bibfnamefont {L.}~\bibnamefont {Eng}},\ }\bibfield  {title} {\enquote
  {\bibinfo {title} {{Heuristic description of magnetoelectricity of
  Cu$_{2}$OSeO$_{3}$}},}\ }\href@noop {} {\bibfield  {journal} {\bibinfo
  {journal} {Nano Lett.}\ }\textbf {\bibinfo {volume} {16}},\ \bibinfo {pages}
  {5612} (\bibinfo {year} {2016})}\BibitemShut {NoStop}%
\bibitem [{\citenamefont {Zhang}\ \emph
  {et~al.}(2016{\natexlab{a}})\citenamefont {Zhang}, \citenamefont {Bauer},
  \citenamefont {Burn}, \citenamefont {Milde}, \citenamefont {Neuber},
  \citenamefont {Eng}, \citenamefont {Berger}, \citenamefont {Pfleiderer},
  \citenamefont {van~der Laan},\ and\ \citenamefont
  {Hesjedal}}]{2016:Zhang:NanoLett}%
  \BibitemOpen
  \bibfield  {author} {\bibinfo {author} {\bibfnamefont {S.~L.}\ \bibnamefont
  {Zhang}}, \bibinfo {author} {\bibfnamefont {A.}~\bibnamefont {Bauer}},
  \bibinfo {author} {\bibfnamefont {D.~M.}\ \bibnamefont {Burn}}, \bibinfo
  {author} {\bibfnamefont {P.}~\bibnamefont {Milde}}, \bibinfo {author}
  {\bibfnamefont {E.}~\bibnamefont {Neuber}}, \bibinfo {author} {\bibfnamefont
  {L.~M.}\ \bibnamefont {Eng}}, \bibinfo {author} {\bibfnamefont
  {H.}~\bibnamefont {Berger}}, \bibinfo {author} {\bibfnamefont
  {C.}~\bibnamefont {Pfleiderer}}, \bibinfo {author} {\bibfnamefont
  {G.}~\bibnamefont {van~der Laan}}, \ and\ \bibinfo {author} {\bibfnamefont
  {T.}~\bibnamefont {Hesjedal}},\ }\bibfield  {title} {\enquote {\bibinfo
  {title} {{Multidomain Skyrmion Lattice State in Cu$_{2}$OSeO$_{3}$}},}\
  }\href {\doibase 10.1021/acs.nanolett.6b00845} {\bibfield  {journal}
  {\bibinfo  {journal} {Nano Lett.}\ }\textbf {\bibinfo {volume} {16}},\
  \bibinfo {pages} {3285} (\bibinfo {year} {2016}{\natexlab{a}})}\BibitemShut
  {NoStop}%
\bibitem [{\citenamefont {Zhang}\ \emph
  {et~al.}(2016{\natexlab{b}})\citenamefont {Zhang}, \citenamefont {Bauer},
  \citenamefont {Berger}, \citenamefont {Pfleiderer}, \citenamefont {van~der
  Laan},\ and\ \citenamefont {Hesjedal}}]{2016:Zhang:PhysRevB}%
  \BibitemOpen
  \bibfield  {author} {\bibinfo {author} {\bibfnamefont {S.~L.}\ \bibnamefont
  {Zhang}}, \bibinfo {author} {\bibfnamefont {A.}~\bibnamefont {Bauer}},
  \bibinfo {author} {\bibfnamefont {H.}~\bibnamefont {Berger}}, \bibinfo
  {author} {\bibfnamefont {C.}~\bibnamefont {Pfleiderer}}, \bibinfo {author}
  {\bibfnamefont {G.}~\bibnamefont {van~der Laan}}, \ and\ \bibinfo {author}
  {\bibfnamefont {T.}~\bibnamefont {Hesjedal}},\ }\bibfield  {title} {\enquote
  {\bibinfo {title} {{Resonant elastic x-ray scattering from the skyrmion
  lattice in Cu$_{2}$OSeO$_{3}$}},}\ }\href {\doibase
  10.1103/PhysRevB.93.214420} {\bibfield  {journal} {\bibinfo  {journal} {Phys.
  Rev. B}\ }\textbf {\bibinfo {volume} {93}},\ \bibinfo {pages} {214420}
  (\bibinfo {year} {2016}{\natexlab{b}})}\BibitemShut {NoStop}%
\bibitem [{\citenamefont {Stasinopoulos}\ \emph {et~al.}({2017})\citenamefont
  {Stasinopoulos}, \citenamefont {Weichselbaumer}, \citenamefont {Bauer},
  \citenamefont {Waizner}, \citenamefont {Berger}, \citenamefont {Maendl},
  \citenamefont {Garst}, \citenamefont {Pfleiderer},\ and\ \citenamefont
  {Grundler}}]{2017:Stasinopoulos:APL}%
  \BibitemOpen
  \bibfield  {author} {\bibinfo {author} {\bibfnamefont {I.}~\bibnamefont
  {Stasinopoulos}}, \bibinfo {author} {\bibfnamefont {S.}~\bibnamefont
  {Weichselbaumer}}, \bibinfo {author} {\bibfnamefont {A.}~\bibnamefont
  {Bauer}}, \bibinfo {author} {\bibfnamefont {J.}~\bibnamefont {Waizner}},
  \bibinfo {author} {\bibfnamefont {H.}~\bibnamefont {Berger}}, \bibinfo
  {author} {\bibfnamefont {S.}~\bibnamefont {Maendl}}, \bibinfo {author}
  {\bibfnamefont {M.}~\bibnamefont {Garst}}, \bibinfo {author} {\bibfnamefont
  {C.}~\bibnamefont {Pfleiderer}}, \ and\ \bibinfo {author} {\bibfnamefont
  {D.}~\bibnamefont {Grundler}},\ }\bibfield  {title} {\enquote {\bibinfo
  {title} {{Low spin wave damping in the insulating chiral magnet
  Cu$_2$OSeO$_3$}},}\ }\href {\doibase {10.1063/1.4995240}} {\bibfield
  {journal} {\bibinfo  {journal} {{Appl. Phys. Lett.}}\ }\textbf {\bibinfo
  {volume} {{111}}} (\bibinfo {year} {{2017}}),\
  {10.1063/1.4995240}}\BibitemShut {NoStop}%
\bibitem [{\citenamefont {Mydosh}(1993)}]{mydosh:1993}%
  \BibitemOpen
  \bibfield  {author} {\bibinfo {author} {\bibfnamefont {J.~A.}\ \bibnamefont
  {Mydosh}},\ }\href@noop {} {\emph {\bibinfo {title} {Spin glasses: an
  experimental introduction}}}\ (\bibinfo  {publisher} {Taylor \& Francis},\
  \bibinfo {year} {1993})\BibitemShut {NoStop}%
\bibitem [{\citenamefont {M{\"u}hlbauer}\ \emph {et~al.}(2016)\citenamefont
  {M{\"u}hlbauer}, \citenamefont {Heinemann}, \citenamefont {Wilhelm},
  \citenamefont {Karge}, \citenamefont {Ostermann}, \citenamefont {Defendi},
  \citenamefont {Schreyer}, \citenamefont {Petry},\ and\ \citenamefont
  {Gilles}}]{muhlbauer:NIaMiPRSAASDaAE:16}%
  \BibitemOpen
  \bibfield  {author} {\bibinfo {author} {\bibfnamefont {S.}~\bibnamefont
  {M{\"u}hlbauer}}, \bibinfo {author} {\bibfnamefont {A.}~\bibnamefont
  {Heinemann}}, \bibinfo {author} {\bibfnamefont {A.}~\bibnamefont {Wilhelm}},
  \bibinfo {author} {\bibfnamefont {L.}~\bibnamefont {Karge}}, \bibinfo
  {author} {\bibfnamefont {A.}~\bibnamefont {Ostermann}}, \bibinfo {author}
  {\bibfnamefont {I.}~\bibnamefont {Defendi}}, \bibinfo {author} {\bibfnamefont
  {A.}~\bibnamefont {Schreyer}}, \bibinfo {author} {\bibfnamefont
  {W.}~\bibnamefont {Petry}}, \ and\ \bibinfo {author} {\bibfnamefont
  {R.}~\bibnamefont {Gilles}},\ }\bibfield  {title} {\enquote {\bibinfo {title}
  {The new small-angle neutron scattering instrument {{SANS}}-1 at
  {{MLZ}}\textemdash{}characterization and first results},}\ }\href {\doibase
  10.1016/j.nima.2016.06.105} {\bibfield  {journal} {\bibinfo  {journal}
  {Nuclear Instruments and Methods in Physics Research Section A: Accelerators,
  Spectrometers, Detectors and Associated Equipment}\ }\textbf {\bibinfo
  {volume} {832}},\ \bibinfo {pages} {297--305} (\bibinfo {year}
  {2016})}\BibitemShut {NoStop}%
\bibitem [{\citenamefont {Aharoni}(1998)}]{1998:Aharoni:JApplPhys}%
  \BibitemOpen
  \bibfield  {author} {\bibinfo {author} {\bibfnamefont {A.}~\bibnamefont
  {Aharoni}},\ }\bibfield  {title} {\enquote {\bibinfo {title} {{Demagnetizing
  factors for rectangular ferromagnetic prisms}},}\ }\href {\doibase
  10.1063/1.367113} {\bibfield  {journal} {\bibinfo  {journal} {J. Appl.
  Phys.}\ }\textbf {\bibinfo {volume} {83}},\ \bibinfo {pages} {3432} (\bibinfo
  {year} {1998})}\BibitemShut {NoStop}%
\bibitem [{\citenamefont {Youssif}\ \emph {et~al.}(2000)\citenamefont
  {Youssif}, \citenamefont {A.},\ and\ \citenamefont {A.}}]{2000:youssif}%
  \BibitemOpen
  \bibfield  {author} {\bibinfo {author} {\bibfnamefont {M.~I.}\ \bibnamefont
  {Youssif}}, \bibinfo {author} {\bibfnamefont {Bahgat~A.}\ \bibnamefont {A.}},
  \ and\ \bibinfo {author} {\bibfnamefont {Ali~I.}\ \bibnamefont {A.}},\
  }\bibfield  {title} {\enquote {\bibinfo {title} {{AC Magnetic Susceptibility
  Technique for the Characterization of High Temperature Superconductors}},}\
  }\href@noop {} {\bibfield  {journal} {\bibinfo  {journal} {Egypt. J. Sol.}\
  }\textbf {\bibinfo {volume} {23}},\ \bibinfo {pages} {231} (\bibinfo {year}
  {2000})}\BibitemShut {NoStop}%
\bibitem [{\citenamefont {Bauer}\ \emph {et~al.}(2017)\citenamefont {Bauer},
  \citenamefont {Chacon}, \citenamefont {Wagner}, \citenamefont {Halder},
  \citenamefont {Georgii}, \citenamefont {Rosch}, \citenamefont {Pfleiderer},\
  and\ \citenamefont {Garst}}]{2017:Bauer:PRB}%
  \BibitemOpen
  \bibfield  {author} {\bibinfo {author} {\bibfnamefont {A.}~\bibnamefont
  {Bauer}}, \bibinfo {author} {\bibfnamefont {A.}~\bibnamefont {Chacon}},
  \bibinfo {author} {\bibfnamefont {M.}~\bibnamefont {Wagner}}, \bibinfo
  {author} {\bibfnamefont {M.}~\bibnamefont {Halder}}, \bibinfo {author}
  {\bibfnamefont {R.}~\bibnamefont {Georgii}}, \bibinfo {author} {\bibfnamefont
  {A.}~\bibnamefont {Rosch}}, \bibinfo {author} {\bibfnamefont
  {C.}~\bibnamefont {Pfleiderer}}, \ and\ \bibinfo {author} {\bibfnamefont
  {M.}~\bibnamefont {Garst}},\ }\bibfield  {title} {\enquote {\bibinfo {title}
  {Symmetry breaking, slow relaxation dynamics, and topological defects at the
  field-induced helix reorientation in mnsi},}\ }\href {\doibase
  10.1103/PhysRevB.95.024429} {\bibfield  {journal} {\bibinfo  {journal} {Phys.
  Rev. B}\ }\textbf {\bibinfo {volume} {95}},\ \bibinfo {pages} {024429}
  (\bibinfo {year} {2017})}\BibitemShut {NoStop}%
\bibitem [{\citenamefont {Balanda}(2013)}]{2013:balanda}%
  \BibitemOpen
  \bibfield  {author} {\bibinfo {author} {\bibfnamefont {M.}~\bibnamefont
  {Balanda}},\ }\bibfield  {title} {\enquote {\bibinfo {title} {{AAC
  Susceptibility Studies of Phase Transitions and Magnetic Relaxation:
  Conventional, Molecular and Low-Dimensional Magnets}},}\ }\href@noop {}
  {\bibfield  {journal} {\bibinfo  {journal} {Acta Physics Polonica A}\
  }\textbf {\bibinfo {volume} {124}},\ \bibinfo {pages} {964} (\bibinfo {year}
  {2013})}\BibitemShut {NoStop}%
\bibitem [{\citenamefont {Wild}\ \emph {et~al.}(2017)\citenamefont {Wild},
  \citenamefont {Meier}, \citenamefont {Poellath}, \citenamefont {Kronseder},
  \citenamefont {Bauer}, \citenamefont {Chacon}, \citenamefont {Halder},
  \citenamefont {Schowalter}, \citenamefont {Rosenauer}, \citenamefont {Zweck},
  \citenamefont {M\"uller}, \citenamefont {Rosch}, \citenamefont {Pfleiderer},\
  and\ \citenamefont {Back}}]{2017:Wild:SciAdv}%
  \BibitemOpen
  \bibfield  {author} {\bibinfo {author} {\bibfnamefont {J.}~\bibnamefont
  {Wild}}, \bibinfo {author} {\bibfnamefont {T.N.G.}\ \bibnamefont {Meier}},
  \bibinfo {author} {\bibfnamefont {S.}~\bibnamefont {Poellath}}, \bibinfo
  {author} {\bibfnamefont {M.}~\bibnamefont {Kronseder}}, \bibinfo {author}
  {\bibfnamefont {A.}~\bibnamefont {Bauer}}, \bibinfo {author} {\bibfnamefont
  {A.}~\bibnamefont {Chacon}}, \bibinfo {author} {\bibfnamefont
  {M.}~\bibnamefont {Halder}}, \bibinfo {author} {\bibfnamefont
  {M.}~\bibnamefont {Schowalter}}, \bibinfo {author} {\bibfnamefont
  {A.}~\bibnamefont {Rosenauer}}, \bibinfo {author} {\bibfnamefont
  {J.}~\bibnamefont {Zweck}}, \bibinfo {author} {\bibfnamefont
  {J.}~\bibnamefont {M\"uller}}, \bibinfo {author} {\bibfnamefont
  {A.}~\bibnamefont {Rosch}}, \bibinfo {author} {\bibfnamefont
  {C.}~\bibnamefont {Pfleiderer}}, \ and\ \bibinfo {author} {\bibfnamefont
  {C.~H.}\ \bibnamefont {Back}},\ }\bibfield  {title} {\enquote {\bibinfo
  {title} {Entropy-limited topological protection of skyrmions},}\ }\href@noop
  {} {\bibfield  {journal} {\bibinfo  {journal} {Science Advances}\ }\textbf
  {\bibinfo {volume} {3}},\ \bibinfo {pages} {e1701704} (\bibinfo {year}
  {2017})}\BibitemShut {NoStop}%
\bibitem [{201(2018)}]{2018:Qian:arxiv}%
  \BibitemOpen
  \href@noop {} {} (\bibinfo {year} {2018}),\ \bibinfo {note} {{F. Qian, L.J.
  Bannenberg, H. Wilhelm, G. Chaboussant, L. M. Debeer-Schmitt, M. P. Schmidt,
  A. Aqeel, T.T.M. Palstra, E. Br\"uck, A.J.E. Lefering, C. Pappas, M.
  Mostovoy, and A. O. Leonov, arXiv/1802.02070v2, unpublished.}}\BibitemShut
  {Stop}%
\bibitem [{\citenamefont {Akulov}(1937)}]{1937:Akulov}%
  \BibitemOpen
  \bibfield  {author} {\bibinfo {author} {\bibfnamefont {N.}~\bibnamefont
  {Akulov}},\ }\href@noop {} {\bibfield  {journal} {\bibinfo  {journal} {Z.
  Phys.}\ }\textbf {\bibinfo {volume} {100}},\ \bibinfo {pages} {197} (\bibinfo
  {year} {1937})}\BibitemShut {NoStop}%
\bibitem [{\citenamefont {Yang}\ \emph {et~al.}({2012})\citenamefont {Yang},
  \citenamefont {Li}, \citenamefont {Lu}, \citenamefont {Whangbo},
  \citenamefont {Wei}, \citenamefont {Gong},\ and\ \citenamefont
  {Xiang}}]{2012:Yang:PRL}%
  \BibitemOpen
  \bibfield  {author} {\bibinfo {author} {\bibfnamefont {J.~H.}\ \bibnamefont
  {Yang}}, \bibinfo {author} {\bibfnamefont {Z.~L.}\ \bibnamefont {Li}},
  \bibinfo {author} {\bibfnamefont {X.~Z.}\ \bibnamefont {Lu}}, \bibinfo
  {author} {\bibfnamefont {M.~H.}\ \bibnamefont {Whangbo}}, \bibinfo {author}
  {\bibfnamefont {Su-Huai}\ \bibnamefont {Wei}}, \bibinfo {author}
  {\bibfnamefont {X.~G.}\ \bibnamefont {Gong}}, \ and\ \bibinfo {author}
  {\bibfnamefont {H.~J.}\ \bibnamefont {Xiang}},\ }\bibfield  {title} {\enquote
  {\bibinfo {title} {{Strong Dzyaloshinskii-Moriya Interaction and Origin of
  Ferroelectricity in Cu$_2$OSeO$_3$}},}\ }\href {\doibase
  {10.1103/PhysRevLett.109.107203}} {\bibfield  {journal} {\bibinfo  {journal}
  {{Phys. Rev. Lett.}}\ }\textbf {\bibinfo {volume} {{109}}} (\bibinfo {year}
  {{2012}}),\ {10.1103/PhysRevLett.109.107203}}\BibitemShut {NoStop}%
\bibitem [{\citenamefont {Grigoriev}\ \emph {et~al.}(2015)\citenamefont
  {Grigoriev}, \citenamefont {Sukhanov},\ and\ \citenamefont
  {Maleyev}}]{2015:Grigoriev:PRB}%
  \BibitemOpen
  \bibfield  {author} {\bibinfo {author} {\bibfnamefont {S.~V.}\ \bibnamefont
  {Grigoriev}}, \bibinfo {author} {\bibfnamefont {A.~S.}\ \bibnamefont
  {Sukhanov}}, \ and\ \bibinfo {author} {\bibfnamefont {S.~V.}\ \bibnamefont
  {Maleyev}},\ }\bibfield  {title} {\enquote {\bibinfo {title} {From spiral to
  ferromagnetic structure in {B20} compounds: Role of cubic anisotropy},}\
  }\href {\doibase 10.1103/PhysRevB.91.224429} {\bibfield  {journal} {\bibinfo
  {journal} {Phys. Rev. B}\ }\textbf {\bibinfo {volume} {91}},\ \bibinfo
  {pages} {224429} (\bibinfo {year} {2015})}\BibitemShut {NoStop}%
\end{thebibliography}


%

\end{document}